\documentclass[aps,prD,reprint,groupedaddress]{revtex4-2}

\usepackage{siunitx}
\usepackage{subcaption}
\usepackage{graphicx}% Include figure files
\usepackage{dcolumn}% Align table columns on decimal point
\usepackage{bm}% bold math
\usepackage{xcolor}
\usepackage{placeins}
\usepackage{comment}
\usepackage{float}
\definecolor{darkgreen}{RGB}{0,100,0}
\definecolor{darkred}{RGB}{139,0,0}
\definecolor{darkorange}{RGB}{255,140,0}

\definecolor{viridis1}{RGB}{70,12,94}
\definecolor{viridis2}{RGB}{32,144,140}
\definecolor{orange}{RGB}{255,128,0}

\usepackage{hyperref}
\hypersetup{
    colorlinks=true,
    linkcolor=blue,
    filecolor=blue,      
    urlcolor=blue,
    citecolor=blue,}

\definecolor{myblue}{rgb}{0.267, 0.005, 0.329}
\definecolor{mybrown}{rgb}{0.545, 0.271, 0.075}
\definecolor{mygreen}{rgb}{0.678, 0.864, 0.190}
\definecolor{myyellow}{RGB}{200,160,0}

\usepackage{orcidlink}

\begin{document}

%\preprint{APS/123-QED}

\title{\textbf{Analysis of LIGO glitches during the first part of the fourth LIGO-Virgo-KAGRA observing run} 
}% 
\author{Tabata Aira Ferreira\,\orcidlink{0000-0002-1166-2005}}
\author{Gabriela González\,\orcidlink{0000-0003-0199-3158}}
\author{Osvaldo Salas\,\orcidlink{0009-0003-3915-8147}}

\affiliation{Louisiana State University, Baton Rouge, LA 70803, USA}



\begin{abstract}
This paper presents an analysis of noise transients observed in LIGO data during the first part of the fourth observing run, using the unsupervised machine learning technique t-distributed Stochastic Neighbor Embedding (t-SNE) to examine the behavior of glitch groups. Based on the t-SNE output, we apply Agglomerative Clustering in combination with the Silhouette Score to determine the optimal number of groups. We then track these groups over time and investigate correlations between their occurrence and environmental or instrumental conditions. At the Livingston observatory, the most common glitches during O4a were seasonal and associated with ground motion, whereas at Hanford, the most prevalent glitches were related to instrumental conditions.
\end{abstract}

\maketitle

%\tableofcontents

\section{Introduction}
Gravitational waves, predicted by Einstein's general theory of relativity, have revolutionized our understanding of astrophysical phenomena by providing a new observational window into the Universe. Since the first direct detection of gravitational waves in 2015~\cite{LIGOScientific:2014pky}, ground-based detectors have been actively searching for gravitational-wave signals during dedicated observing runs. Instruments such as the Laser Interferometer Gravitational-Wave Observatory (LIGO)~\cite{ligo2015advanced}, Virgo in Italy~\cite{VIRGO:2014yos}, and KAGRA in Japan~\cite{akutsu2021overview} operate as part of a global network. Together, they form the LVK collaboration, bringing together a large community of researchers working toward the common goal of detecting an increasing number of gravitational-wave events.

LIGO consists of two observatories designed to detect these spacetime ripples: one located in Hanford, Washington (LHO), and the other in Livingston, Louisiana (LLO). The detectors operate continuously during observing runs to collect data for gravitational-wave searches, with planned commissioning breaks between runs to upgrade instrumentation and improve sensitivity~\cite{capote2025advanced}. The first detection occurred during the first observing run (O1), which lasted from September 12, 2015, to January 2016. On January 16, 2024, LIGO concluded the first part of its fourth observing run (O4a), which had started on May 24, 2023. The fourth observing run (O4) is now complete. As registered in the Gravitational-Wave Candidate Event Database (GraceDB)~\cite{GraceDB}, more than two hundred significant detection candidates have been identified during O4, produced by the mergers of compact objects such as black holes and neutron stars.

LIGO also records various noise transients arising from instrumental or environmental factors, including ground motion, wind, and other disturbances~\cite{davis2021ligo, soni2025ligo}. These non-astrophysical transient signals are commonly referred to as \textit{glitches}. Glitches can exhibit a wide range of durations, from milliseconds to several seconds, and frequencies across the sensitive band of the detector. In this work, we focus on transients with peak frequencies between 10 and 2048~Hz.

Glitches are non-stationary artifacts that contribute to the background in gravitational-wave searches~\cite{glanzer2023data, guimaraes2023bridging, yarbrough2025pinch} and, in some cases, can closely mimic signals from compact binary coalescences (CBC). To mitigate their impact, data quality (DQ) procedures are first applied to identify and exclude contaminated data segments~\cite{davis2021ligo}. Efforts within CBC searches are also essential and rely on different pipelines, such as PyCBC~\cite{usman2016pycbc, dal2021real}, GstLAL~\cite{messick2017analysis, sachdev2019gstlal}, MBTA~\cite{adams2016low, aubin2021mbta}, and SPIIR~\cite{hooper2012summed,chu2017low, chu2022spiir}, which incorporate likelihood weighting and other strategies to reduce the impact of glitches on search sensitivity.  

Glitch morphologies are typically analyzed using spectrograms~\cite{chatterji2004multiresolution}, which facilitate their classification into distinct glitch groups. Different tools have been developed to analyze these unwanted signals~\cite{cornish2015bayeswave, chatterjee2025no, lopez2025simulation, PhysRevD.97.101501, powell2017classification,chan2024gwskynet,razzano2023gwitchhunters, alvarez2024gspynettree}. One example is Gravity Spy, an interdisciplinary project that integrates machine learning, citizen science, and input from LIGO scientists to classify glitches~\cite{zevin2024gravity, zevin2017gravity}. These classes are named based on their morphology and/or known or suspected physical origin. Examples of different spectrogram morphologies can be found in~\cite{zevin2017gravity, glanzer2023data, zoogravityspy,cabero2019blip}. %Among the identified classes are \textit{Power Line}, associated with electrical noise from the 60 Hz alternating current in the United States; \textit{Extremely Loud}, referring to high-amplitude glitches potentially related to scattered light in the detectors~\cite{alog72968,alog73447}; and \textit{Blip}, a short-duration glitch whose physical origin remains unclear~\cite{cabero2019blip}.

The main motivation behind these efforts is to support the overarching goal of glitch studies: identifying their physical origins to eliminate or mitigate their impact on gravitational-wave detection. Here, we analyze the main glitch groups observed during O4a and investigate potential correlations with environmental and/or instrumental variations, following the methodology detailed in~\cite{ferreira2025using}.

To identify patterns among glitches, we use information provided by \textit{Omicron}~\cite{robinet2016omicron, robinet2020omicron}, the primary tool used in LIGO to detect noise transients in detector data. Omicron uses the $Q$-transform~\cite{brown1991calculation} to decompose the data into time-frequency tiles, each characterized by its central time, central frequency, and quality factor ($Q$). Tiles exhibiting excess power above a predefined signal-to-noise ratio (SNR) threshold are recorded as unclustered triggers. The Omicron configuration specifies a frequency range of 4--8192 Hz and a $Q$ range of 3.3166--150. Each plane has its own logarithmically spaced frequency grid, with smaller spacing at higher $Q$ values. Unclustered triggers can be grouped based on their proximity in time to form clustered triggers representing individual transient events. Each clustered trigger is represented by the tile with the highest SNR. We use clustered triggers to characterize the overall glitch population and unclustered triggers to construct a high-dimensional representation of each glitch. After preparing these data, we employ t-distributed Stochastic Neighbor Embedding (t-SNE)~\cite{van2008visualizing}, a dimensionality reduction technique, to project the high-dimensional representation into a two-dimensional space.

As an unsupervised technique, t-SNE does not require prior knowledge or a training dataset. Instead, it relies on probability-based similarity between data points to preserve local relative distances in the lower-dimensional representation. This makes t-SNE particularly useful for uncovering patterns in data without predefined assumptions. The resulting projections often reveal meaningful structures; as shown in~\cite{ferreira2025using, ferreira2022comparison}, cross-checking the clusters obtained from t-SNE with Gravity Spy labels demonstrates that, when applied to unclustered Omicron information, t-SNE produces groups that closely correspond to known glitch classes. 

In this work, the primary objective is to track the temporal evolution of glitches during the O4a observing run. To facilitate this analysis, we first apply t-SNE to project the high-dimensional data into a two-dimensional space and then use \textit{Agglomerative Clustering}~\cite{xu2005survey} to group similar glitches and assign labels. The \textit{Silhouette Score}~\cite{rousseeuw1987silhouettes} is used to evaluate clustering quality and determine the optimal number of groups. These group labels serve primarily as a means to organize and visualize the temporal behavior of distinct glitch populations, to support their possible association with environmental or instrumental conditions, and to enable the classification of previously unseen data. This approach is entirely unsupervised and based solely on features derived from Omicron triggers. Further details about the method can be found in~\cite{ferreira2025using}.

The paper is structured as follows. Section~\ref{sec:o4a_glitches_llo} presents the glitch population observed at LLO during O4a, together with an analysis of ground motion across different frequency bands. The section also evaluates the correlation between ground motion and glitch rate, highlighting how the low-frequency glitches impacted the interferometer during the analyzed observing period. A parallel analysis for LHO is provided in Section~\ref{sec:o4a_glitches_lho}, focusing on the evolution of broadband glitches during O4a, which were primarily associated with internal instrumentation couplings rather than environmental factors. Finally, Section~\ref{sec:conclusions} summarizes and discusses the key findings of this work.

\section{Glitches during O4a at LLO}
\label{sec:o4a_glitches_llo}

%The interferometer was notably affected by glitches at lower frequencies, as illustrated in Figure~\ref{fig}.

The rate of glitches with a signal-to-noise ratio (SNR) exceeding 10 at LLO was higher during O4a (approximately 31 glitches per hour) compared to the third observing run, O3 (approximately 26 glitches per hour)\cite{soni2025ligo}. The glitch rate during O3 was significantly influenced by the prevalence of scattering glitches, which dominated the transient population in that period. Scattering glitches occur when stray light couples back into the main beam after interacting with surfaces, whose identification is particularly challenging. Substantial efforts during O3 successfully identified and mitigated some of these scattering surfaces, resulting in a considerable reduction in the occurrence of such glitches~\cite{soni2020reducing, soni2024modeling}.

\begin{figure}[!t]
    \centering    \includegraphics[width=\columnwidth]{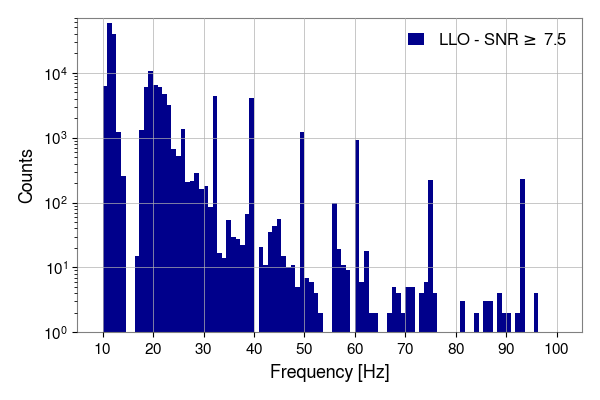}
    \caption{Histogram of the peak frequencies of clustered Omicron triggers during O4a at LLO, with SNR $\geq7.5$}. This shows the range where we observed the most common glitches at LLO during O4a, which occurred between 10 and \SI{100}{\hertz}.
    \label{fig:freq_hist_L1}
\end{figure}

Despite these mitigation efforts, the transient rate during O4a surpassed that of O3; this increase was primarily driven by glitches occurring at low frequencies. Fig.~\ref{fig:freq_hist_L1} illustrates the distribution of glitch frequencies below 100~Hz at LLO during O4a, highlighting this trend. The apparent gap in glitches around \SI{14}{\hertz} is due to masking by noise from the test masses’ roll modes and calibration lines at those frequencies. Additional peaks are also visible in the histogram, notably around 32, 39, 49, 61, 75, and 93 Hz. We compare the Gravity Spy classes~\cite{zevin2017gravity} in each of these frequency bins, and spectrogram analysis shows that the \textit{Tomte} class accounts for more than 50\% of glitches with peak frequencies near 32 and 39 Hz, while the 49 Hz peak consists of approximately 50\% \textit{Blip Low Frequency}. At 61 Hz, the dominant class is \textit{Blip Low Frequency}; the peak around 75 Hz is primarily associated with \textit{Koi Fish}, and the 93 Hz region includes a mix of \textit{Koi Fish} and other classes. Some classes can share similar peak frequencies, but due to the complex nature of glitches, peak frequency alone is usually insufficient to distinguish between certain groups. This underscores the importance of analyzing glitches based on their morphology.

%Besides the high concentration of glitches at lower frequencies, additional peaks can also be observed in the histogram, notably around 32, 39, 49, 61, 75 and 93 Hz. After analyzing their spectrograms, we find that the \textit{Tomte} class accounts for more than 50\% of the glitches with peak frequencies near 32 and 39 Hz, followed by the \textit{Extremely Loud}, \textit{Koi Fish}, and \textit{Blip Low Frequency} classes. The 49 Hz peak consists of approximately 50\% \textit{Blip Low Frequency}, with \textit{Extremely Loud} and \textit{Tomte} making up the remainder. At 61 Hz, the dominant class is \textit{Blip Low Frequency}. The peak around 75 Hz is primarily associated with \textit{Koi Fish}, while the 93 Hz region includes a mix of \textit{Koi Fish} and other classes. Some classes may often have similar peak frequencies, but due to the complex nature of glitches, peak frequency alone is usually not sufficient to distinguish between certain groups. This underscores the importance of analyzing glitches based on their morphology.

\begin{figure}[!htbp]
    \centering
    \includegraphics[width=0.95\columnwidth]{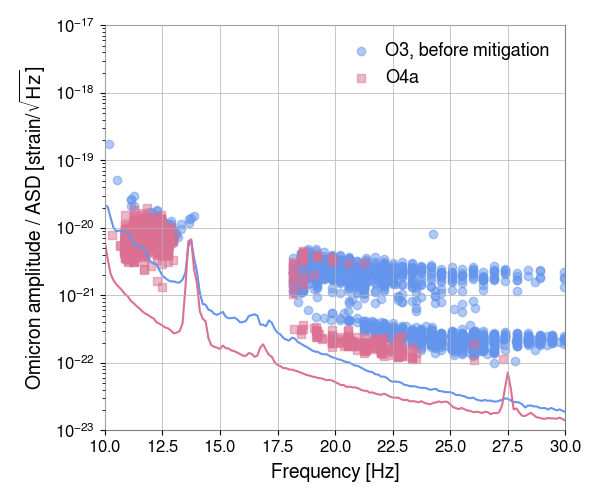}
    \caption{Amplitude spectral densities (ASDs) during O3 in blue and O4a in pink, measured under similar microseismic motion amplitudes (or velocities). The blue circles represent Omicron scattering-glitch amplitudes during O3, while the pink squares represent those during O4a. The reduction in scattering glitches in the 18--30~Hz frequency band during O4a is associated with scattering mitigation~\cite{soni2020reducing}. 
    The ASD curves are shown as a reference for the detector noise level; both the ASD and Omicron amplitudes are expressed in strain/$\sqrt{\mathrm{Hz}}$, but are not directly equivalent. Their ratio is approximately proportional to the trigger SNR, assuming a nearly constant noise spectral density over the corresponding frequency band.
}
    \label{fig:L1sensitivities}
\end{figure}

The increase in the number of detected glitches at lower frequencies may be partly associated with the improved sensitivity of the interferometer in this frequency range~\cite{soni2025ligo}. Figure~\ref{fig:L1sensitivities} illustrates this by comparing the amplitude spectral density (ASD) at low frequencies (10--30~Hz) during O3, shown in blue, with that of O4a, shown in pink. The spectral features between approximately 13 and 17 Hz are associated with known instrumental lines, including test-mass roll modes and calibration lines, while those around 27--28 Hz are associated with suspension bounce modes~\cite{goetz2024o4alines}. The blue dots represent the amplitudes reported by Omicron for scattering glitches over an eight-hour period on December 27, 2019 (during O3). The Omicron trigger amplitude is calculated from the SNR and a band-dependent noise normalization associated with the Q-transform. Dimensional analysis of the Omicron implementation indicates that these amplitudes have the same units as the ASD (strain/$\sqrt{\mathrm{Hz}}$). The ASD curves are therefore included as a reference for the detector noise level, rather than as quantities directly equivalent to the glitch amplitudes. Since these glitches were associated with relatively high ground motion, we selected a day during O4a with similar ground motion amplitude, and also plotted the corresponding glitches, represented by pink squares. There are fewer glitches in O4 between 18~Hz and 30~Hz due to the scattering mitigation~\cite{soni2020reducing}, but there is a higher rate in O4 at lower frequencies. While some high-amplitude glitches between 10 and 13 Hz were detectable during O3, the improved low-frequency sensitivity during O4a may have enabled the detection of additional lower-frequency glitches that would have been more difficult to identify during O3.

%Figures~\ref{fig:spect_o3}) and~\ref{fig:spect_o4}) show examples of low-frequency glitch spectrograms from O3 and O4a, respectively, revealing remarkably similar features. However, due to the limited number of such glitches observed during O3, it is not possible to conclusively determine whether they came from the same origin. The sensitivity of the interferometer during O3 was insufficient to detect a greater number of these glitches unless they were exceptionally strong. Nevertheless, the possibility that these glitches originate from the same source and were present during O3, but remained undetected due to the interferometer's lower sensitivity, cannot be excluded.

%Although it is understood that the high number of glitches is associated with improved sensitivity at low frequencies

%One of the primary questions remains to identify the origins of these glitches. 

\begin{figure}[t]
    \centering
    \includegraphics[width=\columnwidth]{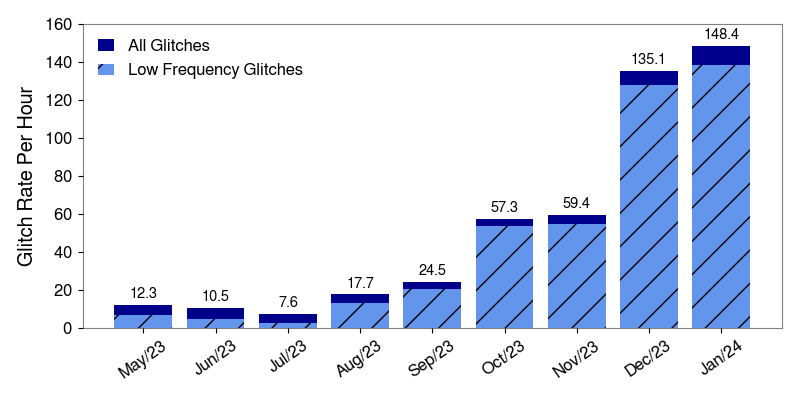}
    \caption{Average hourly glitch rates for each month during O4a, computed over observing time. Dark blue bars, annotated with numerical values, represent the total rate of clustered Omicron triggers with frequencies between \SI{10} and \SI{2048}{\hertz} and a minimum SNR of 7.5. Light blue bars with hatching indicate the subset of glitches with frequencies below \SI{25}{\hertz}.}
    \label{fig:glitches_per_hour_per_month_l1}
\end{figure}

Figure~\ref{fig:glitches_per_hour_per_month_l1} presents the average number of glitches per hour for each month during O4a, represented by the dark blue bars, considering glitches in the 10 to 2048~Hz range with a minimum SNR of 7.5. The glitch rate exhibited a general increase over time, primarily driven by the low-frequency (below 25~Hz) glitch rate, represented by the light blue bars with diagonal hatching.

\begin{figure*}[ht!]
    \begin{subfigure}{0.32\textwidth}
        \centering
        \caption{May}
        \includegraphics[width=\textwidth]{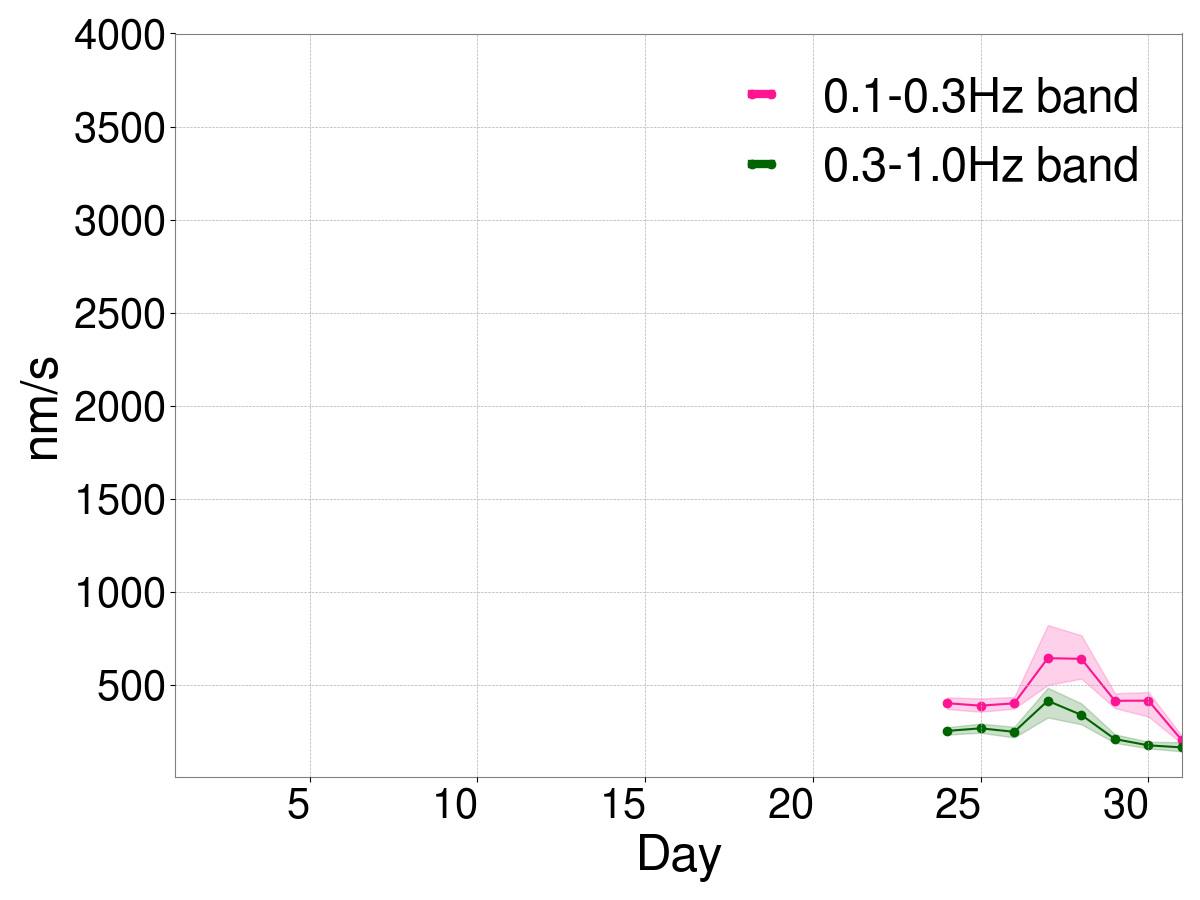}
        \label{fig:may_l1_groundmotion}
    \end{subfigure}
    \begin{subfigure}{0.32\textwidth}
        \centering
        \caption{June}
        \includegraphics[width=\textwidth]{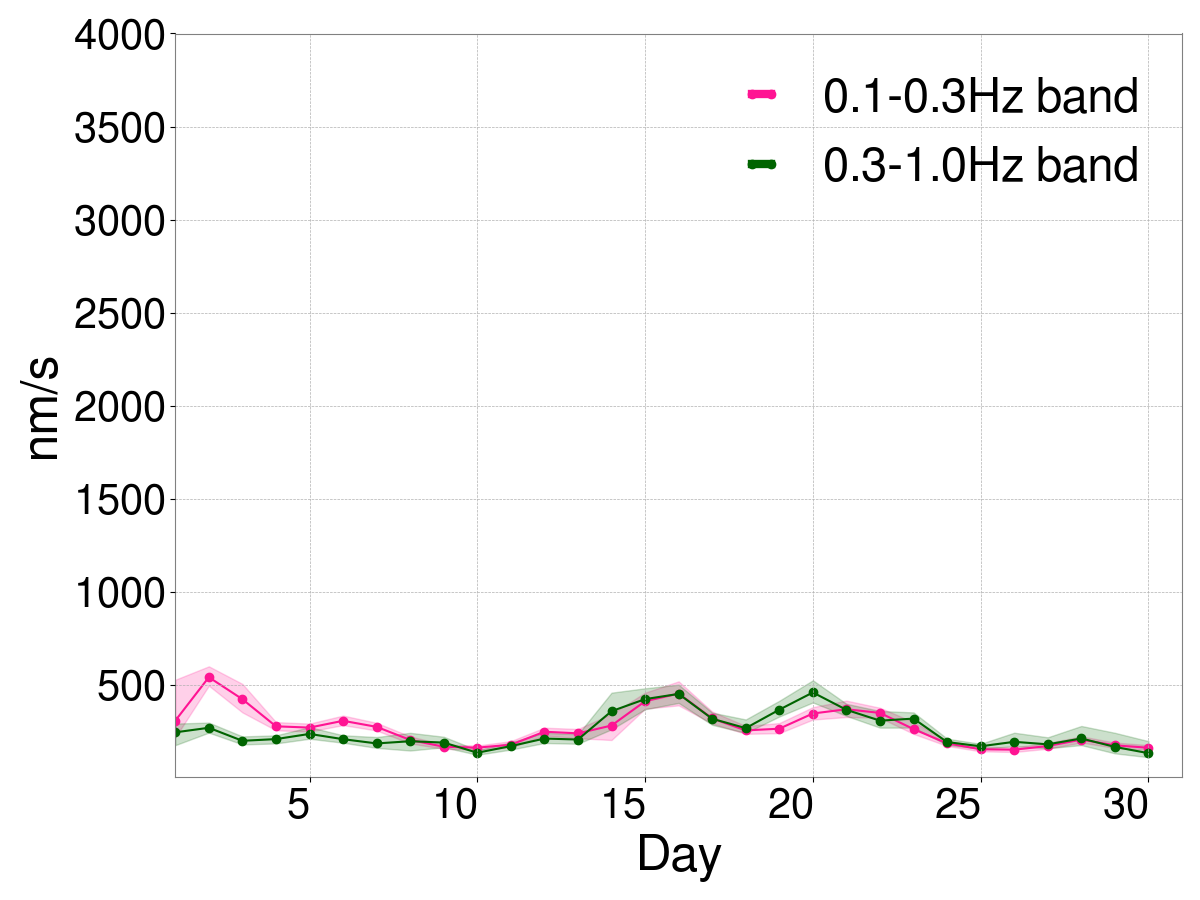}
        \label{fig:june_l1_groundmotion}
    \end{subfigure}
    \begin{subfigure}{0.32\textwidth}
        \centering
        \caption{July}
        \includegraphics[width=\textwidth]{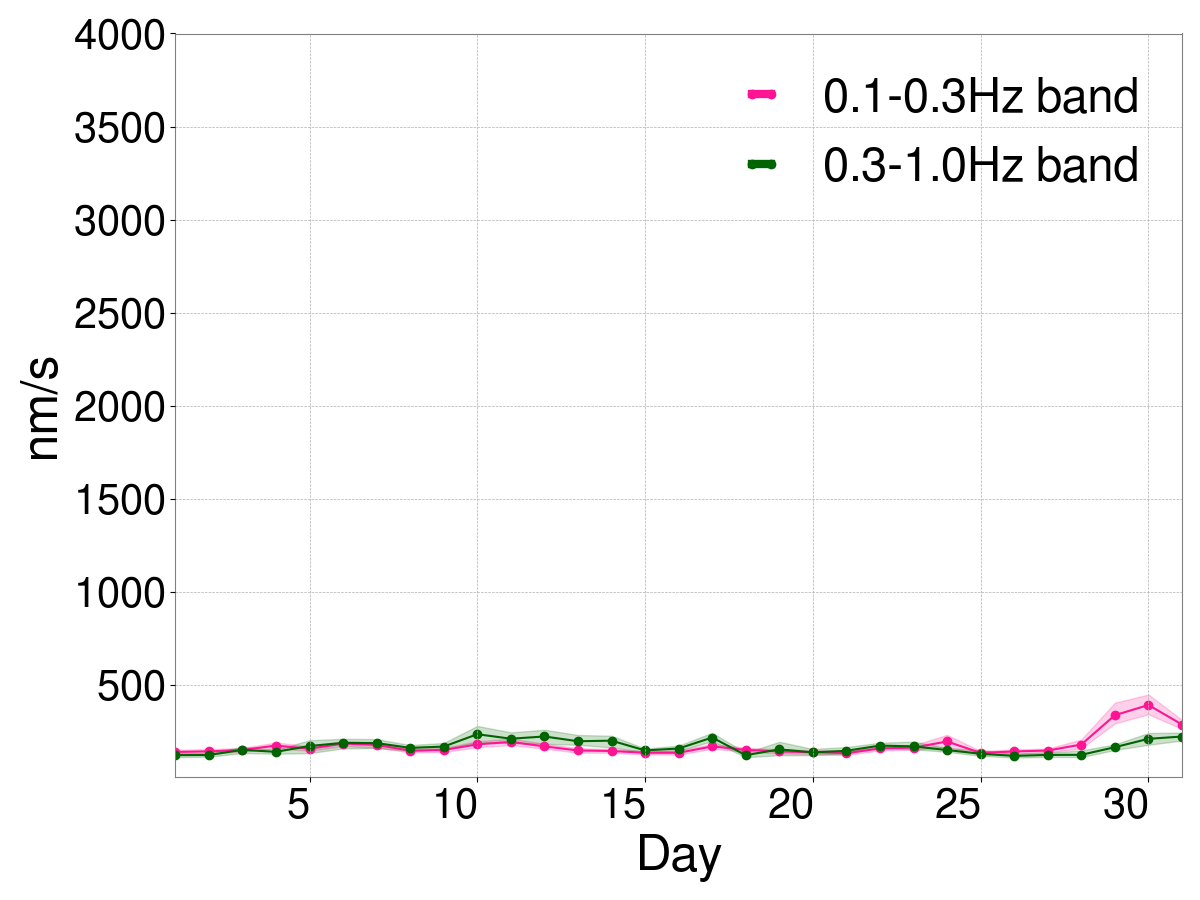}
        \label{fig:july_l1_groundmotion}
    \end{subfigure}
    
    \medskip
    
    \begin{subfigure}{0.32\textwidth}
        \centering
        \caption{August}
        \includegraphics[width=\textwidth]{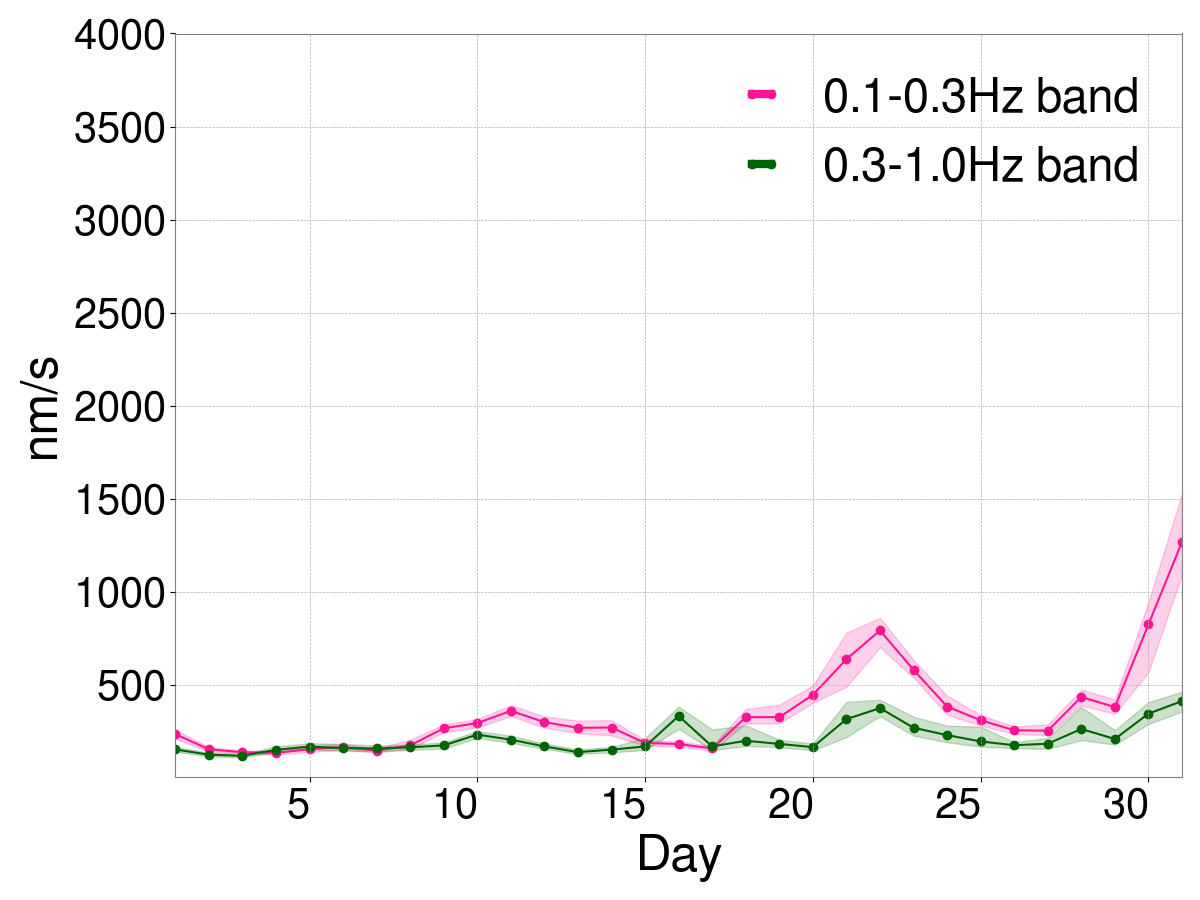}
        \label{fig:august_l1_groundmotion}
    \end{subfigure}
    \begin{subfigure}{0.32\textwidth}
        \centering
        \caption{September}
        \includegraphics[width=\textwidth]{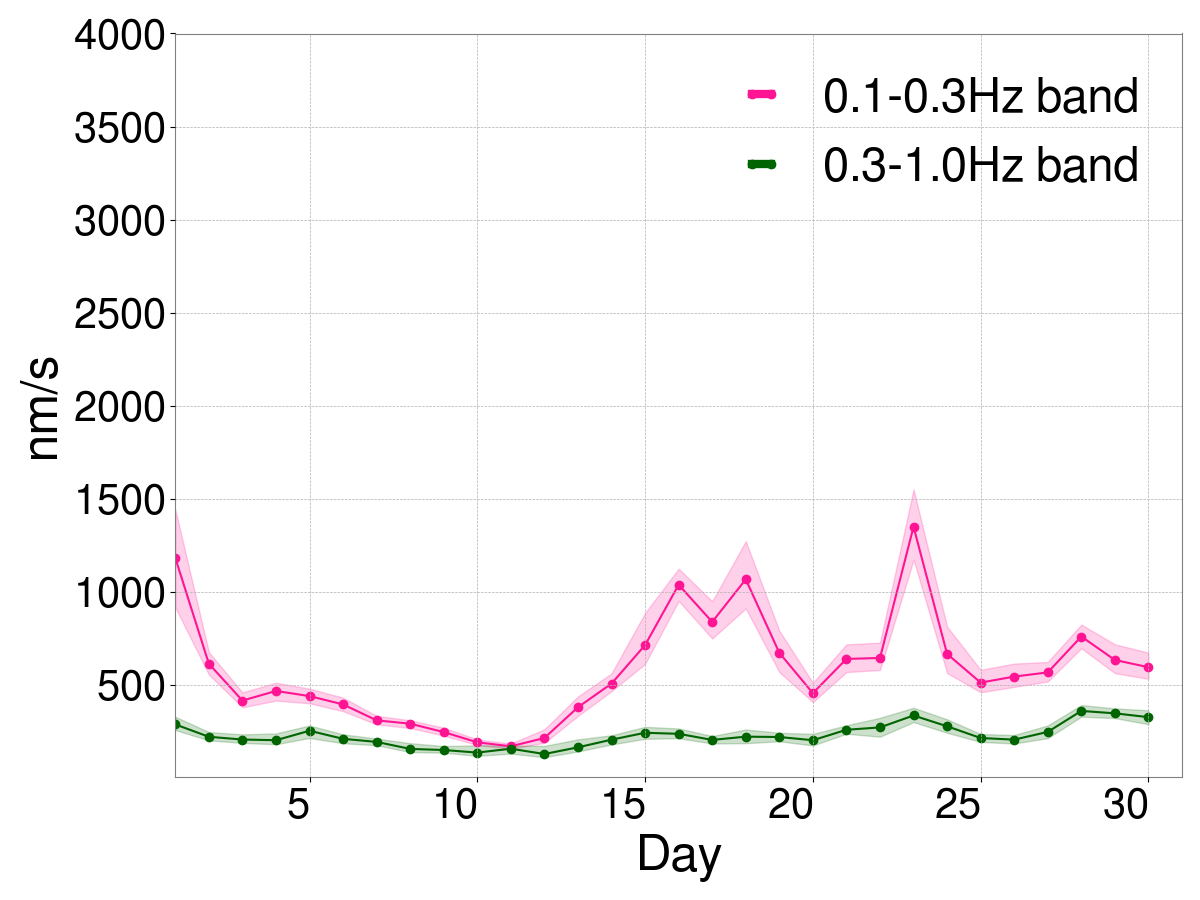}
        \label{fig:september_l1_groundmotion}
    \end{subfigure}
    \begin{subfigure}{0.32\textwidth}
        \centering
        \caption{October}
        \includegraphics[width=\textwidth]{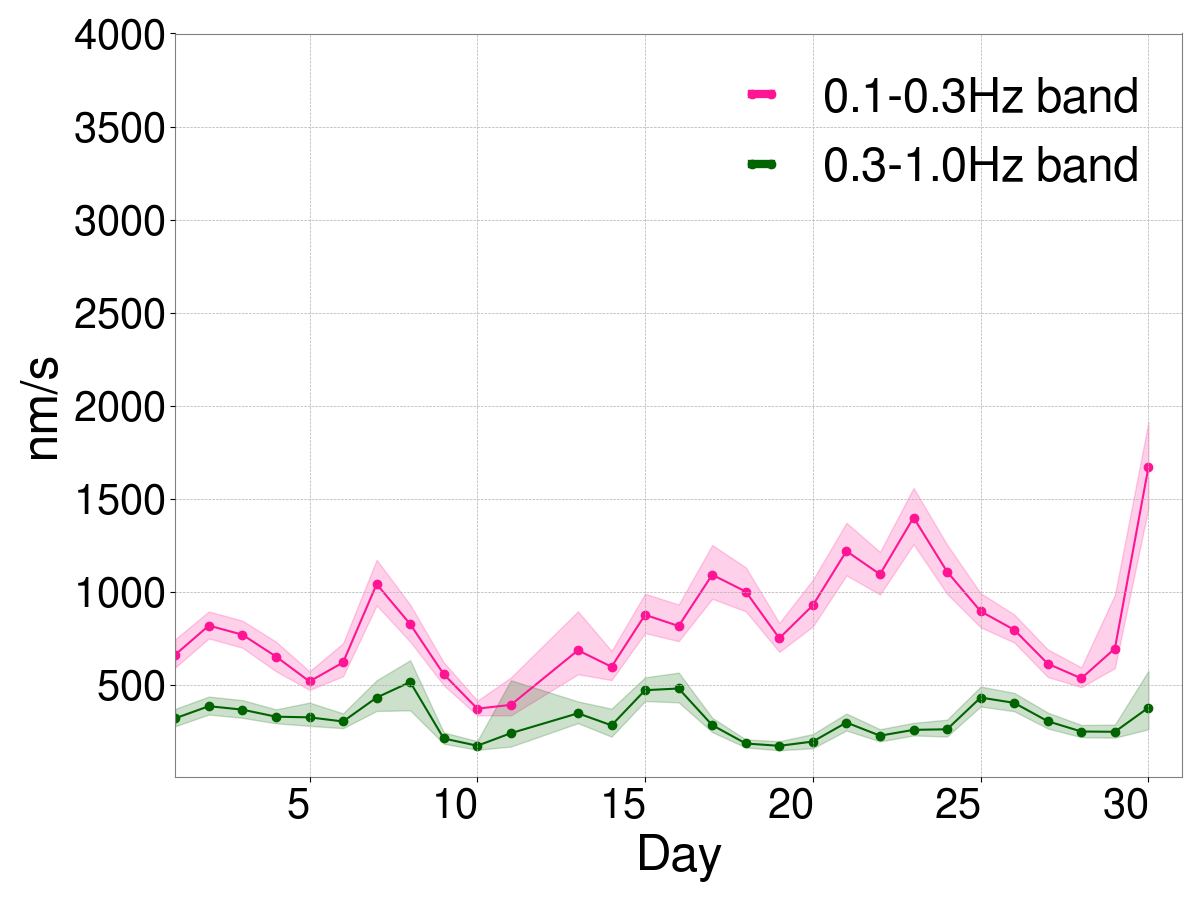}
        \label{fig:october_l1_groundmotion}
    \end{subfigure}
    
    \medskip
    
    \begin{subfigure}{0.32\textwidth}
        \centering
        \caption{November}
        \includegraphics[width=\textwidth]{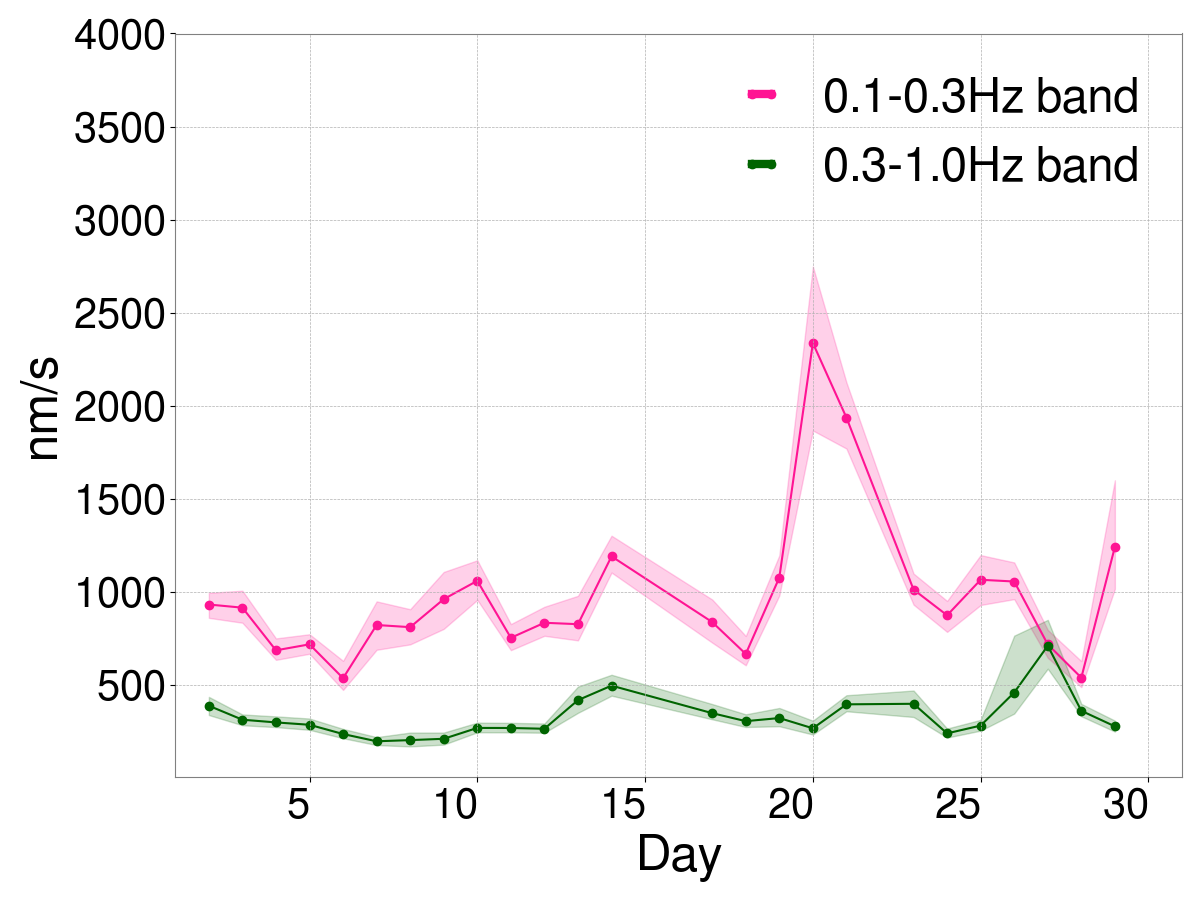}
        \label{fig:november_l1_groundmotion}
    \end{subfigure}
    \begin{subfigure}{0.32\textwidth}
        \centering
        \caption{December}
        \includegraphics[width=\textwidth]{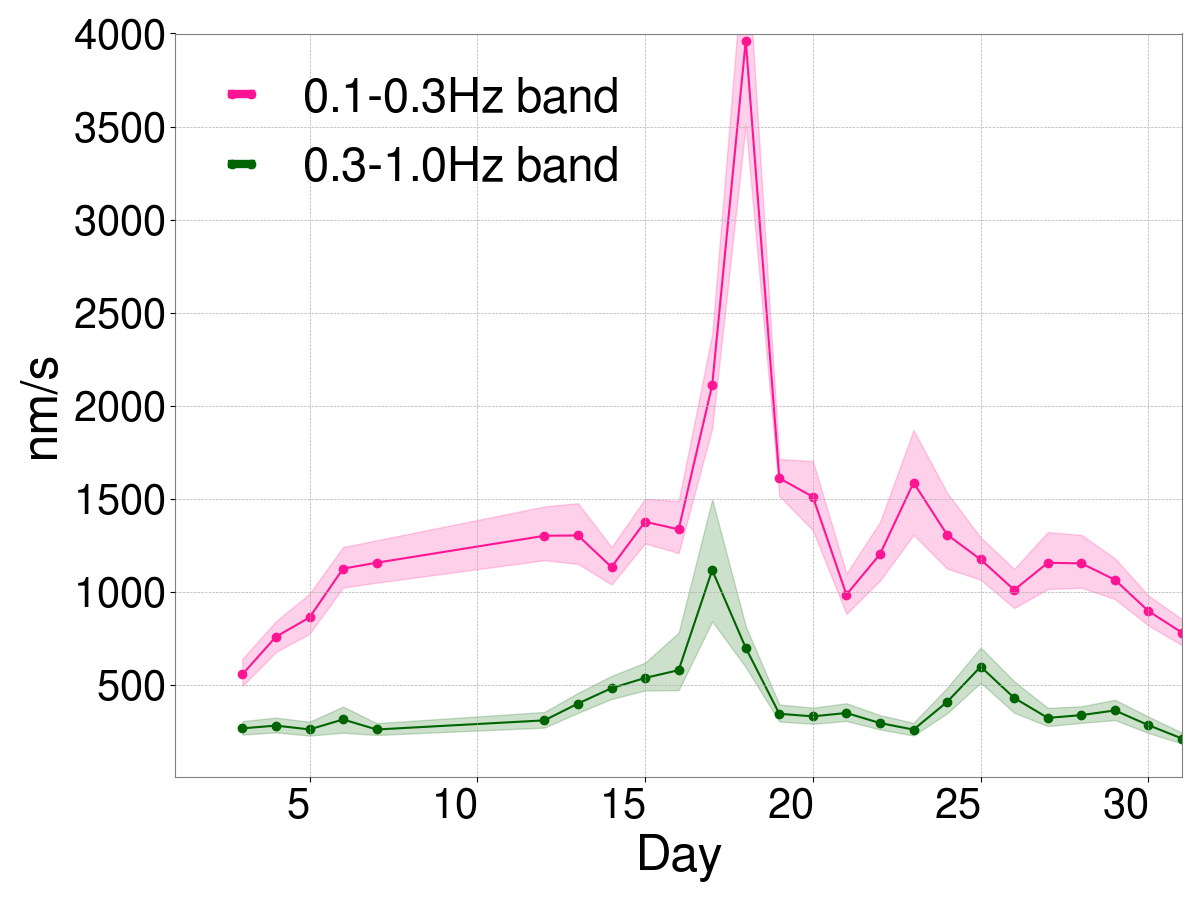}
        \label{fig:december_l1_groundmotion}
    \end{subfigure}
    \begin{subfigure}{0.32\textwidth}
        \centering
            \caption{January}
        \includegraphics[width=\textwidth]{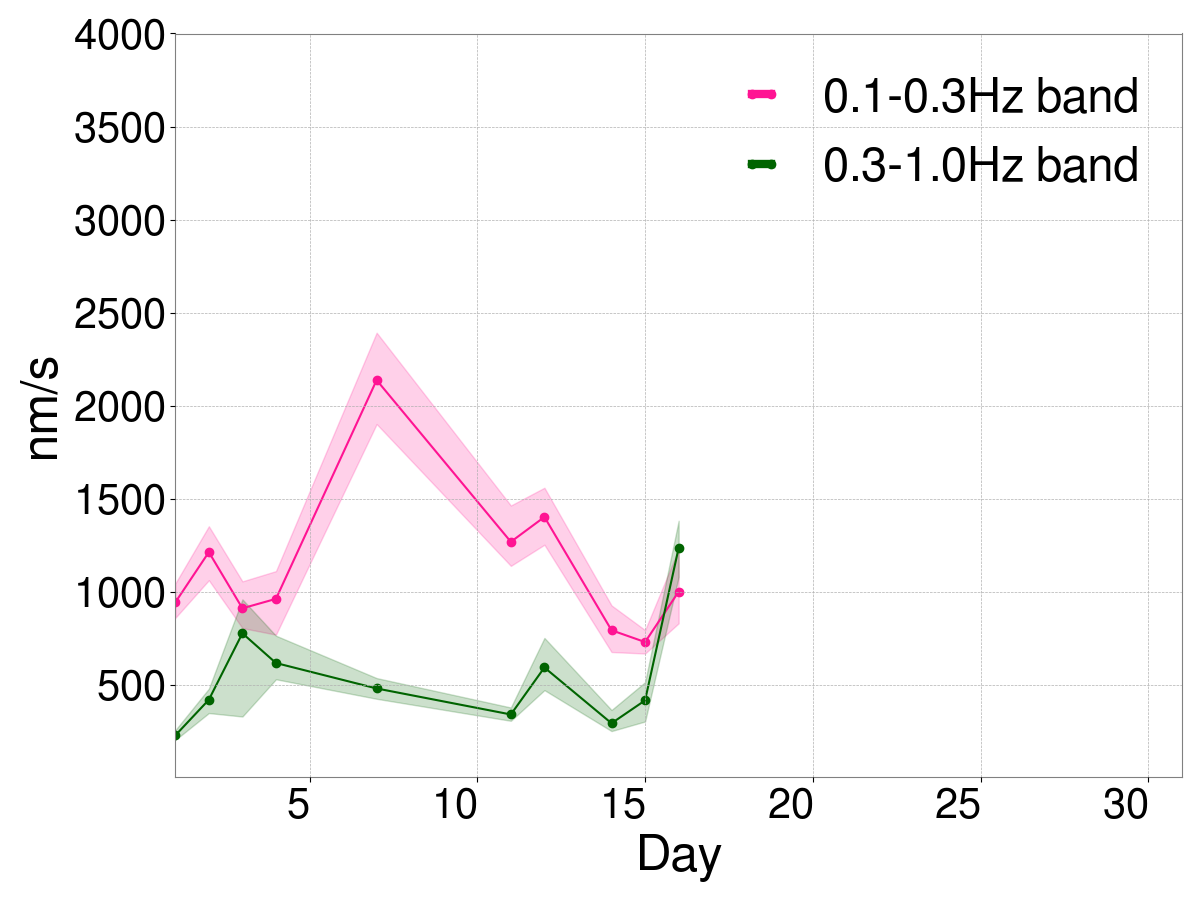}
        \label{fig:january_l1_groundmotion}
    \end{subfigure}
    
    \caption{Daily median ground velocity (in the $\hat{x}$ direction) in the lower microseismic band (magenta) and the higher microseismic band (green) for each month during the O4a observing run. The shaded areas represent the first and third quartiles.}
    \label{fig:mm_vs_gm_by_month}
\end{figure*}

As mentioned earlier, high ground motion (or elevated ground velocities) was the primary cause of scattering transients during O3. One class of these transients, referred to as \textit{Scattered Light} or \textit{slow scattering}~\cite{soni2020reducing,accadia2010noise}, was characterized by arch-like features in their spectrograms. Given this, and based on the seasonal pattern in which ground motion in specific frequency bands tends to exhibit higher amplitudes during the winter months, we identified a potential correlation between such low-frequency glitches observed during O4a and elevated ground motion levels~\cite{alog68208}. This behavior is further supported by the ground motion data shown later in Figure~\ref{fig:mm_vs_gm_by_month}.

%The plot also shows two groups of glitches during O4a between \SI{18}{\hertz} and \SI{24}{\hertz}: one with low amplitudes and another one with higher amplitudes, forming a shoulder shape similar to some glitches observed during O3. Again, we still do not know if they originate from the same source. 

Ground motion at LIGO is measured by seismometers positioned at various locations around the interferometer, which record ground velocity, typically expressed in nanometers per second (nm/s). Measurements are taken along three axes: $\hat{x}$, aligned with the main interferometer beam direction toward the beam splitter, $\hat{y}$, perpendicular to $\hat{x}$; and $\hat{z}$, representing vertical motion orthogonal to the x–y plane.

We analyze the Root Mean Square (RMS) of the ground motion within specific frequency ranges, known as Band-Limited RMS (BLRMS)~\cite{daw2004long}. The frequency bands considered correspond to the microseismic band-usually associated with ocean waves-and the anthropogenic band, which is generally linked to human activities such as passing trains or nearby vehicular traffic. The microseismic band is further subdivided into two distinct sub-bands. The ground motion is measured using the ETMY seismometer (as discussed later in the text). In this context, the analysis focuses on the following:

\begin{itemize}
  \item \textbf{Lower MicroSeismic (LMS) band:} ground motion between $0.1$ and \SI{0.3}{Hz}, measured in the $\hat{x}$ direction.
  \item \textbf{Higher MicroSeismic (HMS) band:} ground motion between $0.3$ and \SI{1}{Hz}, also measured in the $\hat{x}$ direction.
  \item \textbf{Anthropogenic band:} ground motion between $1$ and \SI{3}{Hz}, measured in the $\hat{z}$ direction.
\end{itemize}

%To further investigate the hypothesis linking the low-frequency glitches to ground motion, we perform a t-SNE analysis to examine the monthly presence of glitch clusters.

\begin{figure*}[!t]
    \centering

    \begin{subfigure}[t]{0.48\textwidth}
        \centering
        \caption{Omicron triggers}
        \includegraphics[height=5.5cm]{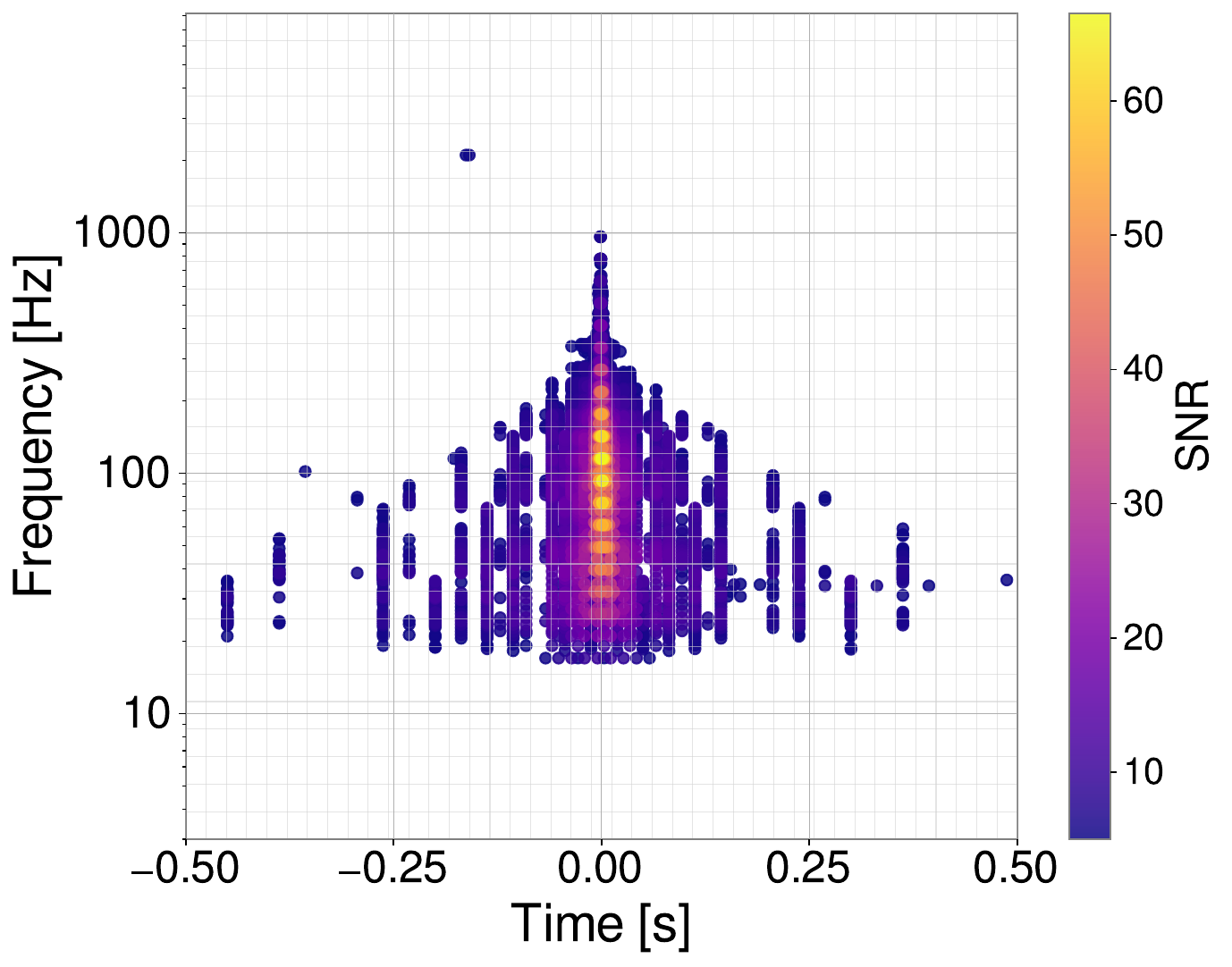}
        \label{fig:glitchgram_scatter}
    \end{subfigure}
    \hfill
    \begin{subfigure}[t]{0.48\textwidth}
        \centering
        \caption{Pixelized 30$\times$41 matrix}
        \includegraphics[height=5.5cm]{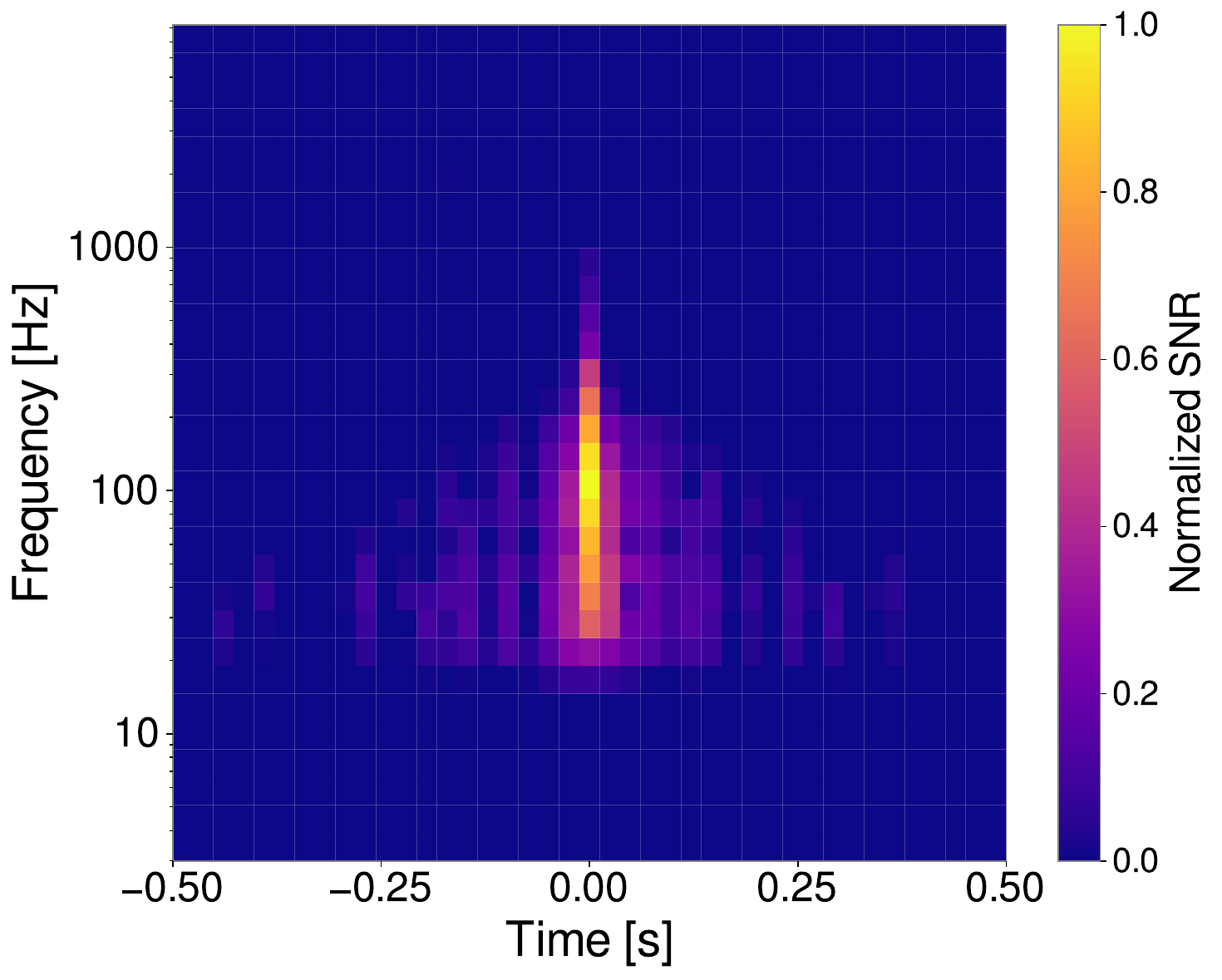}
        \label{fig:glitchgram_matrix}
    \end{subfigure}

    \vspace{0.2cm}

    % --- Subfigura (c) embaixo ---
    \begin{subfigure}[t]{1\textwidth}
        \centering
        \includegraphics[width=\textwidth]{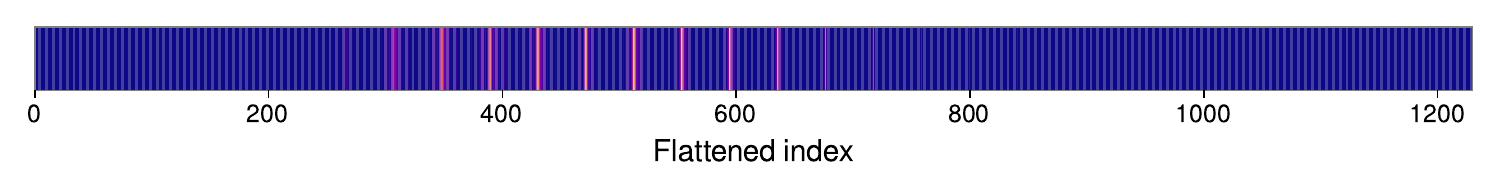}
        \caption{Flattened 1,230-dimensional vector.}
        \label{fig:glitchgram_vector}
    \end{subfigure}

    % ----- Caption geral -----
    \caption{
        Workflow for constructing the vectors used as input to t-SNE, illustrated for one example glitch.
        (a) Unclustered Omicron triggers in a $\pm0.5$~s window with the 30$\times$41 time–frequency grid.
        (b) Pixelized glitchgram matrix with normalized SNR.
        (c) Flattened 1,230-dimensional vector, shown using the same colormap scale as in panel (b), used as input to t-SNE and clustering.
    }
    \label{fig:glitchgram_pipeline}
\end{figure*}

As proposed in~\cite{ferreira2025using}, we began by selecting a dataset containing 1,500 randomly sampled glitches per month during O4a (from May 2023 to January 2024). Since each month contains a different number of glitches, not applying a monthly filter would result in overrepresentation of glitches from months with higher glitch rates, which differ substantially from others, as shown in Figure~\ref{fig:glitches_per_hour_per_month_l1}. Although this corresponds to a limited fraction of the total observing time in each month, the random sampling ensures a representative subset of the glitch population while maintaining a balanced dataset across months. This approach is intended to highlight the dominant glitch families present throughout O4a on a month-by-month basis, while potentially reducing sensitivity to transient populations that appear only during limited time intervals, which were explored in our previous day-by-day analysis~\cite{ferreira2025using}.

For each selected clustered Omicron trigger, we retrieve the unclustered Omicron triggers within a $\pm 0.5$~s window centered on its time, as part of our procedure for constructing the glitch representation. Each glitch is then represented by a $30\times41$ time--frequency--$Q$ matrix in which each bin stores the maximum normalized SNR of the triggers assigned to that bin (panel~\ref{fig:glitchgram_matrix}). This matrix is then flattened into a 1,230-dimensional feature vector following a row-major ordering
(panel~\ref{fig:glitchgram_vector}). The final dataset, therefore, contains 1,500 glitches per month over 9 months, totaling 13,500 samples, each described by 1,230 features. To prevent glitches with very high SNR from dominating the \mbox{t-SNE} output, we applied min-max normalization to the SNR components of each vector.

\begin{figure*}[ht!]
    \centering
    \begin{subfigure}{0.49\textwidth}
        \centering
        \caption{}
        \includegraphics[width=\textwidth]{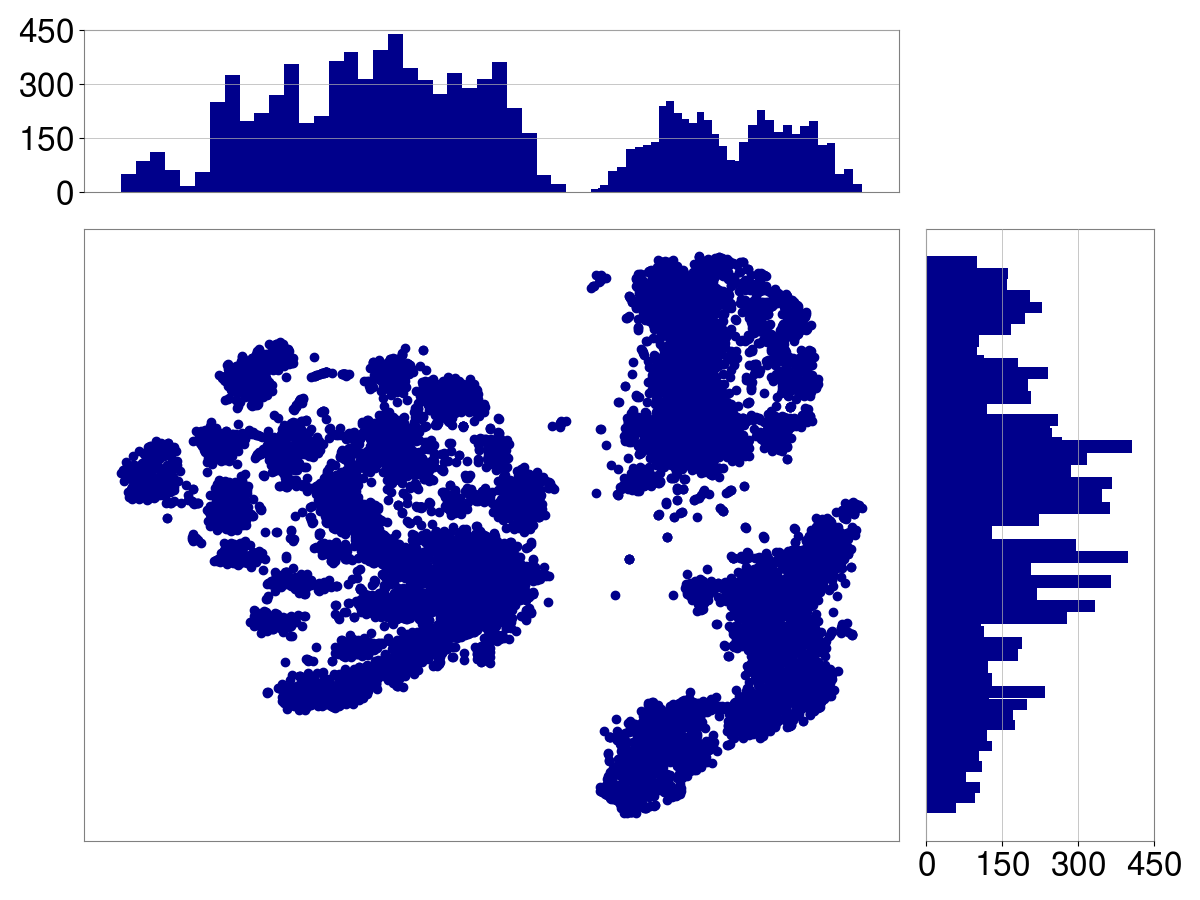}
        \label{fig:o4a_tsne_l1_blue}
    \end{subfigure}
    \hfill
    \begin{subfigure}{0.49\textwidth}
        \centering
        \caption{}
        \includegraphics[width=\textwidth]{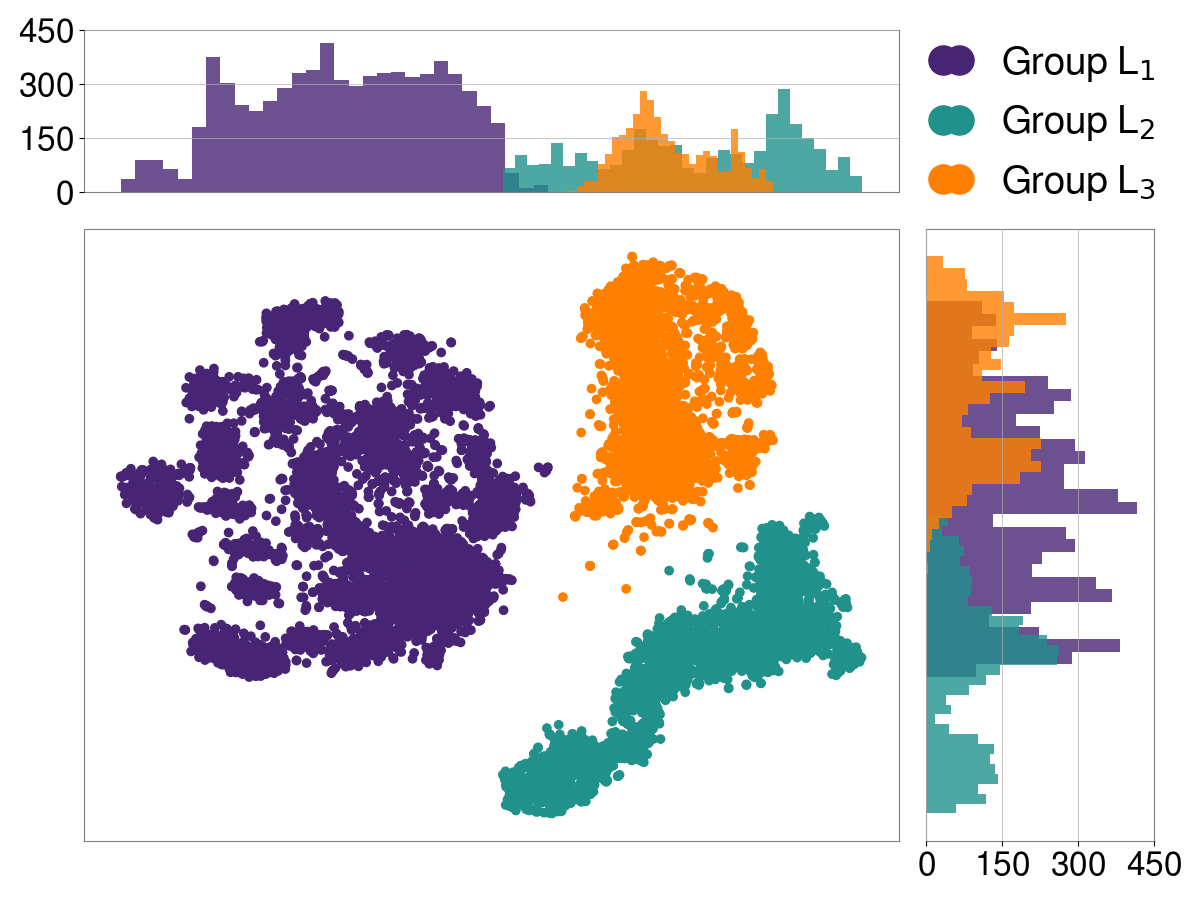}
        \label{fig:o4a_tsne_l1_colored}
    \end{subfigure}
    
    \caption{(a) Scatter plot of the 2D t-SNE output coordinates representing 13,500 data points, each originally embedded in a 1,230-dimensional space derived from unclustered Omicron triggers. Each point corresponds to a glitch. Histograms along the top and right axes illustrate the distribution of glitches along each coordinate. (b) The same t-SNE output, with points colored according to the classes assigned by Agglomerative Clustering.}
    \label{fig:o4a_tsne_l1}
\end{figure*}

The t-SNE algorithm reduced the original 1,230-dimensional vector representation to two dimensions, enabling the visualization of 13,500 glitches on a 2D plane. This representation is shown in Figure~\ref{fig:o4a_tsne_l1_blue} as a scatter plot of the two \mbox{t-SNE} output coordinates, where each point corresponds to a single glitch. Three major groups are identifiable: the largest cluster is positioned on the left, the second cluster is located at the top right, and the third cluster appears at the bottom right. 

%The distributions of the data points associated with each group are also shown in the histograms positioned along the top and right axes of the plot. %Details about the techniques can be found in~\cite{ferreira2025using}.
%Accompanying the scatter plot, histograms along the top and right axes illustrate the distribution of glitches across the t-SNE coordinates. The histogram on the right reveals a higher concentration of glitches near the center, while the histogram on the top indicates a more evenly distributed number of glitches on the left side and two distinct peaks on the right. 

Figure~\ref{fig:o4a_tsne_l1_colored} presents the clusters identified via Agglomerative Clustering, where the number of clusters was determined through a systematic scan by maximizing the \textit{Silhouette Score}. Consistent with the visual structure observed in the t-SNE representation, this procedure robustly selects three clusters, referred to here as \textit{Group L$_1$} (purple), \textit{Group L$_2$} (green), and \textit{Group L$_3$} (orange). Both techniques were implemented with the \textit{Scikit-learn} library~\cite{pedregosa2011scikit}.

\begin{figure*}[p]
    \centering
    \begin{subfigure}{0.32\textwidth}
        \centering
        \caption{May}
        \textbf{\textcolor{viridis1}{2.4/h} \hspace{0.4em} \textcolor{viridis2}{7.6/h} \hspace{0.4em} \textcolor{orange}{2.3/h}}
        \fbox{\includegraphics[width=0.9\textwidth]{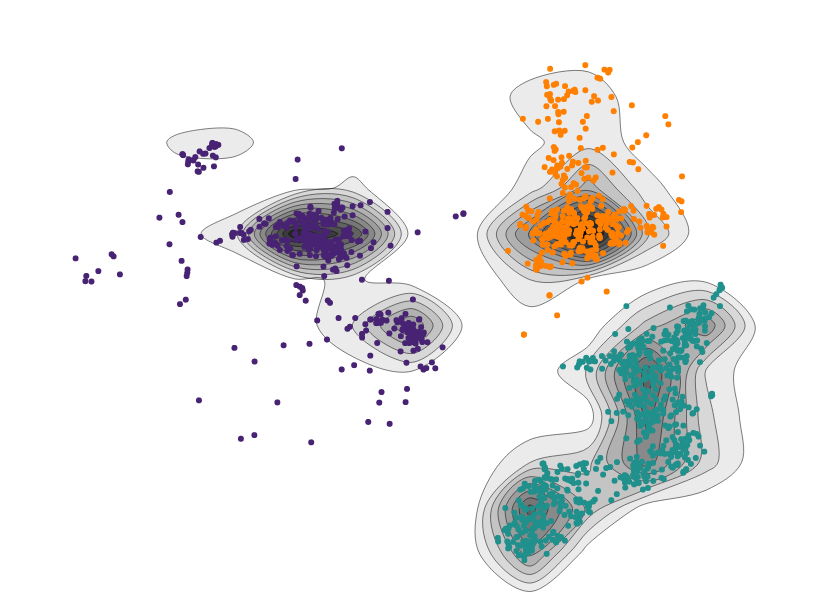}}
        \textbf{\textcolor{viridis1}{19.6\%} \hspace{0.4em} \textcolor{viridis2}{61.8\%} \hspace{0.4em} \textcolor{orange}{18.6\%}}
        \label{fig:May_L1}
    \end{subfigure}
    \vspace{0.3cm}
    \begin{subfigure}{0.32\textwidth}
        \centering
        \caption{June}
        \textbf{\textcolor{viridis1}{1.5/h} \hspace{0.4em} \textcolor{viridis2}{7.3/h} \hspace{0.4em} \textcolor{orange}{1.7/h}}
        \fbox{\includegraphics[width=0.9\textwidth]{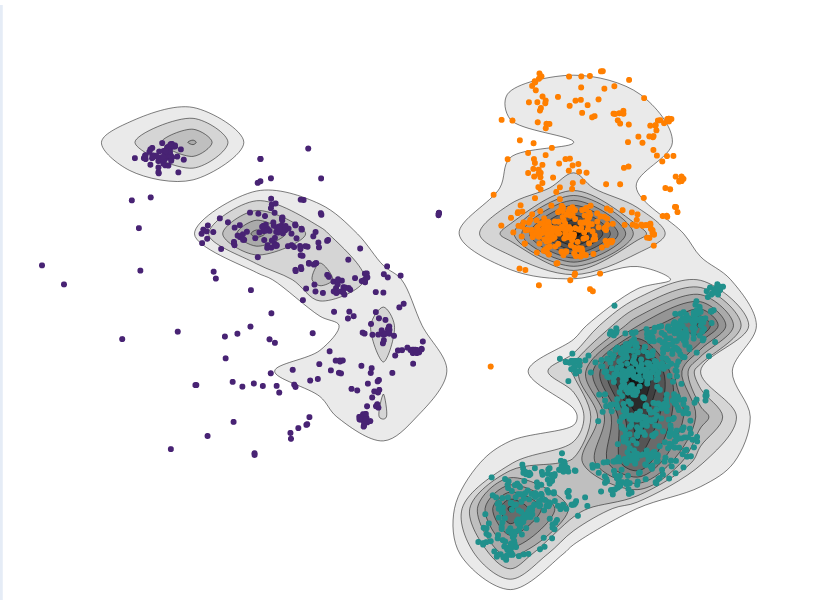}}
        \textbf{\textcolor{viridis1}{14.2\%} \hspace{0.4em} \textcolor{viridis2}{69.2\%} \hspace{0.4em} \textcolor{orange}{16.6\%}}
        \label{fig:June_L1}
    \end{subfigure}
    \vspace{0.3cm}
    \begin{subfigure}{0.32\textwidth}
        \centering
        \caption{July}
        \textbf{\textcolor{viridis1}{1.4/h} \hspace{0.4em} \textcolor{viridis2}{5.5/h} \hspace{0.4em} \textcolor{orange}{0.8/h}}
        \fbox{\includegraphics[width=0.9\textwidth]{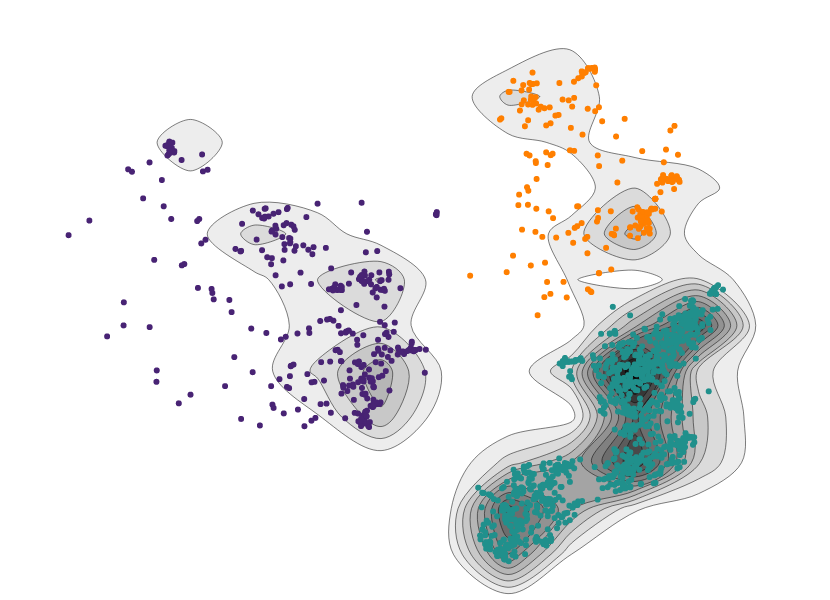}}
        \textbf{\textcolor{viridis1}{18.4\%} \hspace{0.4em} \textcolor{viridis2}{71.5\%} \hspace{0.4em} \textcolor{orange}{10.1\%}}
        \label{fig:July_L1}
    \end{subfigure}
    
    \medskip
    
    \begin{subfigure}{0.32\textwidth}
        \centering
        \caption{August}
        \textbf{\textcolor{viridis1}{7.9/h} \hspace{0.4em} \textcolor{viridis2}{5.7/h} \hspace{0.4em} \textcolor{orange}{4.1/h}}
        \fbox{\includegraphics[width=0.9\textwidth]{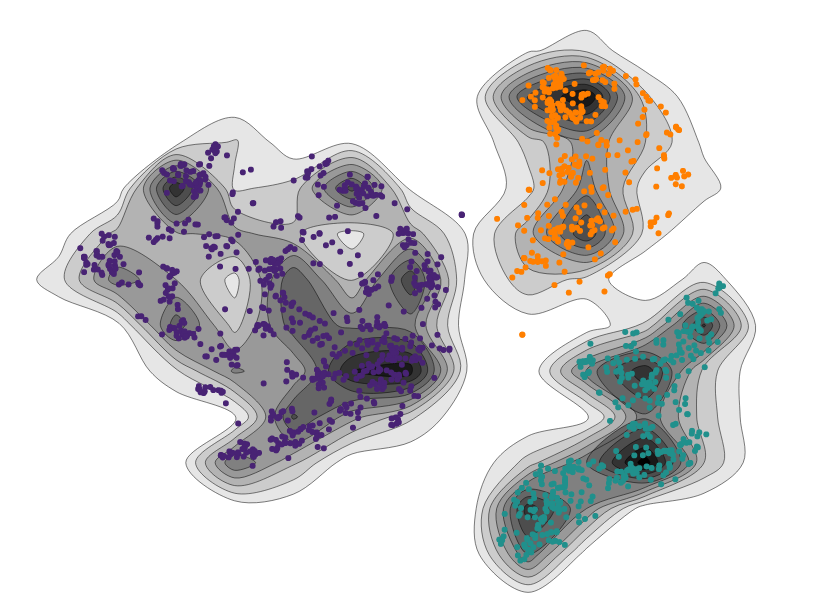}}
        \textbf{\textcolor{viridis1}{44.6\%} \hspace{0.4em} \textcolor{viridis2}{32.2\%} \hspace{0.4em} \textcolor{orange}{23.2\%}}
        \label{fig:August}
    \end{subfigure}
    \vspace{0.3cm}
    \begin{subfigure}{0.32\textwidth}
        \centering
        \caption{September}
        \textbf{\textcolor{viridis1}{9.8/h} \hspace{0.4em} \textcolor{viridis2}{6.3/h} \hspace{0.4em} \textcolor{orange}{8.3/h}}
        \fbox{\includegraphics[width=0.9\textwidth]{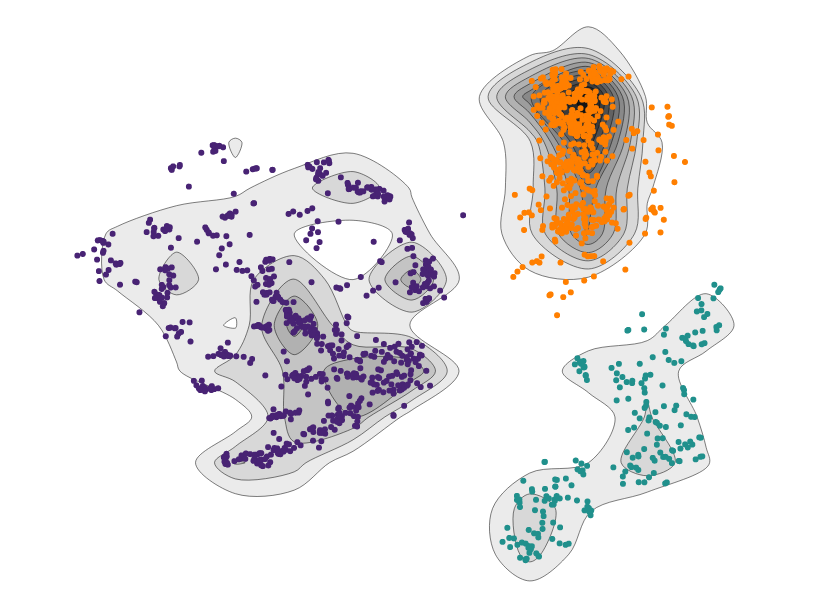}}
        \textbf{\textcolor{viridis1}{40.2\%} \hspace{0.4em} \textcolor{viridis2}{25.8\%} \hspace{0.4em} \textcolor{orange}{34.0\%}}
        \label{fig:September_L1}
    \end{subfigure}
    \vspace{0.3cm}
    \begin{subfigure}{0.32\textwidth}
        \centering
        \caption{October}
        \textbf{\textcolor{viridis1}{33.6/h} \hspace{0.4em} \textcolor{viridis2}{8.1/h} \hspace{0.4em} \textcolor{orange}{15.6/h}}
        \fbox{\includegraphics[width=0.9\textwidth]{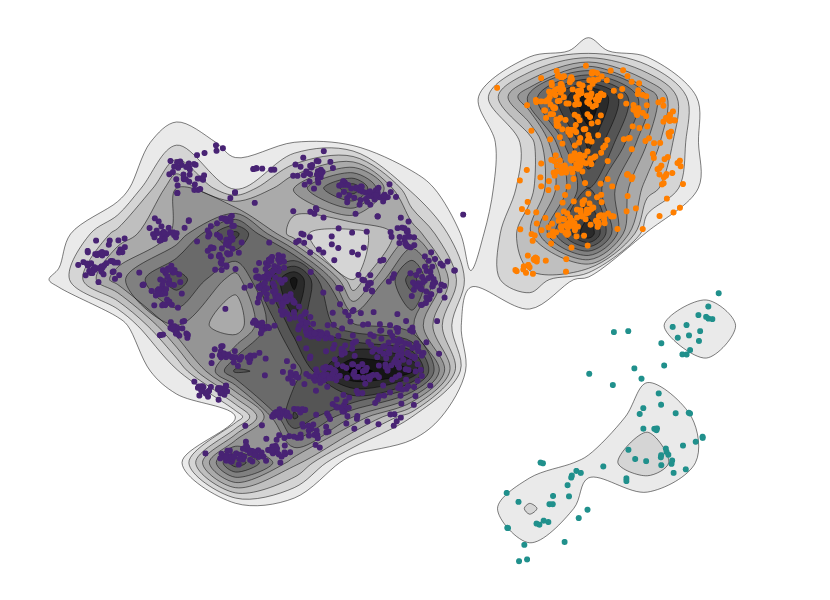}}
        \textbf{\textcolor{viridis1}{58.7\%} \hspace{0.4em} \textcolor{viridis2}{14.1\%} \hspace{0.4em} \textcolor{orange}{27.2\%}}
        \label{fig:October_L1}
    \end{subfigure}

    \medskip

    \begin{subfigure}{0.32\textwidth}
        \centering
        \caption{November}
        \textbf{\textcolor{viridis1}{37.3/h} \hspace{0.3em} \textcolor{viridis2}{7.3/h} \hspace{0.3em} \textcolor{orange}{14.7/h}}
        \fbox{\includegraphics[width=0.9\textwidth]{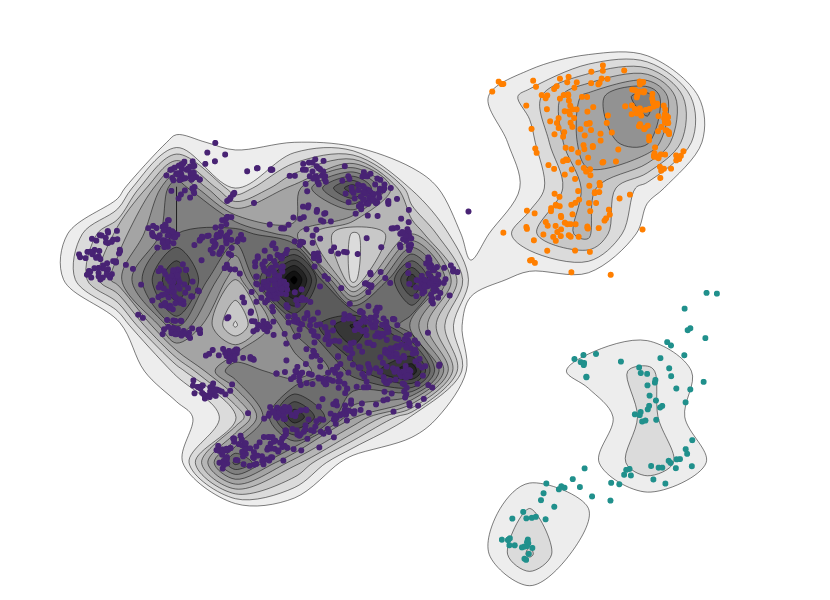}}
        \textbf{\textcolor{viridis1}{62.9\%} \hspace{0.4em} \textcolor{viridis2}{12.4\%} \hspace{0.4em} \textcolor{orange}{24.7\%}}
        \label{fig:November_L1}
    \end{subfigure}
    \begin{subfigure}{0.32\textwidth}
        \centering
        \caption{December}
        \textbf{\textcolor{viridis1}{75.1/h} \hspace{0.1em} \textcolor{viridis2}{21.2/h} \hspace{0.1em} \textcolor{orange}{38.9/h}}
        \fbox{\includegraphics[width=0.9\textwidth]{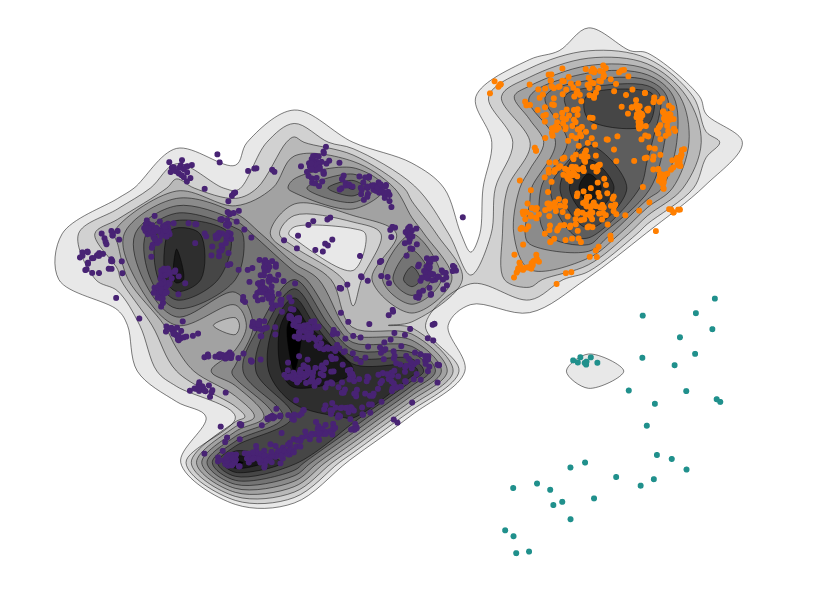}}
        \textbf{\textcolor{viridis1}{55.5\%} \hspace{0.4em} \textcolor{viridis2}{15.7\%} \hspace{0.4em} \textcolor{orange}{28.8\%}}
        \label{fig:December_L1}
    \end{subfigure}
    \begin{subfigure}{0.32\textwidth}
        \centering
        \caption{January}
        \textbf{\textcolor{viridis1}{84.5/h} \hspace{0.1em} \textcolor{viridis2}{17.0/h} \hspace{0.1em} \textcolor{orange}{46.8/h}}
        \fbox{\includegraphics[width=0.9\textwidth]{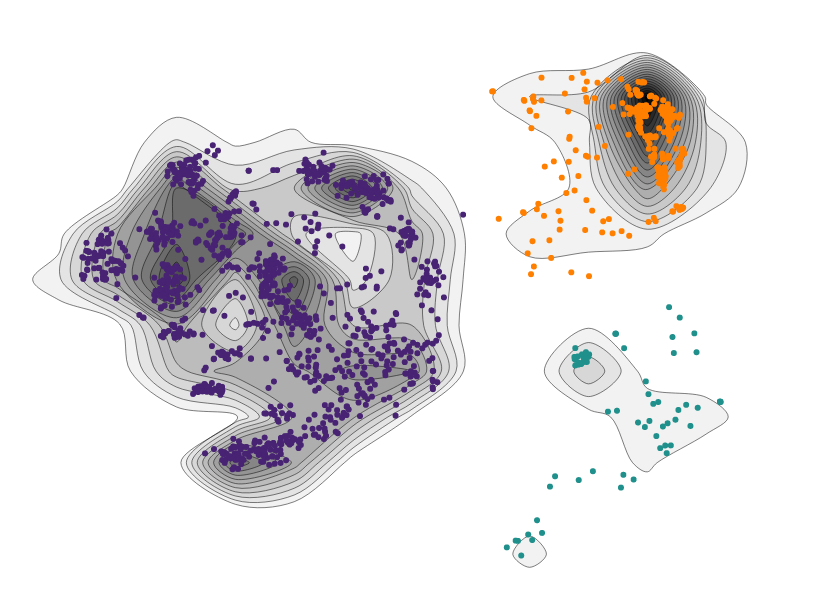}}
        \textbf{\textcolor{viridis1}{57.0\%} \hspace{0.4em} \textcolor{viridis2}{11.5\%} \hspace{0.4em} \textcolor{orange}{31.5\%}}
        \label{fig:January_L1}
    \end{subfigure}
    
    \medskip
    
    \begin{subfigure}{0.9\textwidth}
        \centering
        \includegraphics[width=0.9\textwidth]{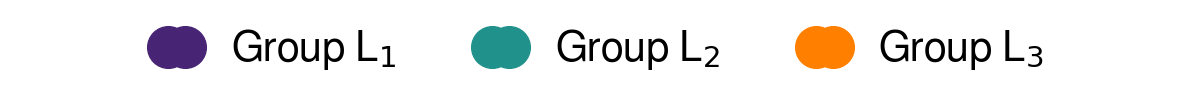}
        %\label{fig:legend}
    \end{subfigure}
    
    \caption{Monthly distribution of glitches in the
    2D t-SNE space at LLO, showing the evolution of the
    groups identified using Agglomerative Clustering.
    Each panel shows the distribution of these groups
    during O4a, from May 2023 to January 2024.
    The values shown above each panel indicate the
    duty-cycle-corrected glitch rates (in glitches per hour)
    for each group, while the values below indicate the
    corresponding fraction of glitches assigned to each
    group using the full classified dataset.}    \label{fig:tsne_by_month_l1}
\end{figure*}

An important advantage of this method is its ability to analyze the temporal evolution of glitches. Figure~\ref{fig:tsne_by_month_l1} presents the same t-SNE embedding shown in Figure~\ref{fig:o4a_tsne_l1}, now filtered month by month during O4a, providing an overview of which group is more prominent in each month. As mentioned, we selected the same number of glitches per month, enabling a balanced temporal analysis and preventing the dominance of months with higher glitch rates. Although this approach introduces an artificial uniformity across months, it allows for a clearer comparison of cluster behavior over time at this scale. If we had instead selected a total of 13,500 glitches randomly, without monthly filtering, months with higher glitch rates (see Figure~\ref{fig:glitches_per_hour_per_month_l1}) would have dominated the sample. A more detailed discussion of this selection strategy is provided in~\cite{ferreira2025using}.

The numbers displayed below each scatter plot represent the percentage contribution of each group for the respective month. Comparing May, June, and July reveals a high similarity in cluster distribution, with Group L$_2$ (green) being the most prominent during all three months. The gray contours behind the data points indicate the estimated density, computed using Kernel Density Estimation (KDE) as implemented in Seaborn~\cite{waskom2021seaborn}, where darker regions correspond to areas with higher data point concentrations.

In August, Group L$_1$ begins to exhibit higher densities, accounting for more than $50\%$ of the data. It slightly contracts in September (when an increase in the glitch rate from Group L$_3$ is observed) but remains the dominant group in subsequent months, reaching nearly $80\%$ of the data in November. Group L$_3$ fluctuates over time but remains consistently present, with its highest predominance occurring in September. Conversely, from August onward, Group L$_2$ declines due to the increasing prominence of other clusters. In December, for instance, the low density of points in this region results in an almost complete absence of KDE contours. This does not imply the disappearance of Group L$_2$, but rather that it was no longer among the most significant groups. %This effect arises because the data for each month were randomly selected, allowing glitch types with higher occurrence rates to dominate the dataset. 

\begin{figure*}[ht!]
    \centering
    \begin{subfigure}{0.32\textwidth}
        \includegraphics[width=\textwidth]{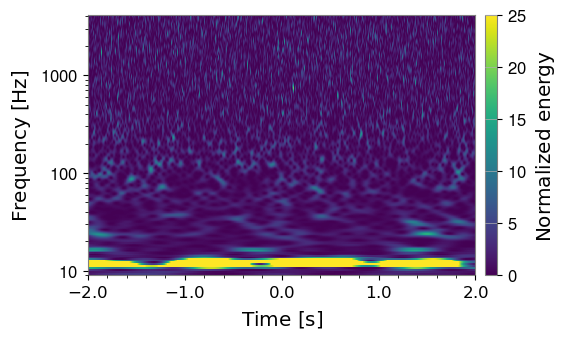}
    \end{subfigure}
    \begin{subfigure}{0.32\textwidth}
        \includegraphics[width=\textwidth]{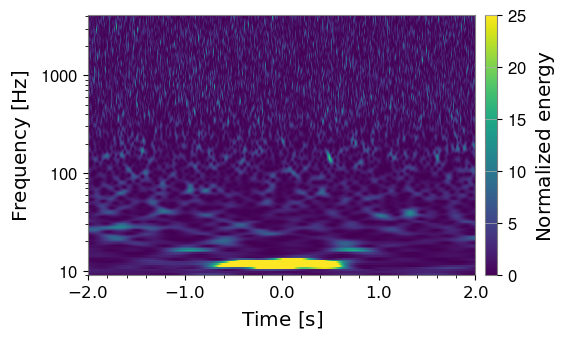}
    \end{subfigure}
    \begin{subfigure}{0.32\textwidth}
        \includegraphics[width=\textwidth]{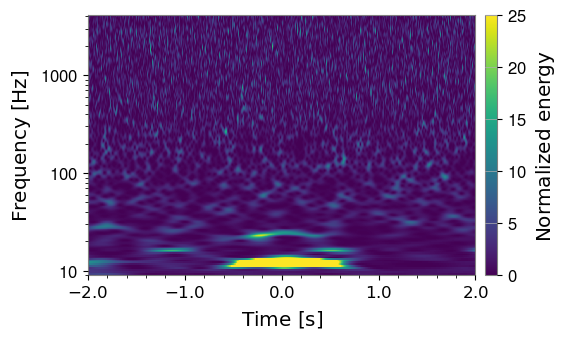}
    \end{subfigure}
    \caption{Examples of spectrograms present in Group L$_1$. The spectrograms exhibit frequencies around \SI{12}{Hz}, with variations in morphology.}
    \label{fig:spectrograms_group01}
\end{figure*}

\begin{figure*}[ht!]
    \centering
    \begin{subfigure}{0.32\textwidth}
        \includegraphics[width=\textwidth]{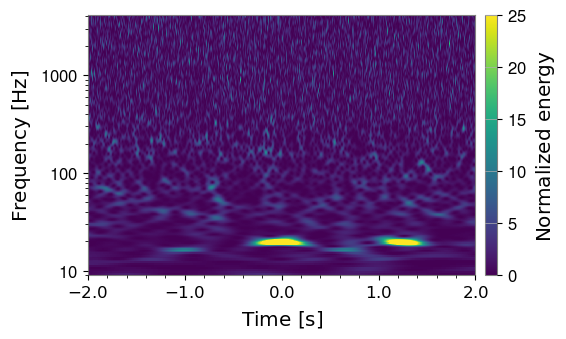}
    \end{subfigure}
    \begin{subfigure}{0.32\textwidth}
        \includegraphics[width=\textwidth]{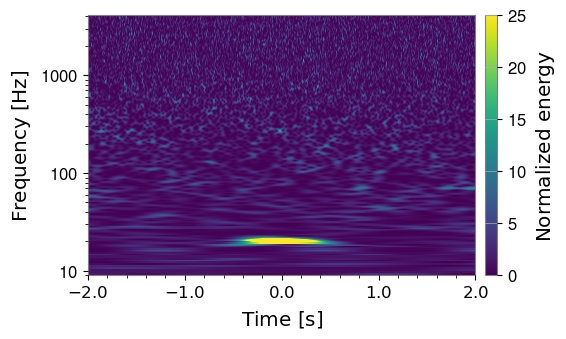}
    \end{subfigure}
    \begin{subfigure}{0.32\textwidth}
        \includegraphics[width=\textwidth]{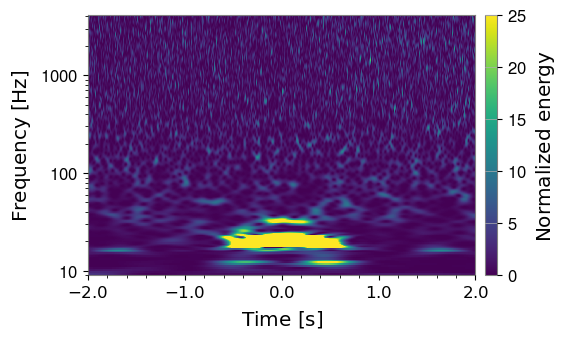}
    \end{subfigure}
    \caption{Examples of spectrograms present in Group L$_3$. The spectrograms exhibit frequencies around \SI{20}{Hz}, with variations in morphology.}
    \label{fig:spectrograms_group03}
\end{figure*}

By visually inspecting a representative subset of spectrograms from each group, we observe that Group L$_1$ is dominated by glitches with arch-like structures at a frequency of approximately \SI{12}{Hz}; examples are shown in Figure~\ref{fig:spectrograms_group01}. Some of these glitches exhibit longer durations, others show repetitions over time, and some vary in the number of harmonics. Despite occurring at similar frequencies, these morphological differences explain the formation of mini-clusters within the group (see Figure~\ref{fig:o4a_tsne_l1}). Glitches with similar morphological features (arch-like patterns) are observed in Group L$_3$, but with a primary frequency around \SI{20}{Hz}; examples of their spectrograms are presented in Figure~\ref{fig:spectrograms_group03}.

\begin{figure*}[ht!]
    \centering
    \begin{subfigure}{0.32\textwidth}
        \includegraphics[width=\textwidth]{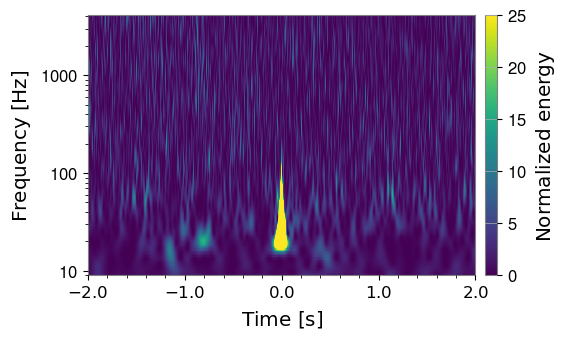}
    \end{subfigure}
    \begin{subfigure}{0.32\textwidth}
        \includegraphics[width=\textwidth]{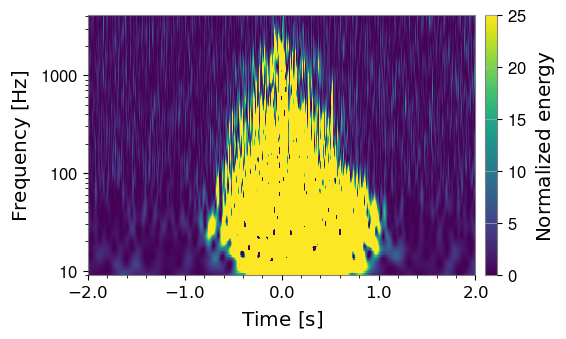}
    \end{subfigure}
    \begin{subfigure}{0.32\textwidth}
        \includegraphics[width=\textwidth]{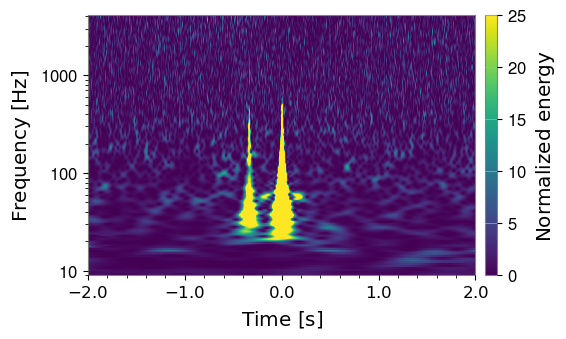}
    \end{subfigure}
    \caption{Examples of spectrograms present in Group L$_2$.}
    \label{fig:spectrograms_group02}
\end{figure*}

Group L$_2$ primarily consists of glitch classes identified by Gravity Spy~\cite{wu2025advancing} as \textit{Extremely Loud}, \textit{Blip}, \textit{Koi Fish}, \textit{Tomte}, and others (examples are given in Figure~\ref{fig:spectrograms_group02}). These glitches do not have a significant seasonal pattern (at least not as prominently as those in Group L$_1$ and Group L$_3$), which results in Group L$_2$ gradually losing structure in the t-SNE output over time and being more dominant at the beginning of O4a, when the other groups were still less prominent. To further subdivide Group L$_2$ into its specific subclasses, it would be necessary to rerun t-SNE on this subset. However, as previously mentioned, the primary objective of this study is to understand the behavior of the most frequent glitches that most affected the detector during O4a and to investigate potential correlations with environmental sensors surrounding the interferometer.

Given the previous insights from the t-SNE output, we investigate possible relationships between the identified groups and seismic noise. In general, amplitude variations in ground motion across the three directions exhibit similar overall trends. Appendix~\ref{app:ground_motion} presents a study of seismometers installed at different locations, illustrating this behavior and providing pairwise comparisons between them. The ETMY location exhibits a higher amplitude in the anthropogenic band, which led to its seismometer being chosen as the reference in this analysis --- although any of the seismometers could have been selected.

\begin{figure*}[ht!]
    \centering
    \begin{subfigure}{0.32\textwidth}
        \centering
        \caption{June}
        \includegraphics[width=\textwidth]{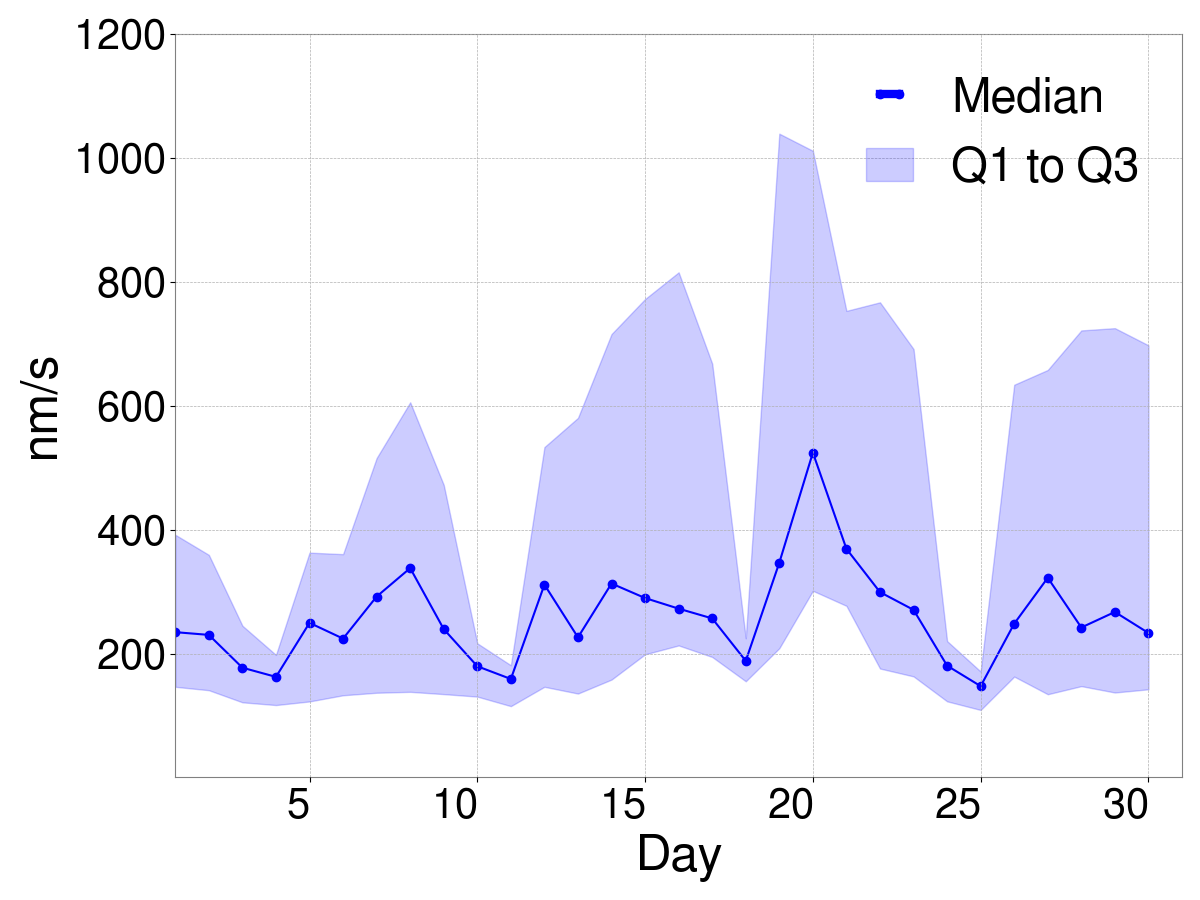}
        \label{fig:June_ETMY_Z_BLRMS_1_3}
    \end{subfigure}    
    \begin{subfigure}{0.32\textwidth}
        \centering
        \caption{October}
        \includegraphics[width=\textwidth]{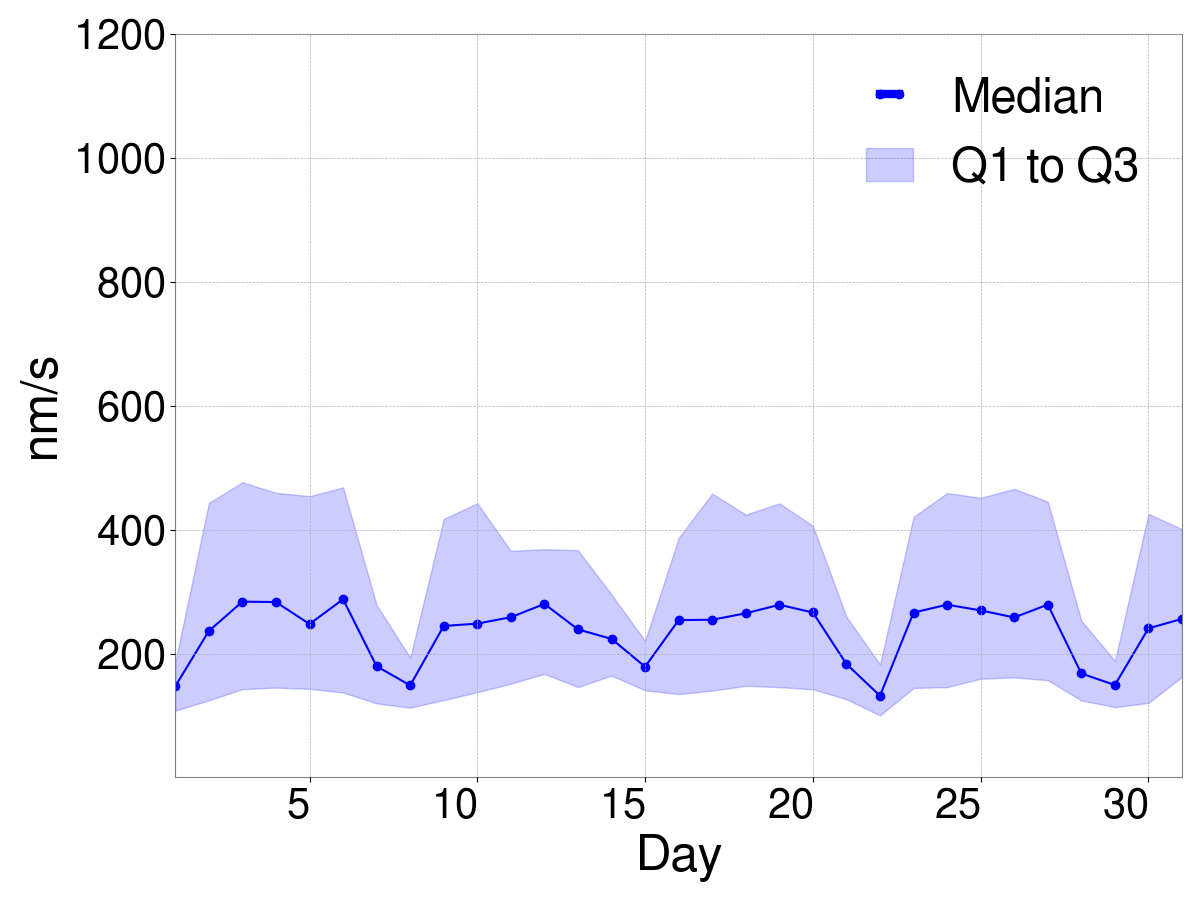}
        \label{fig:October_ETMY_Z_BLRMS_1_3}
    \end{subfigure}
    \begin{subfigure}{0.32\textwidth}
        \centering
        \caption{December}
        \includegraphics[width=\textwidth]{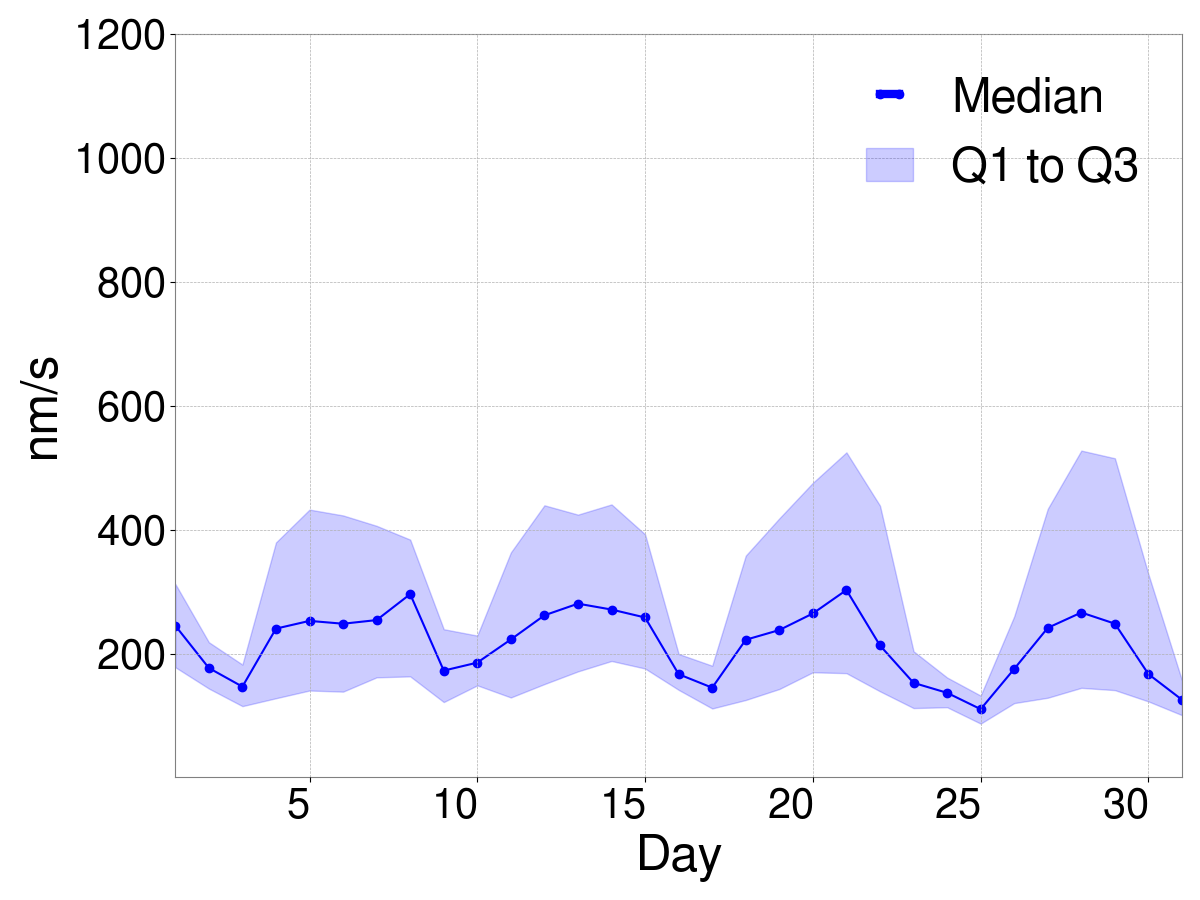}
        \label{fig:December_ETMY_Z_BLRMS_1_3}
    \end{subfigure}
    
    \caption{Daily median ground velocity in the anthropogenic band, measured at ETMY in the $\hat{z}$ direction during three months of O4a, where shaded regions indicate the first (Q1) and third (Q3) quartiles. The periodic reduction in noise corresponds to weekends, reflecting lower human activity.}
    \label{fig:antmotion_by_month}
\end{figure*}

Figure~\ref{fig:antmotion_by_month} shows the ground motion measured by the seismometer at ETMY (in the $\hat{z}$ direction) within the anthropogenic band, which does not present a clear correlation with the glitches identified in the t-SNE output. The figure displays the daily median motion in the anthropogenic band for June, October, and December, with the corresponding first and third quartiles indicated by the shaded regions. The remaining months follow a similar pattern, characterized by increased motion on weekdays-primarily due to nearby traffic-and significantly reduced motion on weekends, resulting in a visible seven-day modulation in the data.

Figure~\ref{fig:mm_vs_gm_by_month} presents the monthly median of the daily ground motion in the $\hat{x}$ direction for the two additional frequency bands defined in this study. The velocity in the LMS band is shown in magenta, while the HMS band is represented in green. In this case, data are plotted exclusively for observing times --- that is, the daily median velocity was calculated only during periods when LIGO was operational, to facilitate correspondence with glitch activity. Among all months, May, June, and July exhibited the lowest ground motion, with only slight fluctuations on specific days. July, in particular, had consistently low values throughout the entire month, with the highest number of days exhibiting low ground motion. The t-SNE plots for these same months --- Figures~\ref{fig:May_L1}, \ref{fig:June_L1}, and \ref{fig:July_L1} --- show a predominance of \textit{Group L$_2$} across all three, with the highest percentage occurring, coincidentally, in July, where \textit{Group L$_2$} accounts for $71.5\%$ of the data. These results suggest a weak correlation between ground motion and the presence of \textit{Group L$_2$}.

Beginning in August, ground motion started to increase significantly, reaching higher velocity peaks, particularly in November, December, and January (winter). This trend aligns with the emergence and persistence of the other two clusters (Groups L$_1$ and L$_3$) in the t-SNE output. In August, the first month to show a noticeable increase in ground motion on some days, we observe a transition: Group L$_2$ is no longer dominant, and its presence continues to decline as ground motion increases. This temporal evolution reinforces the weak relationship between Group L$_2$ and ground motion, while emphasizing the seasonal behavior of Groups L$_1$ and L$_3$, which show a strong correlation with increased ground motion.

%Before running t-SNE, the unclustered Omicron data was normalized. Figure~\ref{L1:tsne_not_normalized} presents a KDE plot when the SNR is not normalized. %On December 18th, an outlier with exceptionally high ground velocity (approximately 3000 $nm/s$) was observed. On this day, observations were limited to only a few hours. However, due to the elevated ground motion, the interferometer experienced a lock loss - a condition in which it loses stability in controlling the test masses. This results in a complete loss of sensitivity, forcing the instrument out of its observing mode.

\begin{figure*}[ht!]    
    \begin{subfigure}{0.49\textwidth}
        \centering
        \caption{Lower Microseismic band}
        \includegraphics[width=\textwidth]{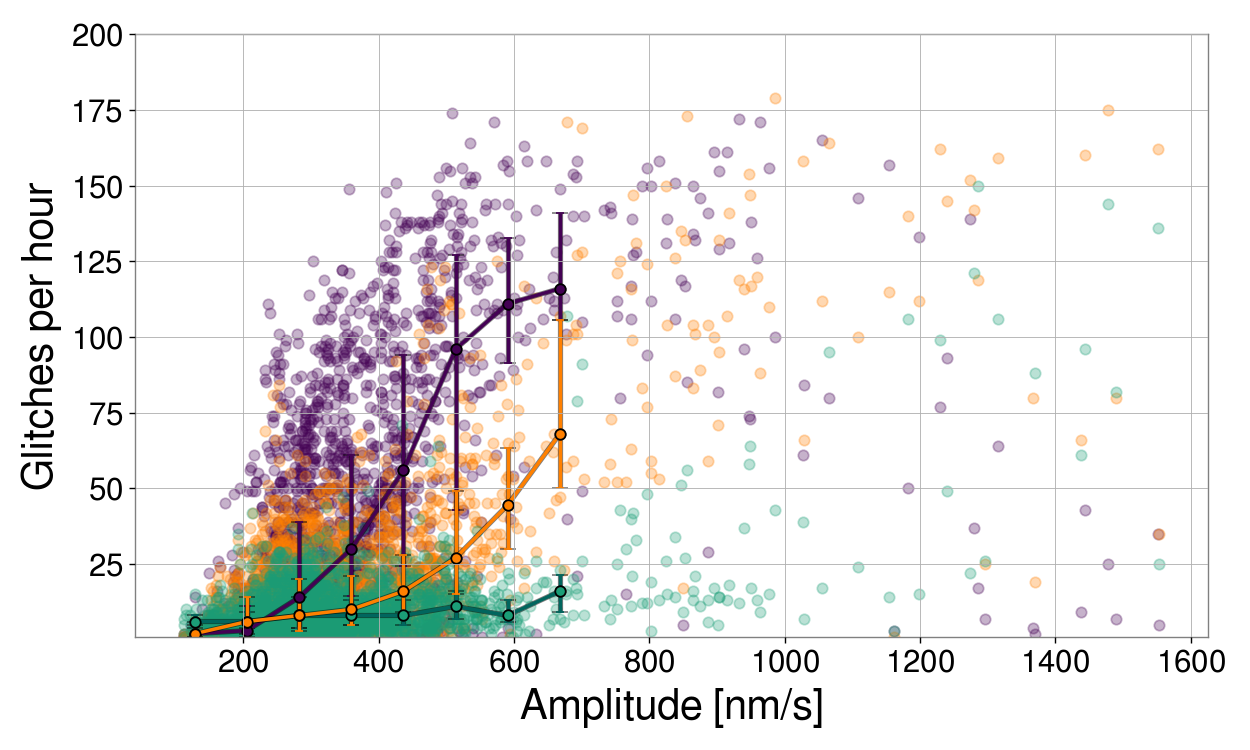}
        \label{fig:lmx_vs_groups}
    \end{subfigure}
    \begin{subfigure}{0.49\textwidth}
        \centering
        \caption{Higher Microseismic band}
        \includegraphics[width=\textwidth]{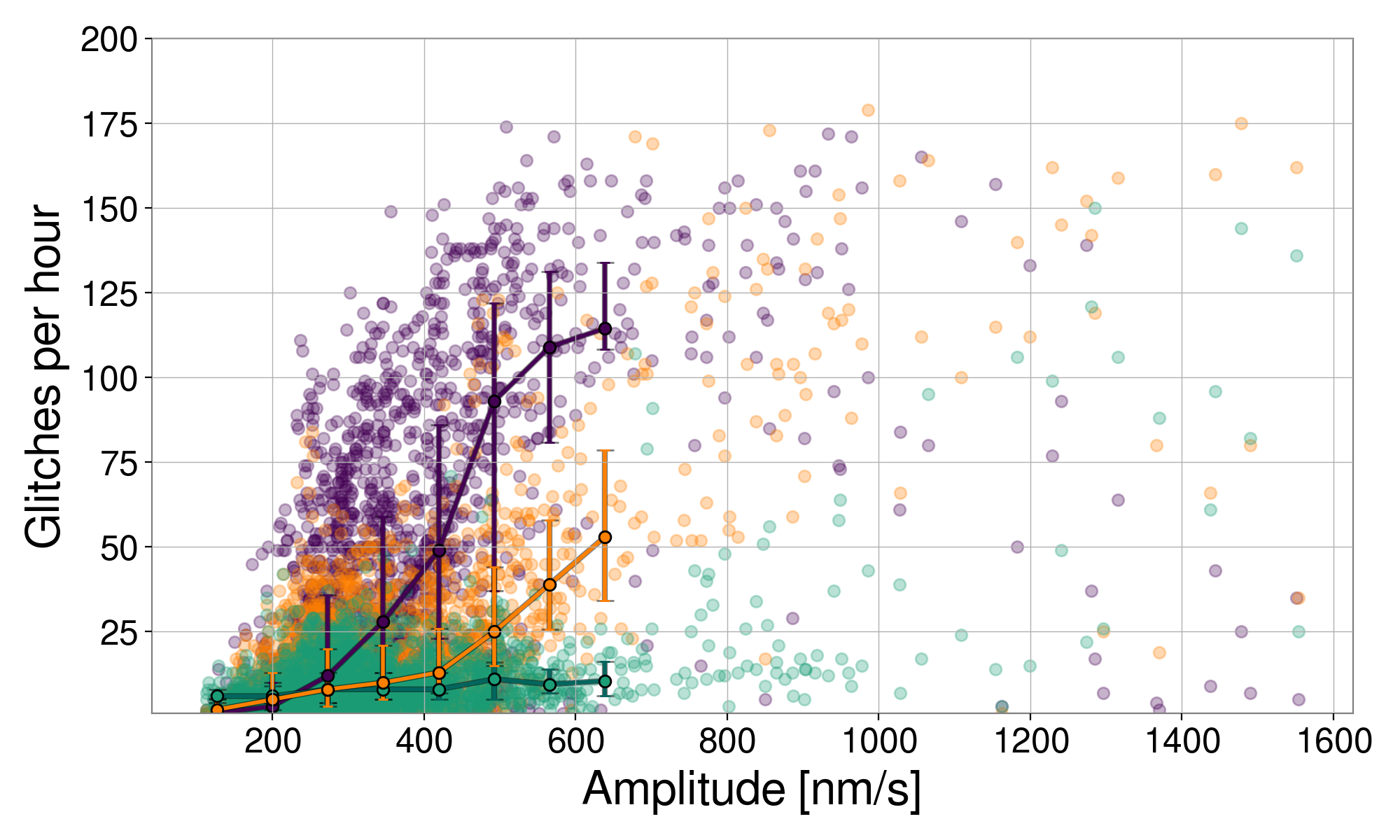}
        \label{fig:hmx_vs_groups}
    \end{subfigure}

    \medskip

    \centering
    \begin{subfigure}{0.49\textwidth}
        \centering
        \caption{Anthropogenic band}
        \includegraphics[width=\textwidth]{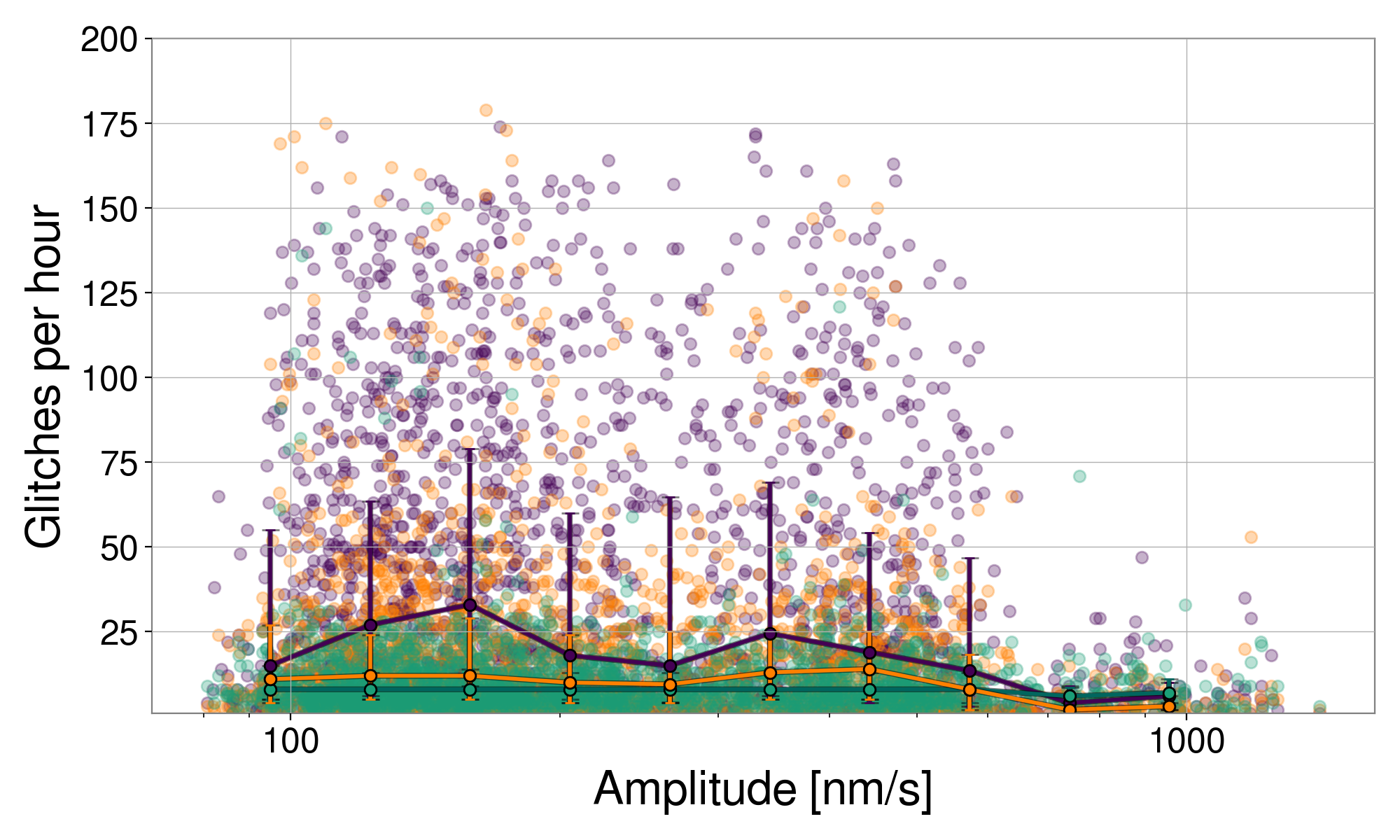}
        \label{fig:amz_vs_groups}
    \end{subfigure}

    \medskip
    
    \begin{subfigure}{0.9\textwidth}
        \centering
        \includegraphics[width=0.9\textwidth]{figures/legend.png}
        %\label{fig:legend}
    \end{subfigure}

    \centering

    \caption{Number of glitches as a function of ground motion amplitude for the three different bands, separated according to the groups identified using t-SNE and Agglomerative Clustering. Each dot represents a measurement interval, color-coded by group. Solid lines indicate the median number of glitches per amplitude bin (computed using the central 98\% of the data), with error bars showing the interquartile range.}
    \label{fig:mm_vs_gm_by_groups}
\end{figure*}

To further support the main conclusions regarding the relationship between the high rate of low-frequency glitches (associated with Groups L$_1$ and L$_3$) and ground motion, we classified all glitches detected at LLO during O4a with frequencies between 10 and 2048 Hz and a minimum SNR of 7.5, assigning each to one of the three previously identified groups. In total, over one hundred thousand glitches were classified. After classifying the glitches, the data were organized to show hourly count of glitches from each class, across all days. For each corresponding hour, the median ground motion amplitude was also computed for the three defined frequency bands. Figures~\ref{fig:lmx_vs_groups}, \ref{fig:hmx_vs_groups}, and \ref{fig:amz_vs_groups} present the hourly rate of each group as a function of the median ground motion in the LMS, HMS, and anthropogenic bands, respectively.

As shown by the distribution of data points, the glitch rates for Group L$_1$ (purple) and Group L$_3$ (orange) exhibit a positive correlation with increasing amplitudes in the LMS and HMS bands (Figures~\ref{fig:lmx_vs_groups} and~\ref{fig:hmx_vs_groups}). In contrast, this trend is not observed for the anthropogenic band, suggesting no significant correlation in that case (Figures~\ref{fig:amz_vs_groups}). During the O4a period, some anthropogenic activities, such as logging~\cite{alog65597,alog65627} took place near the interferometer, leading to unusually high amplitudes in this band (exceeding \SI{600}{nm/s}); this effect was observed, for example, in June, as illustrated by the variations shown in Figure~\ref{fig:June_ETMY_Z_BLRMS_1_3}. Coincidentally, microseismic noise was low at that time, and since most glitches were associated with microseismic motion, the overall glitch rate remained low. As a result, the region on the right side of Figure~\ref{fig:amz_vs_groups} behaves as an outlier when compared to the main distribution. This observation reinforces the conclusion that high amplitudes in the anthropogenic band do not significantly affect the glitch rate, confirming that this band did not contribute meaningfully to glitches with SNR $>$ 7.5, as also suggested in~\cite{alog77176}.

The curves in the figures represent the median glitch rate within amplitude bins, and are shown only when each bin contains at least 20 data points. This criterion helps to exclude sparsely populated bins, minimizing the influence of statistical fluctuations or outliers. To further reduce the influence of extreme values, the median glitch rates were computed using the central 98\% of the data, excluding the upper and lower 1\% of the distribution. Group L$_1$ appears to exhibit a flatter trend or even a decrease in glitch rate at very high amplitudes of microseismic motion. This effect is attributed to a shift in the peak frequency of the glitches at elevated ground motion levels. Since Group L$_1$ is typically characterized by a peak frequency around \SI{12}{Hz}, higher ground motion can shift this peak upward, causing the classifier to instead assign these glitches to Group L$_3$, which is defined by a primary peak frequency around 20~Hz. This transition tends to occur when the ground motion amplitude becomes significantly higher. 

We quantified the correlation between each identified group and each ground-motion band using the Pearson correlation coefficient ($r$)~\cite{benesty2009pearson}, which measures the strength of the linear relationship between two variables---in this case, the glitch rate and ground motion in different bands. Table~\ref{tab:pearson_llo} summarizes the correlation coefficients for each case, computed from more than 2,000 valid data points, each corresponding to a one-hour interval. Overall, the results confirm a strong correlation between Groups L$_1$ and L$_3$ and both the LMS and HMS bands, indicating a clear dependence of their occurrence on ground motion; their corresponding $p$-values (where smaller values mean the correlation is less likely to be due to chance) were below $10^{-100}$. Group L$_2$ exhibits a weak correlation with the LMS band, which may partially arise from classification overlaps or a subclass with such dependence. As expected, none of the groups shows a significant correlation with the anthropogenic band. However, due to high-amplitude outliers associated with anthropogenic activities (previously mentioned), a small negative correlation appears for all three groups, with $p$-values equal to or below $0.05$. Despite these small $p$-values (likely driven by the large sample size), the corresponding correlation coefficients are very close to zero when considering only ground-motion amplitudes below $600$ nm/s (see Table~\ref{tab:pearson_llo}), indicating no meaningful linear relationship.

\begin{table*}[!tb]
\centering
\caption{Pearson correlation between LMS, HMS, and anthropogenic bands for Groups L$_1$, L$_2$, and L$_3$.}
\label{tab:pearson_llo}

\begin{tabular}{lccc}
\hline
Correlation & LMS & HMS & Ant. band \\
\hline
Group L$_1$ & 0.64 & 0.63 & -0.03 \\
Group L$_2$ & 0.32 & 0.17 & -0.04 \\
Group L$_3$ & 0.53 & 0.47 & -0.05 \\
\hline
\end{tabular}
\end{table*}

In conclusion, applying t-SNE to unclustered Omicron data from LLO enabled the identification of two glitch groups that follow the seasonal behavior of microseismic ground motion, providing strong support for the discussion reported in~\cite{soni2025ligo}.

\section{Glitches during O4a at LHO}
\label{sec:o4a_glitches_lho}

\begin{figure}[ht] 
    \centering 
    \includegraphics[width=\columnwidth]{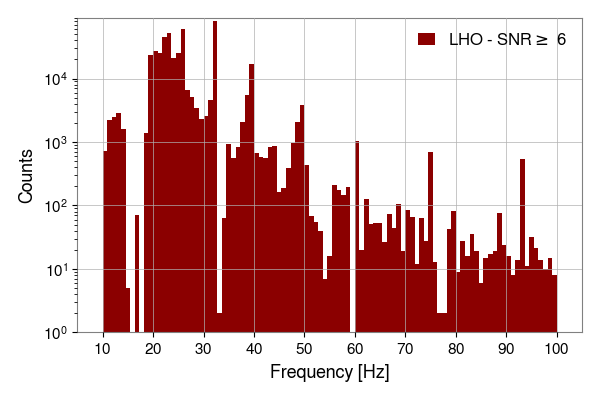} \caption{Histogram of the peak frequencies of clustered Omicron triggers during O4a at LHO. The plot highlights the most common glitch frequency range, between 20 and \SI{40}{Hz}, considering only glitches with SNR greater than or equal to 6.} 
    \label{fig:freq_hist_H1} 
\end{figure}

The rate of glitches with an SNR greater than 10 was significantly lower at LHO during O4a (approximately 6 per hour) compared to O3 (around 17 per hour). However, LHO was notably affected by glitches with lower SNR. For this reason, while the LLO analysis used a minimum SNR of 7.5, we lowered the threshold to 6 at LHO to retain this low-SNR broadband glitch population. As reported in Ref.~\cite{soni2025ligo}, when considering glitches with an SNR exceeding 6.5, Hanford experienced an average of 77 glitches per hour during O4a --- more than double the glitch rate observed during O3, which was approximately 33 per hour. This increase is attributed to the presence of broadband transient noise between 20–40 Hz during O4a, which occasionally extended to peak frequencies around 50 Hz. The predominance of this frequency range is evident in Figure~\ref{fig:freq_hist_H1}, which shows the histogram of glitches during O4a with an SNR greater than or equal to 6.

\begin{figure}{ht!}
    \centering
    \begin{subfigure}{0.475\textwidth}
        \centering
        \caption{Low amplitude motion in the LMS band}
        \includegraphics[width=\textwidth]{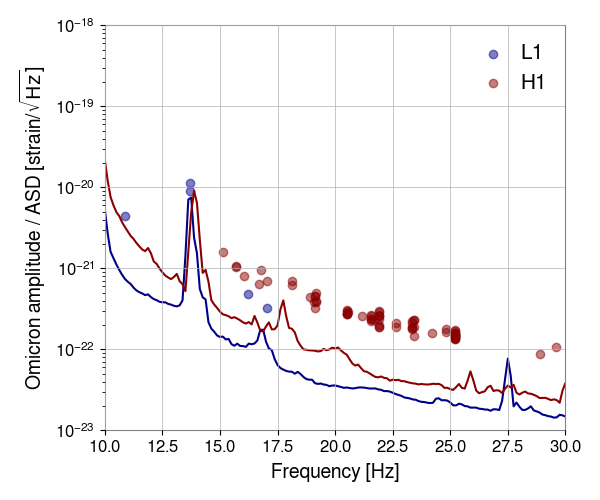}
        \label{fig:charstrain_h1_lowmseismic}
    \end{subfigure}
    \hfill
    \begin{subfigure}{0.475\textwidth}
        \centering
        \caption{High amplitude motion in the LMS band}
        \includegraphics[width=\textwidth]{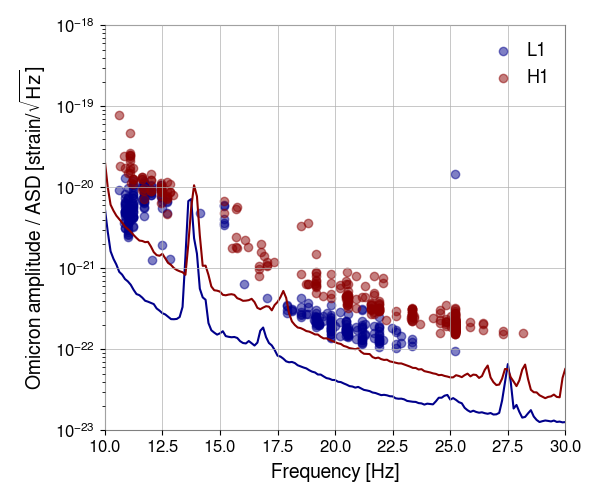}
        \label{fig:charstrain_h1_highmseismic}
    \end{subfigure}

    \caption{Amplitude spectral densities of Hanford (dark red) and Livingston (dark blue) strain channels during selected two-hour periods characterized by (a) low and (b) high amplitude ground motion in the low microseismic (LMS) band, approximately \SI{150}{nm/s} and \SI{1000}{nm/s}, respectively. The dots represent the Omicron amplitudes and peak frequencies of glitches with SNR~$\geq~6.0$ detected during the corresponding periods. The ASD curves are shown as references for the detector noise levels and are not directly equivalent to the Omicron trigger amplitudes.}
    \label{fig:charcteristic_strain_h1}
\end{figure}

In comparison to LLO (Figure~\ref{fig:freq_hist_L1}), LHO exhibits noticeably fewer glitches below \SI{20}{Hz}, even when lower-SNR glitches are considered. This difference may be partly attributed to the better low-frequency sensitivity of the Livingston detector. The contrasting sensitivities of the two detectors are illustrated in Figures~\ref{fig:charstrain_h1_lowmseismic} and~\ref{fig:charstrain_h1_highmseismic}, which compare the amplitude spectral densities of LHO (dark red) and LLO (dark blue) during independently selected two-hour periods of O4a, chosen to represent comparable levels of microseismic motion at both observatories under relatively low and high microseismic activity, respectively. The narrow spectral features visible in Figure~\ref{fig:charcteristic_strain_h1} are also associated with known instrumental lines. In particular, the features around 14 Hz are associated with the test masses' roll modes, those between approximately 15 and 17 Hz include calibration lines, and the features around 27--28 Hz are associated with suspension bounce modes~\cite{goetz2024o4alines}. The dots represent the Omicron amplitudes and peak frequencies of glitches with SNR $\geq 6.0$ detected during the same periods. During low microseismic activity (Figure~\ref{fig:charstrain_h1_lowmseismic}), fewer glitches are observed at both observatories, although some broadband glitches remain present at Hanford. In contrast, during periods of elevated microseismic motion (Figure~\ref{fig:charstrain_h1_highmseismic}), both observatories exhibit a higher occurrence of transient noise spanning a broader frequency range, including frequencies below \SI{13}{Hz}.

\begin{figure}[ht] 
    \centering 
    \includegraphics[width=\columnwidth]{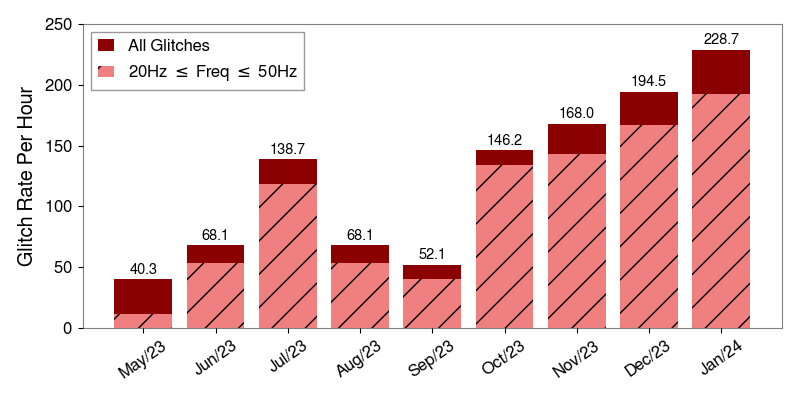} 
    \caption{Average glitch rate per hour for each month during O4a at LHO. All clustered Omicron triggers between \SI{10} and \SI{2048}{\hertz} with SNR greater than or equal to 6 are shown in solid dark red bars, while broadband glitches are indicated by hatched bars.} \label{fig:glitches_per_hour_per_month_h1} 
\end{figure}

The monthly glitch rate in Hanford data during O4a is shown in Figure~\ref{fig:glitches_per_hour_per_month_h1}. The overall trend closely resembles that observed at LLO, with a notable increase beginning in October, but with an additional rise at LHO in July. The broadband glitches in the 20 to \SI{50}{Hz} range are highlighted by the red bars with diagonal hatching, illustrating their predominance throughout the O4a run, except during May. %Note that, since LHO was more affected by low-SNR glitches, the analyses presented here consider glitches with a minimum SNR of 6, whereas for LLO, we use 7.5.

\begin{figure*}[ht!]
    \centering
    \begin{subfigure}{0.49\textwidth}
        \centering
        \caption{}
        \includegraphics[width=\textwidth]{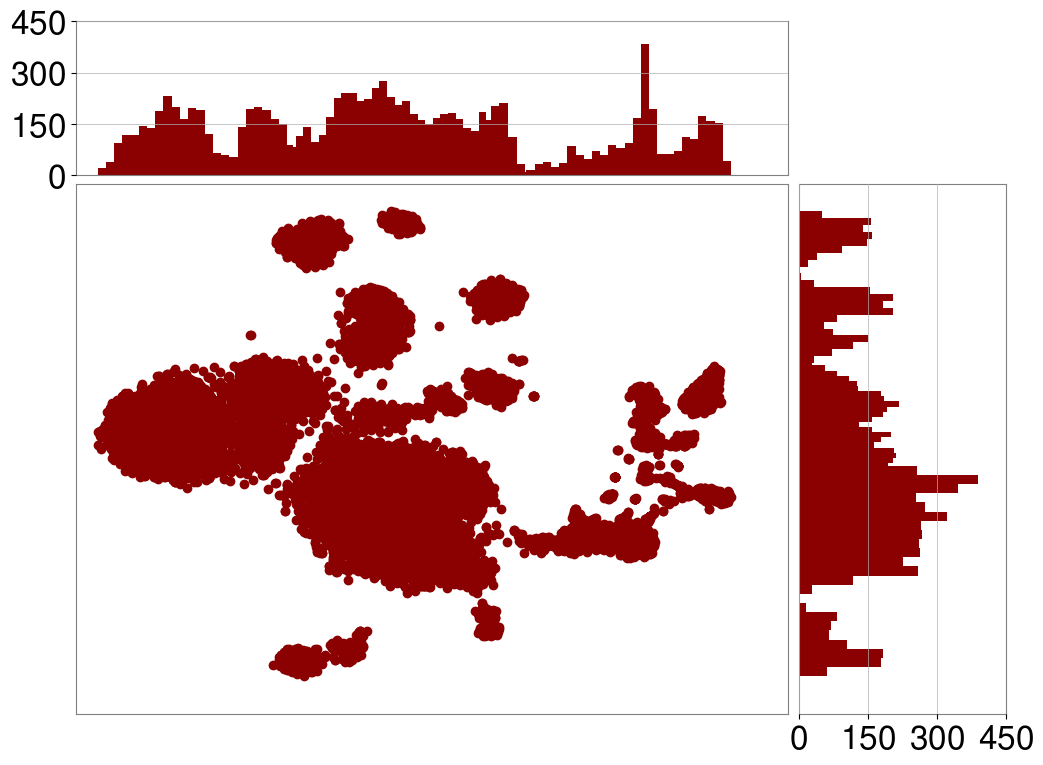}
        \label{fig:o4a_tsne_h1_unclassified}
    \end{subfigure}
    \hfill
    \begin{subfigure}{0.49\textwidth}
        \centering
        \caption{}
        \includegraphics[width=\textwidth]{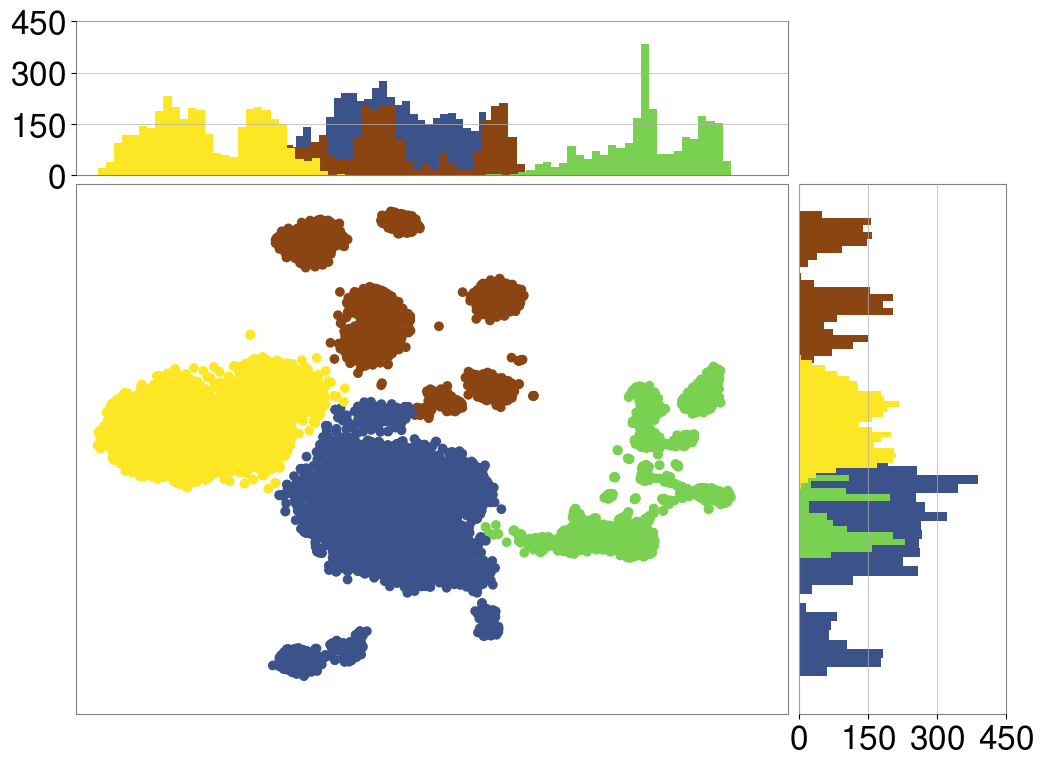}
        \label{fig:o4a_tsne_h1_classified}
    \end{subfigure}

    \begin{subfigure}{\textwidth}
        \centering
        \includegraphics[width=1.0\textwidth]{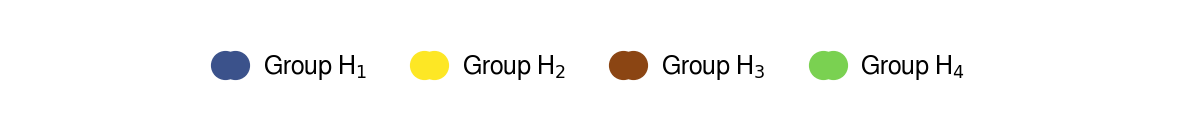}
        %\label{fig:legend_six_groups}
    \end{subfigure}
    
    \caption{(a) Scatter plot of the 2D t-SNE output coordinates
    representing 13,500 data points from LHO, each originally
    in a 1,230-dimensional space derived from unclustered
    Omicron triggers. Each point corresponds to a glitch.
    The histograms on the top and right axes illustrate the
    distribution of glitches along each coordinate.
    (b) The same t-SNE representation, with points
    colored according to the groups assigned by
    Agglomerative Clustering.}
    \label{fig:o4a_tsne_h1}
\end{figure*}

By employing the same t-SNE-based methodology, we analyzed the transient glitches observed at LHO during the O4a run. Figure~\ref{fig:o4a_tsne_h1_unclassified} shows the two-dimensional t-SNE representation of the glitches before applying Agglomerative Clustering, while Figure~\ref{fig:o4a_tsne_h1_classified} shows the same representation with the glitches colored according to the groups identified by Agglomerative Clustering. The classified output highlights four main groups, labeled Group H1 through Group H4.

%The unclassified data reveal two big clusters located in the bottom-center and left-center regions, along with smaller clusters toward the top and right; The classified output highlights four main groups, labeled Group H$_1$ through Group H$_4$. %In both panels, marginal histograms are shown along the top and right axes, representing the glitch density across the 2D embedding. 

\begin{figure*}[ht!]
    \centering

    \begin{subfigure}[t]{0.32\textwidth}
        \centering
        \caption{May}
        {\fontsize{7.5}{9}\selectfont
        \begin{tabular}{c@{\hspace{0.82em}}c@{\hspace{0.82em}}c@{\hspace{0.82em}}c}
        \textbf{\textcolor{myblue}{7.4/h}} &
        \textbf{\textcolor{myyellow}{1.5/h}} &
        \textbf{\textcolor{mybrown}{24.3/h}} &
        \textbf{\textcolor{mygreen}{7.0/h}}
        \end{tabular}
        }

        \fbox{\includegraphics[width=0.9\textwidth]{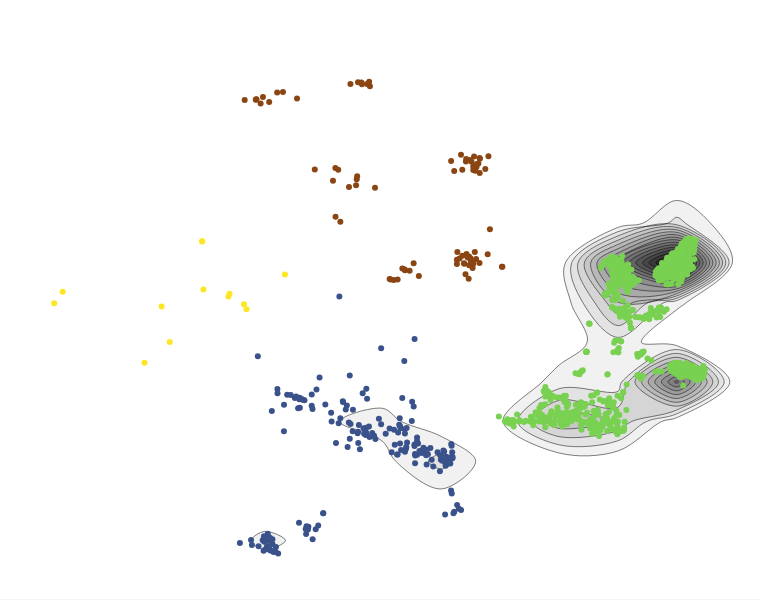}}

        {\fontsize{9.5}{9}\selectfont
        \begin{tabular}{c@{\hspace{1em}}c@{\hspace{1em}}c@{\hspace{1em}}c}
        \textbf{\textcolor{myblue}{18.4\%}} &
        \textbf{\textcolor{myyellow}{3.8\%}} &
        \textbf{\textcolor{mybrown}{60.4\%}} &
        \textbf{\textcolor{mygreen}{17.3\%}}
        \end{tabular}
        }
        \label{fig:May_H1}
    \end{subfigure}
    \hfill
    \begin{subfigure}[t]{0.32\textwidth}
        \centering
        \caption{June}
        {\fontsize{7.5}{9}\selectfont
        \begin{tabular}{c@{\hspace{0.82em}}c@{\hspace{0.82em}}c@{\hspace{0.82em}}c}
        \textbf{\textcolor{myblue}{35/h}} &
        \textbf{\textcolor{myyellow}{6.2/h}} &
        \textbf{\textcolor{mybrown}{21.4/h}} &
        \textbf{\textcolor{mygreen}{5.5/h}}
        \end{tabular}
        }

        \fbox{\includegraphics[width=0.9\textwidth]{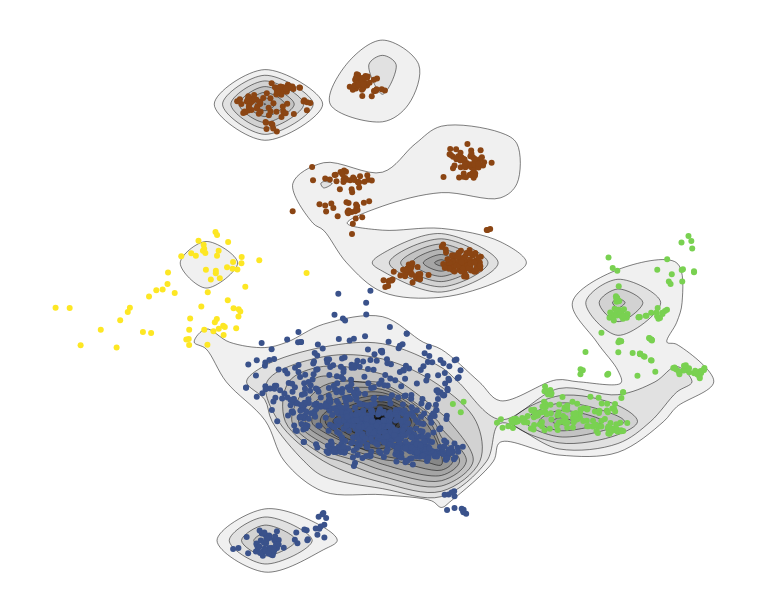}}

        {\fontsize{9.5}{9}\selectfont
        \begin{tabular}{c@{\hspace{1em}}c@{\hspace{1em}}c@{\hspace{1em}}c}
        \textbf{\textcolor{myblue}{51.4\%}} &
        \textbf{\textcolor{myyellow}{9.1\%}} &
        \textbf{\textcolor{mybrown}{31.4\%}} &
        \textbf{\textcolor{mygreen}{8.1\%}}
        \end{tabular}
        }
        \label{fig:June_H1}
    \end{subfigure}
    \hfill
    \begin{subfigure}[t]{0.32\textwidth}
        \centering
        \caption{July}
        {\fontsize{7.5}{9}\selectfont
        \begin{tabular}{c@{\hspace{0.82em}}c@{\hspace{0.82em}}c@{\hspace{0.82em}}c}
        \textbf{\textcolor{myblue}{62.9/h}} &
        \textbf{\textcolor{myyellow}{23.6/h}} &
        \textbf{\textcolor{mybrown}{26.4/h}} &
        \textbf{\textcolor{mygreen}{25.8/h}}
        \end{tabular}
        }
        \fbox{\includegraphics[width=0.9\textwidth]{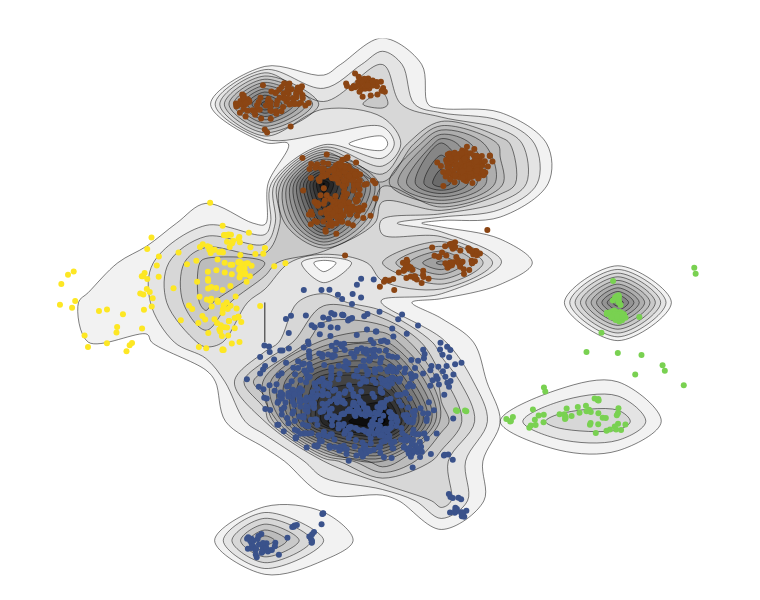}}

        {\fontsize{9.5}{9}\selectfont
        \begin{tabular}{c@{\hspace{1em}}c@{\hspace{1em}}c@{\hspace{1em}}c}
        \textbf{\textcolor{myblue}{45.4\%}} &
        \textbf{\textcolor{myyellow}{17.0\%}} &
        \textbf{\textcolor{mybrown}{19.0\%}} &
        \textbf{\textcolor{mygreen}{18.6\%}}
        \end{tabular}
        }
        \label{fig:July_H1}
    \end{subfigure}

    \medskip

    \begin{subfigure}[t]{0.32\textwidth}
        \centering
        \caption{August}
        {\fontsize{7.5}{9}\selectfont
        \begin{tabular}{c@{\hspace{0.82em}}c@{\hspace{0.82em}}c@{\hspace{0.82em}}c}
        \textbf{\textcolor{myblue}{14.6/h}} &
        \textbf{\textcolor{myyellow}{7.5/h}} &
        \textbf{\textcolor{mybrown}{29.8/h}} &
        \textbf{\textcolor{mygreen}{16.1/h}}
        \end{tabular}
        }

        \fbox{\includegraphics[width=0.9\textwidth]{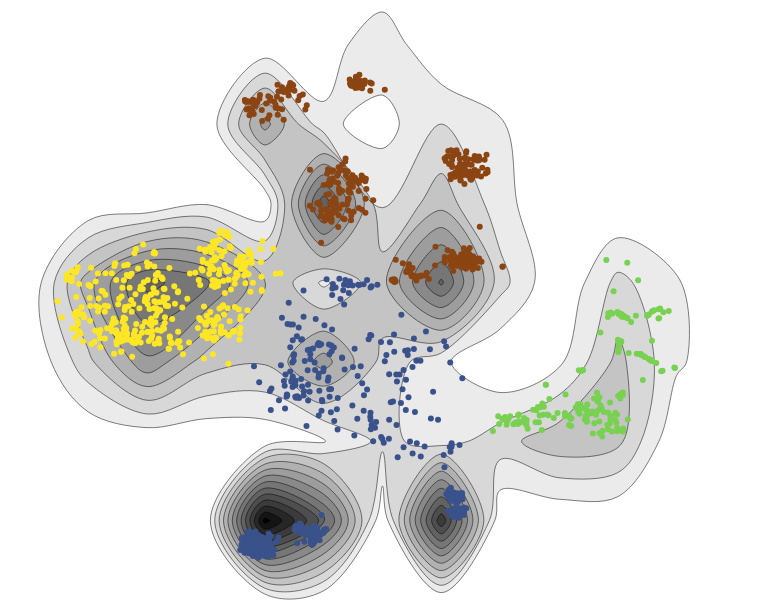}}

        {\fontsize{9.5}{9}\selectfont
        \begin{tabular}{c@{\hspace{1em}}c@{\hspace{1em}}c@{\hspace{1em}}c}
        \textbf{\textcolor{myblue}{21.5\%}} &
        \textbf{\textcolor{myyellow}{11.1\%}} &
        \textbf{\textcolor{mybrown}{43.9\%}} &
        \textbf{\textcolor{mygreen}{23.6\%}}
        \end{tabular}
        }
        \label{fig:August_H1}
    \end{subfigure}
    \hfill
    \begin{subfigure}[t]{0.32\textwidth}
        \centering
        \caption{September}
        {\fontsize{7.5}{9}\selectfont
        \begin{tabular}{c@{\hspace{0.82em}}c@{\hspace{0.82em}}c@{\hspace{0.82em}}c}
        \textbf{\textcolor{myblue}{11.6/h}} &
        \textbf{\textcolor{myyellow}{6.6/h}} &
        \textbf{\textcolor{mybrown}{17.5/h}} &
        \textbf{\textcolor{mygreen}{16.4/h}}
        \end{tabular}
        }

        \fbox{\includegraphics[width=0.9\textwidth]{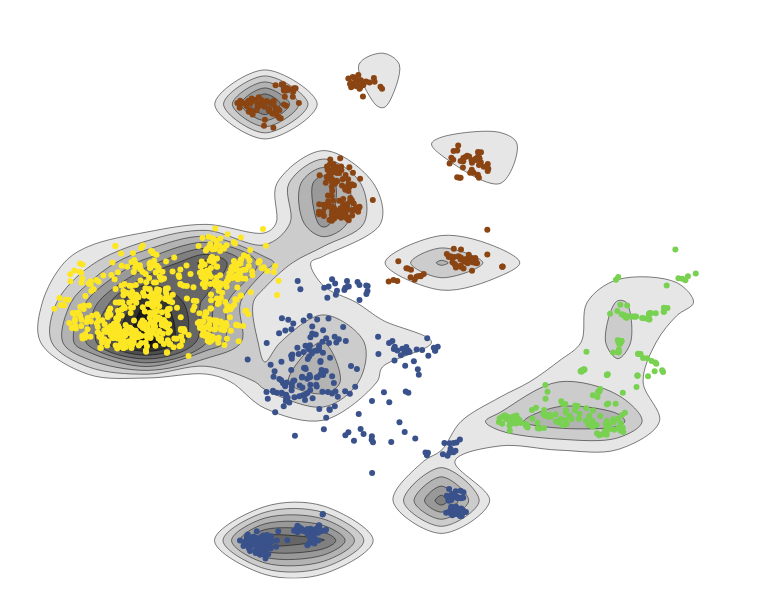}}

        {\fontsize{9.5}{9}\selectfont
        \begin{tabular}{c@{\hspace{1em}}c@{\hspace{1em}}c@{\hspace{1em}}c}
        \textbf{\textcolor{myblue}{22.4\%}} &
        \textbf{\textcolor{myyellow}{12.6\%}} &
        \textbf{\textcolor{mybrown}{33.6\%}} &
        \textbf{\textcolor{mygreen}{31.4\%}}
        \end{tabular}
        }
        \label{fig:September_H1}
    \end{subfigure}
    \hfill
    \begin{subfigure}[t]{0.32\textwidth}
        \centering
        \caption{October}
        {\fontsize{7.5}{9}\selectfont
        \begin{tabular}{c@{\hspace{0.82em}}c@{\hspace{0.82em}}c@{\hspace{0.82em}}c}
        \textbf{\textcolor{myblue}{74.2/h}} &
        \textbf{\textcolor{myyellow}{14.0/h}} &
        \textbf{\textcolor{mybrown}{16.5/h}} &
        \textbf{\textcolor{mygreen}{41.4/h}}
        \end{tabular}
        }

        \fbox{\includegraphics[width=0.9\textwidth]{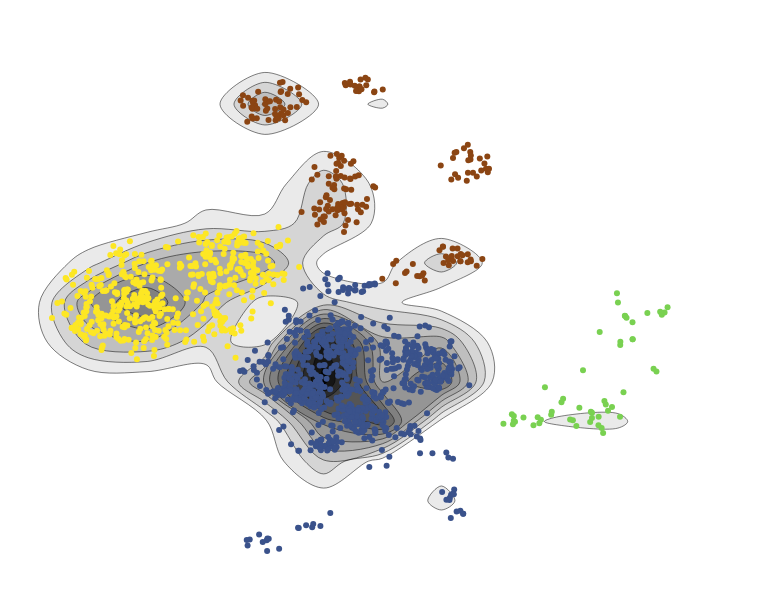}}

        {\fontsize{9.5}{9}\selectfont
        \begin{tabular}{c@{\hspace{1em}}c@{\hspace{1em}}c@{\hspace{1em}}c}
        \textbf{\textcolor{myblue}{50.8\%}} &
        \textbf{\textcolor{myyellow}{9.6\%}} &
        \textbf{\textcolor{mybrown}{11.3\%}} &
        \textbf{\textcolor{mygreen}{28.3\%}}
        \end{tabular}
        }
        \label{fig:October_H1}
    \end{subfigure}

    \medskip

    \begin{subfigure}[t]{0.32\textwidth}
        \centering
        \caption{November}
        {\fontsize{7.5}{9}\selectfont
        \begin{tabular}{c@{\hspace{0.82em}}c@{\hspace{0.82em}}c@{\hspace{0.82em}}c}
        \textbf{\textcolor{myblue}{66.4/h}} &
        \textbf{\textcolor{myyellow}{17.8/h}} &
        \textbf{\textcolor{mybrown}{26.1/h}} &
        \textbf{\textcolor{mygreen}{57.6/h}}
        \end{tabular}
        }

        \fbox{\includegraphics[width=0.9\textwidth]{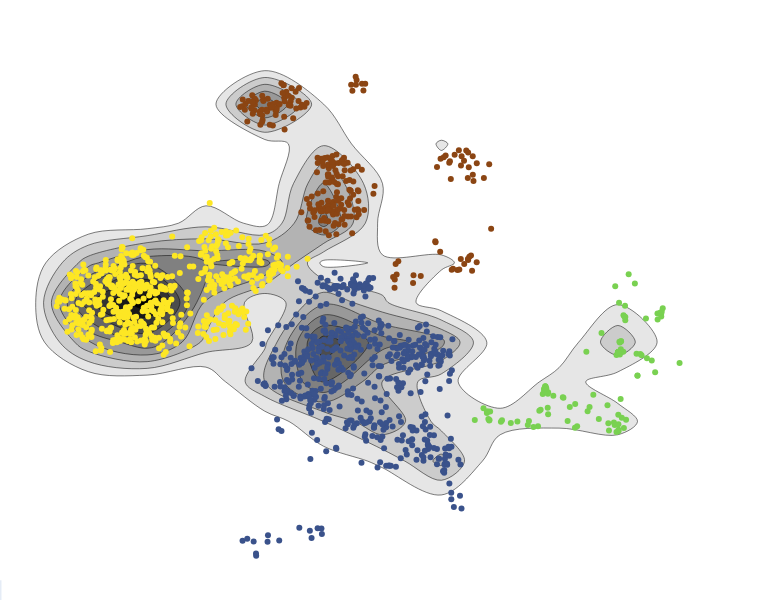}}

        {\fontsize{9.5}{9}\selectfont
        \begin{tabular}{c@{\hspace{1em}}c@{\hspace{1em}}c@{\hspace{1em}}c}
        \textbf{\textcolor{myblue}{39.6\%}} &
        \textbf{\textcolor{myyellow}{10.6\%}} &
        \textbf{\textcolor{mybrown}{15.5\%}} &
        \textbf{\textcolor{mygreen}{34.3\%}}
        \end{tabular}
        }
        \label{fig:November_H1}
    \end{subfigure}
    \hfill
    \begin{subfigure}[t]{0.32\textwidth}
        \centering
        \caption{December}
        {\fontsize{7.5}{9}\selectfont
        \begin{tabular}{c@{\hspace{0.82em}}c@{\hspace{0.82em}}c@{\hspace{0.82em}}c}
        \textbf{\textcolor{myblue}{83.3/h}} &
        \textbf{\textcolor{myyellow}{19.4/h}} &
        \textbf{\textcolor{mybrown}{26.3/h}} &
        \textbf{\textcolor{mygreen}{65.5/h}}
        \end{tabular}
        }

        \fbox{\includegraphics[width=0.9\textwidth]{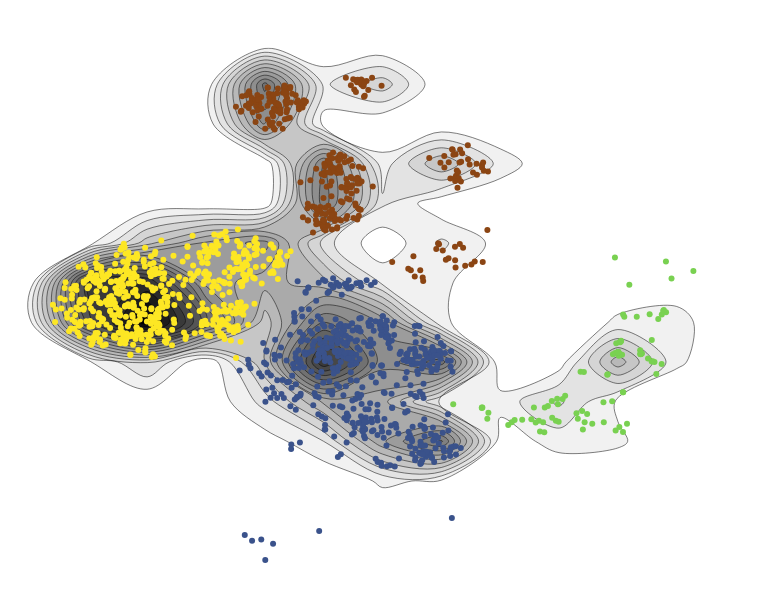}}

        {\fontsize{9.5}{9}\selectfont
        \begin{tabular}{c@{\hspace{1em}}c@{\hspace{1em}}c@{\hspace{1em}}c}
        \textbf{\textcolor{myblue}{42.8\%}} &
        \textbf{\textcolor{myyellow}{10.0\%}} &
        \textbf{\textcolor{mybrown}{13.5\%}} &
        \textbf{\textcolor{mygreen}{33.7\%}}
        \end{tabular}
        }
        \label{fig:December_H1}
    \end{subfigure}
    \hfill
    \begin{subfigure}[t]{0.32\textwidth}
        \centering
        \caption{January}
        {\fontsize{7.5}{9}\selectfont
        \begin{tabular}{c@{\hspace{0.82em}}c@{\hspace{0.82em}}c@{\hspace{0.82em}}c}
        \textbf{\textcolor{myblue}{102.5/h}} &
        \textbf{\textcolor{myyellow}{21.1/h}} &
        \textbf{\textcolor{mybrown}{28.7/h}} &
        \textbf{\textcolor{mygreen}{76.3/h}}
        \end{tabular}
        }

        \fbox{\includegraphics[width=0.9\textwidth]{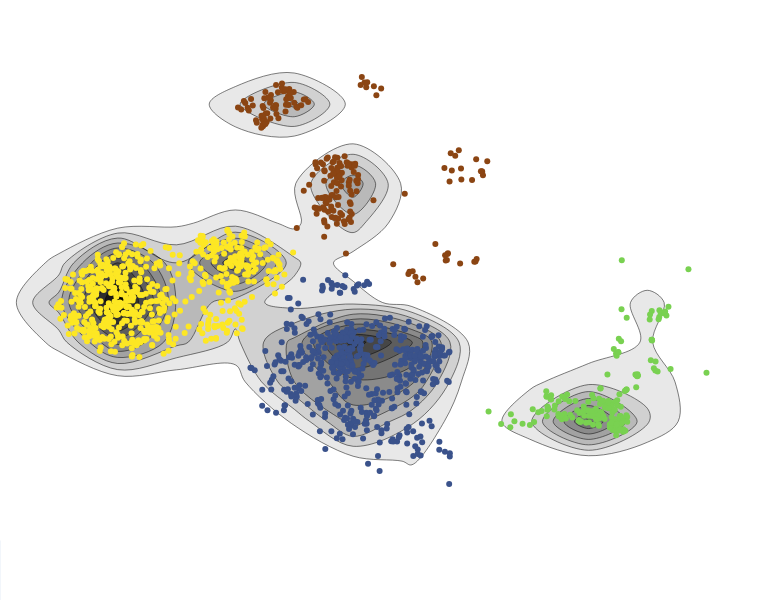}}

        {\fontsize{9.5}{9}\selectfont
        \begin{tabular}{c@{\hspace{1em}}c@{\hspace{1em}}c@{\hspace{1em}}c}
        \textbf{\textcolor{myblue}{44.8\%}} &
        \textbf{\textcolor{myyellow}{9.2\%}} &
        \textbf{\textcolor{mybrown}{12.5\%}} &
        \textbf{\textcolor{mygreen}{33.4\%}}
        \end{tabular}
        }
        \label{fig:January_H1}
    \end{subfigure}

    \begin{subfigure}[t]{\textwidth}
        \centering
        \includegraphics[width=\textwidth]{figures/legend_h1.png}
    \end{subfigure}

    \caption{Monthly distribution of glitches in the
    2D t-SNE space at LHO (May 2023--January 2024),
    showing the evolution of the groups identified
    using Agglomerative Clustering.}
    Values above each panel indicate glitch rates
    (per hour), and values below indicate the fraction
    of glitches in each group.
    \label{fig:h1_tsne_by_month}
\end{figure*}

Figure~\ref{fig:h1_tsne_by_month} presents the t-SNE scatter plots for LHO, filtered by month, enabling us to track the evolution and persistence of glitch clusters over time. In May, less than one week of data was available, during which Group H$_4$ dominated the distribution, accounting for more than 80\% of the glitches. This group primarily consists of glitches analogous to those observed at LLO --- such as \textit{Koi Fish}, \textit{Blip}, and \textit{Extremely Loud} (present in Group L$_2$) --- in addition to low-frequency glitches with a peak frequency around 12~Hz. These low-frequency glitches, which were not as prevalent at LHO as at LLO, occurred over a few hours at the end of May and, due to their low occurrence, they do not form a distinct group at LHO, unlike at LLO. The presence of these glitches in May helps explain the larger discrepancy between the total glitch rate (solid bar) and the rate of broadband glitches (hatched bar) shown in Figure~\ref{fig:glitches_per_hour_per_month_h1}.

In June, the bottom-center region of the \mbox{t-SNE} map becomes increasingly prominent, persists through July, and decreases in August and September, coinciding with the activation of other groups and resulting in a more uniform overall distribution. This decrease was driven by the implementation of an improved alignment control system~\cite{71927}, which substantially reduced the occurrence of glitches with SNR greater than 7.5~\cite{alog71933}. This is also reflected in Figure~\ref{fig:glitches_per_hour_per_month_h1}, which shows a reduction in the glitch rate from 138.7 to 68.1 glitches per hour between July and August.

From October through January, the large cluster H$_1$, together with H$_2$, became the most persistent feature in the \mbox{t-SNE} projection, collectively accounting for at least 70\% of the observed glitches. Although the implementation of the alignment controller substantially contributed to glitch mitigation, an increase in broadband glitches emerged in October and persisted throughout the remainder of the observing run, leading to the predominance of these two clusters. This increase coincided with the relocation of the LSC (Length Sensing and Control) feedforward path from the ETMX (End Test Mass of the X-arm) to the ETMY actuator~\cite{H1alog73420}, which initially reduced noise above \SI{50}{Hz} but introduced additional noise at lower frequencies. Subsequent investigations suggested that the ETMY R0~\footnote{R0 refers to the uppermost stage of the reaction chain~\cite{colgan2023detecting}.} actuator, responsible for length control, may not have been operating properly at that time~\cite{H1alog73447}, potentially generating excess scattered light and contributing to the observed rise in glitch rate.

\begin{figure*}[ht!]
    \centering{\includegraphics[width=1\textwidth]{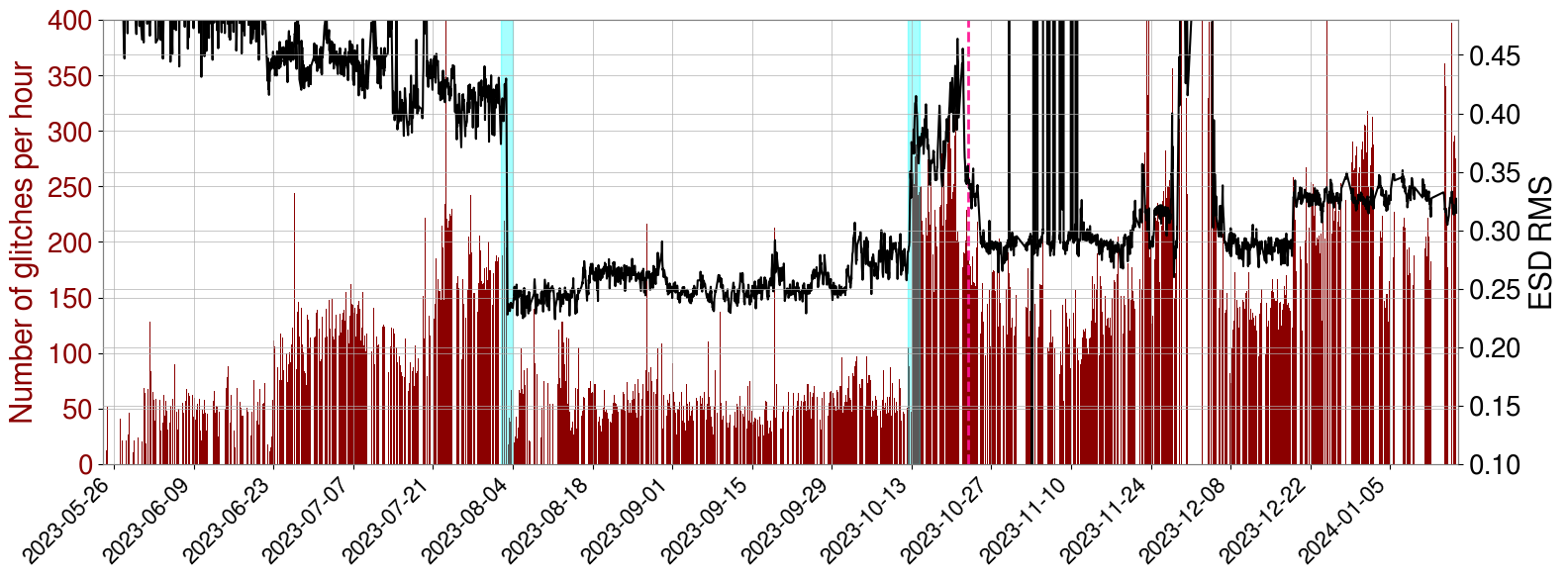}} 
    \caption{Hourly rate of clustered Omicron
    triggers at LHO (dark red) and RMS of the
    electrostatic drive voltage at the ETMX UL
    component (black) during O4a.}
    We selected glitches in the 20--50 Hz frequency
    band and discarded data segments shorter than
    20 minutes to avoid peaks caused by a high
    number of glitches over a short duration.
    The cyan-shaded bars represent the two main
    periods when the ESD RMS (and the glitch rate)
    changed significantly.
    \label{fig:H1_ESD_vs_glitchrate}    
\end{figure*}

A faulty cable connected to the ETMY R0 actuator was identified as the source of the malfunction~\cite{alog73522}; its replacement led to a slight decrease in the glitch rate. This was followed by a new increase, possibly associated with the implementation of an updated feedforward filter in the Signal Recycling Cavity Length (SRCL) control loop of the Length Sensing and Control (LSC) system~\cite{alog73561}, which was later revised, resulting in a marked decrease in the glitch rate from October 23 onward~\cite{alog73662} --- this decrease is highlighted by the vertical pink dashed line in Figure~\ref{fig:H1_ESD_vs_glitchrate}, where the dark red vertical bars represent the evolution of the hourly glitch rate throughout the entire O4a.

A bicoherence~\footnote{Bicoherence~\cite{choudhury2008bispectrum} is a diagnostic tool that assesses nonlinear coupling between two frequencies. If two frequencies, $f_1$ and $f_2$, are present in a signal, bicoherence can indicate whether a new frequency component at $f_1 + f_2$ (or $f_1 - f_2$) is being generated through nonlinear interactions.} analysis revealed that these broadband glitches were related to the electrostatic drive (ESD) system in the detector~\cite{alog71005}. Positioned behind the test mass, the ESD applied controlled forces to the mirror to maintain the interferometer in its locked state --- that is, when all the optical cavities are held on resonance through feedback control --- enabling it to operate in observation mode for gravitational-wave detection. The ESD consists of a constant voltage component --- commonly referred to as the bias --- and multiple lower-voltage components applied at four sections: upper left (UL), upper right (UR), lower left (LL), and lower right (LR). Each of these contributes independently, with voltages that typically vary in amplitude according to the control needs. Together, they enable fine adjustments to the mirror's position along the length degrees of freedom. %both length and angular degrees of freedom, including pitch and yaw. 

Along with the hourly glitch rate during O4a, Figure~\ref{fig:H1_ESD_vs_glitchrate} also presents the RMS value of the ESD voltage for the UL component at ETMX (in black). The impact of the new alignment control system is evident at the beginning of August, with a significant drop in both ESD RMS and glitch rate. Another key moment occurs in October, when both show a substantial increase, related to the reallocation of the LSC feedforward from ETMX to ETMY. The main purpose of this plot is to illustrate that significant changes in the ESD indeed affected the rate of the broadband glitches at LHO during O4a. Although other factors also contributed to transient noise in the detector and influenced the curve behavior~\cite{alog73550,alog73740}, a clear correlation can be observed between variations in the ESD RMS signal and fluctuations in the glitch rate, supporting the hypothesis of nonlinear coupling between the ESD actuation system and the detector strain output.

\begin{figure*}[ht!]
    \centering
    \begin{subfigure}{0.325\textwidth}
        \caption{}
        \includegraphics[width=\textwidth]{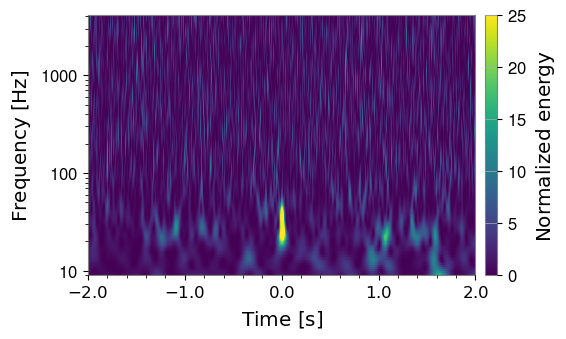}
        \label{fig:spec_H1}
    \end{subfigure}
    \begin{subfigure}{0.325\textwidth}
    \caption{}
        \includegraphics[width=\textwidth]{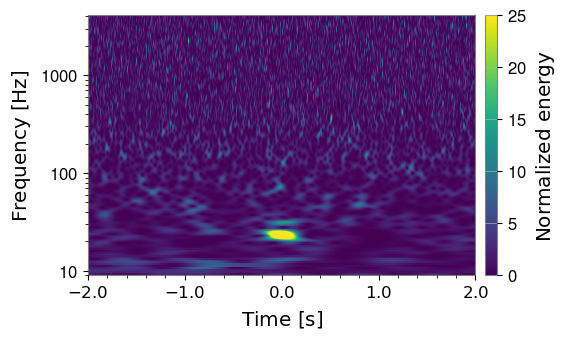}
        \label{fig:spec_H2}    \end{subfigure}
    \begin{subfigure}{0.325\textwidth}
        \caption{}
        \includegraphics[width=\textwidth]{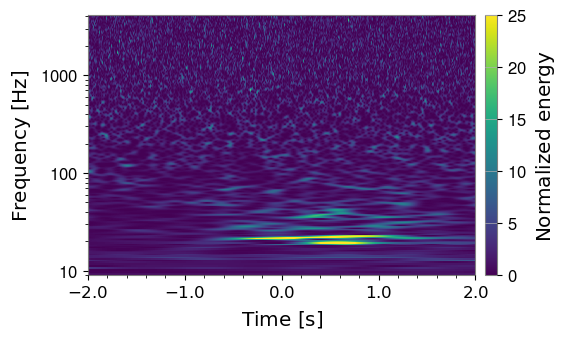}
        \label{fig:spec_H3}    \end{subfigure}
    \caption{Examples of spectrograms for each of the groups H$_1$, H$_2$, and H$_3$ identified at LHO, shown in order.}
    \label{fig:spectrograms_h1}
\end{figure*}

The glitch groups have distinct morphologies, with some becoming more dominant in certain months. The number of classes is determined by their morphological differences, as t-SNE inference primarily relies on the glitch morphology from Omicron. Despite their morphological differences, the analysis of their temporal occurrence suggests that, except for Group H$_4$, all the main glitches are likely associated with the same underlying source. Their differences lie primarily in the recurrence and morphology of specific features: vertical “blips” (Figure~\ref{fig:spec_H1}, predominantly in Group H$_1$), short-duration horizontal lines (Figure~\ref{fig:spec_H2}, common in Group H$_2$), and longer-duration lines that vary in frequency, as well as in the number and timing of repetitions (Figure~\ref{fig:spec_H3}, more frequent in Group H$_3$). These variations also explain the presence of smaller subclusters within Group H$_3$.

Glitches observed at Hanford were particularly challenging to characterize, as many were very short in duration and thus better represented using low Q-values, while others required higher Q-values to adequately capture their morphology. The inclusion of a wide range of Q-values allowed for the identification of groups with distinct morphological features, and the monthly distribution of these groups highlights periods when certain classes were more prevalent. The decrease in the number of glitches in Group H$_1$ during August and September --- associated with the implementation of the new controller --- contrasts with the increase in Group H$_2$ during the same period, reinforcing the importance of distinguishing these two groups. As mentioned before, the apparent disappearance of a glitch group does not indicate its absence; it simply reflects a period of reduced prominence. %The morphologies of Group H$_1$ glitches before and after the ETMX-to-ETMY reallocation in October were found to be very similar, which explains why this group became the most prominent once again.

\begin{figure*}[ht!]
    \begin{subfigure}{0.49\textwidth}
        \centering
        \caption{}
        \includegraphics[width=\textwidth]{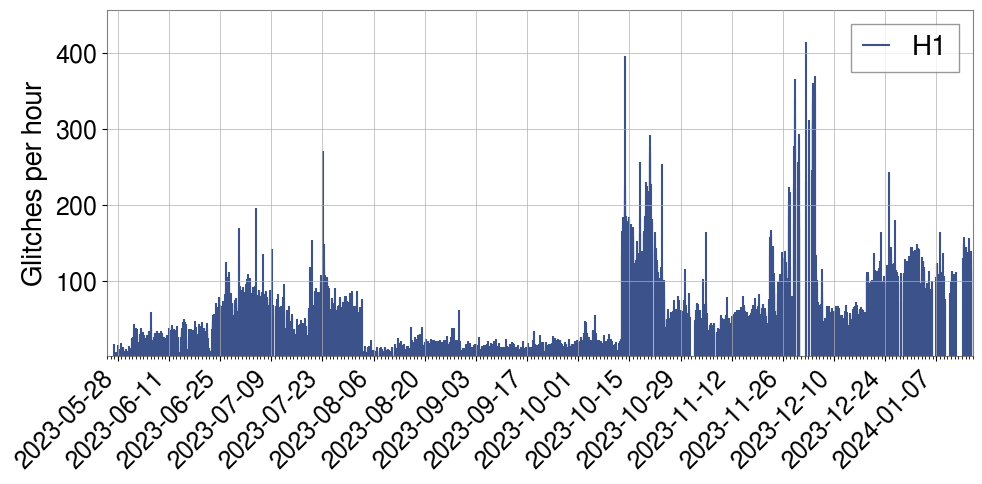}
        \label{fig:grouph1_vs_time}
    \end{subfigure}
    \hfill
    \begin{subfigure}{0.49\textwidth}
        \centering
        \caption{}
        \includegraphics[width=\textwidth]{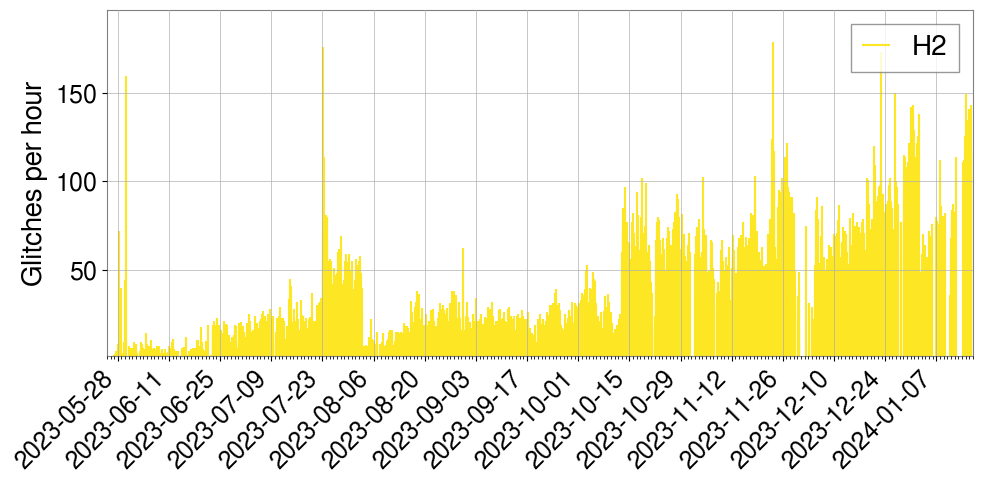}
        \label{fig:grouph2_vs_time}
    \end{subfigure}
    \hfill
    \begin{subfigure}{0.49\textwidth}
        \centering
        \caption{}
        \includegraphics[width=\textwidth]{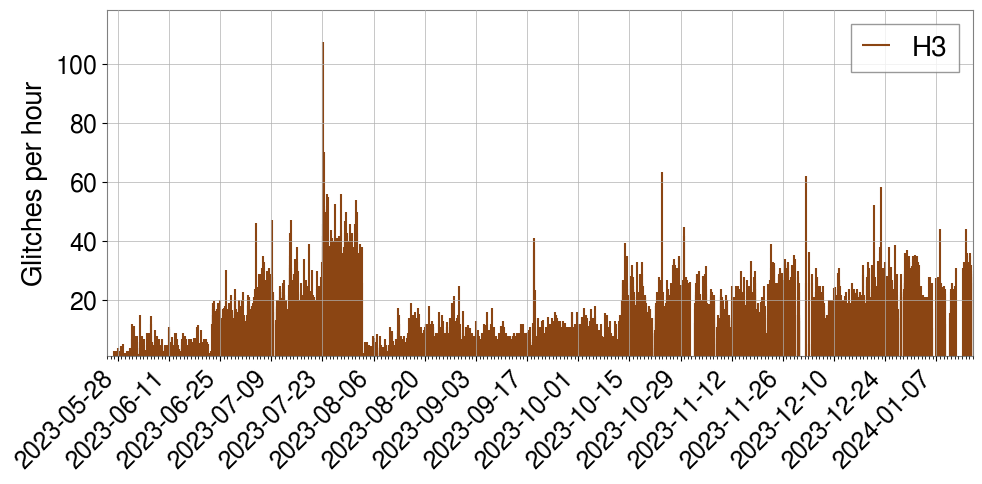}
        \label{fig:grouph3_vs_time}
    \end{subfigure}
    \hfill
    \begin{subfigure}{0.49\textwidth}
        \centering
        \caption{}
        \includegraphics[width=\textwidth]{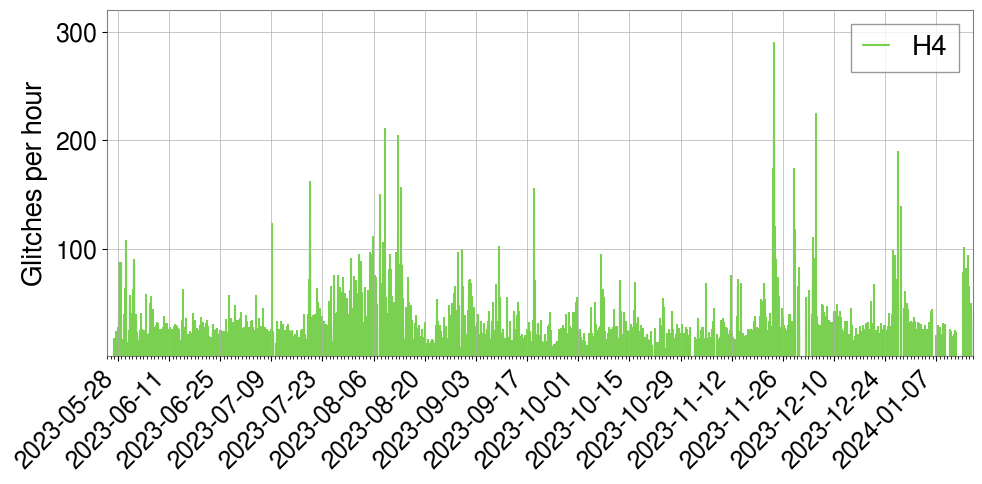}
        \label{fig:grouph4_vs_time}
    \end{subfigure}

    \caption{Hourly glitch rates during O4a at
    LHO for the four groups identified using
    t-SNE and Agglomerative Clustering.
    For this plot, we discarded data segments
    shorter than 20 minutes to avoid peaks caused
    by a high number of glitches over a short
    duration.}
    \label{fig:o4a_H1_and_H4}
\end{figure*}

Figures~\ref{fig:grouph1_vs_time}~to~\ref{fig:grouph4_vs_time} show glitch rates during the entire O4a for each group, considering all glitches detected at LHO with a minimum SNR of 6, which were classified according to the four identified groups. Only Group H$_4$ deviates from the overall trend of broadband glitches shown in Figure~\ref{fig:H1_ESD_vs_glitchrate}, as expected given its morphological composition. The y-axis scales differ across panels, highlighting the predominance of Group H$_1$, followed by Group H$_2$, consistent with the monthly t-SNE behavior.

\begin{figure*}[ht!]
    \begin{subfigure}{0.43\textwidth}
        \centering
        \caption{}
        \includegraphics[width=\textwidth]{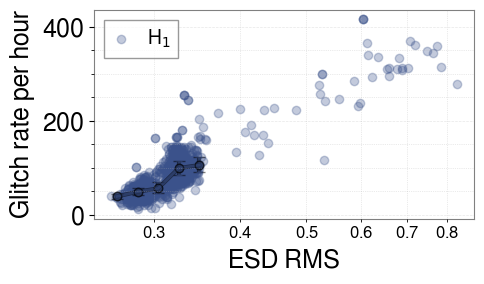}
        \label{fig:scatter_groupH1_esd}
    \end{subfigure}
    \hfill
    \begin{subfigure}{0.43\textwidth}
        \centering
        \caption{}
        \includegraphics[width=\textwidth]{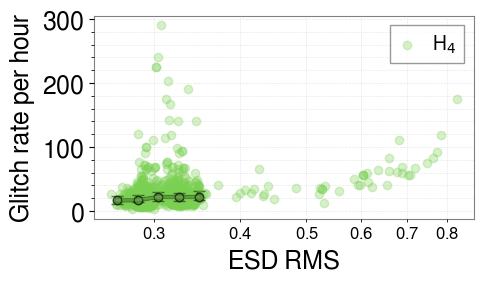}
        \label{fig:scatter_groupH4_esd}
    \end{subfigure}

    \caption{Hourly glitch rates during O4a
    at LHO for glitches assigned to (a) Group H1
    and (b) Group H4 using the t-SNE-based classification method. For these plots, data segments shorter than 20 minutes were discarded.}
    \label{fig:scatter_h1_groups}
\end{figure*}

To support the hypothesis of no correlation between the ESD and Group H$_4$, Figure~\ref{fig:scatter_groupH4_esd} shows a scatter plot of the hourly glitch rate for all O4a data classified as Group H$_4$ versus the median ESD over the same periods, revealing no apparent correlation. The curve representing the median glitch rate is shown only for bins containing at least 20 data points. In contrast, the same type of plot for all glitches classified as H$_1$ --- the most predominant broadband group --- illustrates a clear linear correlation in Figure~\ref{fig:scatter_groupH1_esd}, highlighting the distinct behavior of these two groups. 

The correlation between the glitch rate and ESD RMS for these two groups was also quantitatively estimated using the Pearson correlation coefficient ($r$)~\cite{benesty2009pearson}. The analysis confirmed a lack of correlation for Group $H_4$ ($r = 0.12$) and a strong positive correlation for Group $H_1$ with $r=0.76$. For these plots and numbers, we focus on data collected after October, when the ESD exhibited significant oscillations. Between August and October, the correlation was not evident, as the ESD remained nearly constant (see Figure~\ref{fig:H1_ESD_vs_glitchrate}). Likewise, the same plot using data prior to August does not show this correlation, which remains not fully understood, given that broadband glitches appear to originate from the same key source.

%As mentioned before, during O4a, Hanford data were primarily characterized by broadband glitches. correlated with a spectral peak around 2.6 Hz~\cite{71092} 

%Spikes correspond to loud glitches

Although these glitches affected the entire O4a, dedicated testing and efforts by the LHO team led to the implementation of a new feedback control configuration, which significantly reduced the noise and resulted in a more stable and stationary system by lowering the ESD drive voltage. This improved configuration was adopted for the O4b observing run, which began with no recurrence of these glitches~\cite{alog74887,alog74939,alog76449,alog76459}. 

\clearpage
\FloatBarrier
\section{Conclusions and Discussion}
\label{sec:conclusions}
This paper presents an analysis of the main glitch groups observed during O4a at the LIGO observatories. We employed a fully unsupervised machine learning framework, based on t-distributed Stochastic Neighbor Embedding (t-SNE) followed by Agglomerative Clustering, to investigate their behavior. This approach reduces the high-dimensional representation of glitches, built from unclustered Omicron data across all Q-values, into a two-dimensional space where morphological similarities emerge. By applying this methodology independently to data from LLO and LHO, we were able to identify dominant glitch populations, track their evolution over time, and correlate their occurrence with environmental or instrumental conditions.

At Livingston, the most prominent glitch populations during O4a were closely correlated with seasonal variations in ground motion, particularly within the 0.1–\SI{1}{Hz} frequency range. The applied methodology identified three main groups. Two of them --- characterized by arch-like spectrogram morphologies, with peak frequencies around \SI{12}{Hz} and \SI{20}{Hz}, respectively --- were dominant during months of elevated ground motion, which began to increase in the autumn and peaked during the winter in the United States. These two groups exhibited a strong positive correlation with ground velocity in both the lower (0.1–\SI{0.3}{Hz}) and higher (0.3–\SI{1}{Hz}) microseismic bands, but not in the anthropogenic band (1–\SI{3}{Hz}). By contrast, the third group, composed primarily of known classes such as \textit{Blip}, \textit{Tomte}, and \textit{Koi Fish}, stood out during periods of low ground motion and showed no significant dependence on any of the seismic frequency bands tested.

At Hanford, a different glitch landscape was observed. The detector experienced a high occurrence of broadband glitches in the 20–\SI{50}{Hz} range, particularly at lower SNR. These glitches, persistent throughout O4a, were shown to be associated with the electrostatic drive (ESD) system, as indicated by their correlation with the RMS of the applied ESD voltage.
In total, four groups were identified using our methodology: three were related to the ESD, and one was not --- the latter consisting mainly of known glitch classes such as \textit{Extremely Loud}, \textit{Blip}, and \textit{Low Frequency Lines}. While the three ESD-related groups appeared to share a common instrumental origin, they exhibited distinct morphologies. We also found evidence that some detector tests affected only one of the glitch groups while leaving the others unchanged. This observation highlights the importance of analyzing the groups separately --- an analysis made possible by the methodology applied in this work.

The technique aims to uncover the monthly behavior of glitches in gravitational-wave detectors such as LIGO, Virgo, and KAGRA. The only requirement is access to Omicron data --- or a similar tool capable of identifying triggers that encode glitch morphology. Once the glitches are tracked, the ideal next step --- as demonstrated in this study --- is to examine relevant detector logbook entries that document changes or interventions at the observatories, in order to correlate the behavior of glitch groups with those modifications.
By analyzing monthly trends during O4a, we were able to correlate certain glitch groups with environmental or instrumental conditions, providing strong indications of their possible origin. As discussed in the methods paper~\cite{ferreira2025using}, this approach can be applied at different timescales depending on the specific goals. Although the methodology autonomously identifies glitch groups, human interpretation --- along with association to auxiliary channels and documented changes in the observatories --- remains essential to understanding their physical origin.

%which varies to keep the variation of the glitch rate can be seen , at least in part, to the seasonal increase in ground motion during the winter months. Higher ground motion requires the  ESD to exert greater corrective forces, which may, in turn, introduce additional noise into the detector data.  (...)

\begin{acknowledgments}
We would like to thank Elenna Capote and Jess McIver, for their very useful insights, comments, and suggestions. We express our gratitude to the members of the LIGO Detector Characterization Group and the LIGO-Virgo-KAGRA (LVK) collaboration for their valuable discussions. This work was supported by the National Science Foundation under grant numbers PHY-2110509 and PHY-2409740. The material is based upon work supported by NSF's LIGO Laboratory, a major facility fully funded by the National Science Foundation. Additionally, the work utilizes the LIGO computing clusters and data from the Advanced LIGO detectors; the authors are grateful for the computational resources provided by the LIGO Laboratory, supported by National Science Foundation Grants PHY-0757058 and PHY-0823459.
\end{acknowledgments}

\appendix
\section[\appendixname~\thesection]{Ground motion measurements}
\label{app:ground_motion}
This appendix presents a comparison of measurements from seismometers located near the test masses at the LIGO Livingston observatory. Two months were selected for this analysis: June, which exhibits relatively low microseismic motion, and December, which shows high ground motion. The seismometer channels follow the naming convention shown in Table~\ref{tab:seismometer}, where each channel name consists of the component identifier (e.g., ETM), representing the selected test mass, followed by the corresponding arm, and the motion direction: $\hat{x}$ for motion along the laser direction, $\hat{y}$ for motion perpendicular to the laser, and  $\hat{z}$ for vertical motion, perpendicular to the $x$–$y$ plane. 

\begin{table}[ht]
    \centering
    \caption{Seismometer channels and corresponding position, arm, and motion directions.}
    \begin{tabular}{lccc}
        \hline
        \textbf{Channel Name} & \textbf{Position} & \textbf{Arm} & \textbf{Motion Direction} \\
        \hline
        ITM\textcolor{violet}{X}\_\textcolor{darkorange}{X}/\textcolor{darkgreen}{Y}/\textcolor{darkred}{Z} & Input Test Mass & \textcolor{violet}{X} & \textcolor{darkorange}{$\hat{\textbf{x}}$} / \textcolor{darkgreen}{$\hat{\textbf{y}}$} / \textcolor{darkred}{$\hat{\textbf{z}}$} \\
        ITM\textcolor{violet}{Y}\_\textcolor{darkorange}{X}/\textcolor{darkgreen}{Y}/\textcolor{darkred}{Z} & Input Test Mass & \textcolor{violet}{Y} & \textcolor{darkorange}{$\hat{\textbf{x}}$} / \textcolor{darkgreen}{$\hat{\textbf{y}}$} / \textcolor{darkred}{$\hat{\textbf{z}}$} \\
        ETM\textcolor{violet}{X}\_\textcolor{darkorange}{X}/\textcolor{darkgreen}{Y}/\textcolor{darkred}{Z} & End Test Mass   & \textcolor{violet}{X} & \textcolor{darkorange}{$\hat{\textbf{x}}$} / \textcolor{darkgreen}{$\hat{\textbf{y}}$} / \textcolor{darkred}{$\hat{\textbf{z}}$} \\
        ETM\textcolor{violet}{Y}\_\textcolor{darkorange}{X}/\textcolor{darkgreen}{Y}/\textcolor{darkred}{Z} & End Test Mass   & \textcolor{violet}{Y} & \textcolor{darkorange}{$\hat{\textbf{x}}$} / \textcolor{darkgreen}{$\hat{\textbf{y}}$} / \textcolor{darkred}{$\hat{\textbf{z}}$} \\
        \hline
    \end{tabular}
    \label{tab:seismometer}
\end{table}

\begin{figure*}[htbp]
    \centering
    {\large\textbf{ITMX: Comparison of motion directions}\par\vspace{0.7em}}
    
    \begin{subfigure}[b]{0.32\textwidth}
        \caption{June: 0.1–0.3 Hz}
        \centering
        \includegraphics[width=\textwidth]{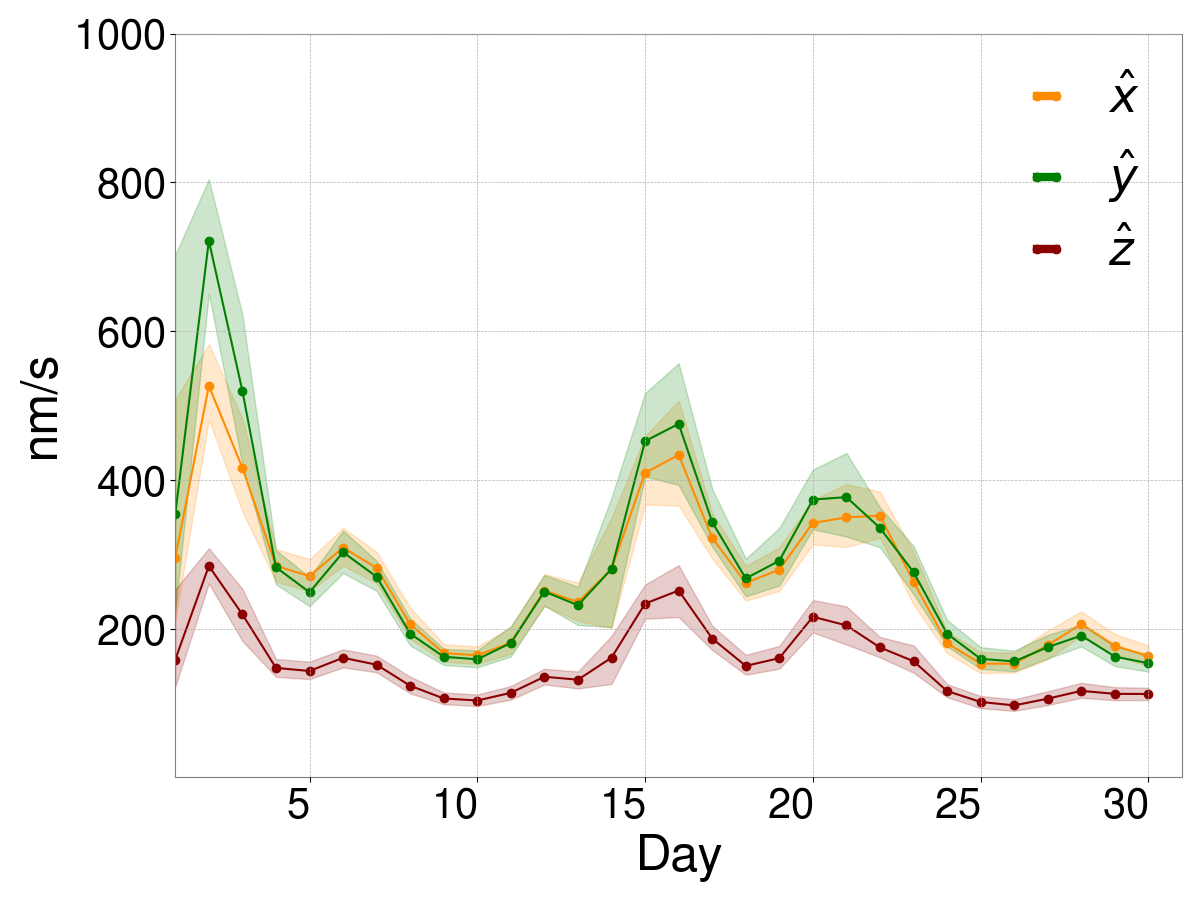}
    \end{subfigure}
    \hfill
    \begin{subfigure}[b]{0.32\textwidth}
        \caption{June: 0.3–1.0 Hz}
        \centering
        \includegraphics[width=\textwidth]{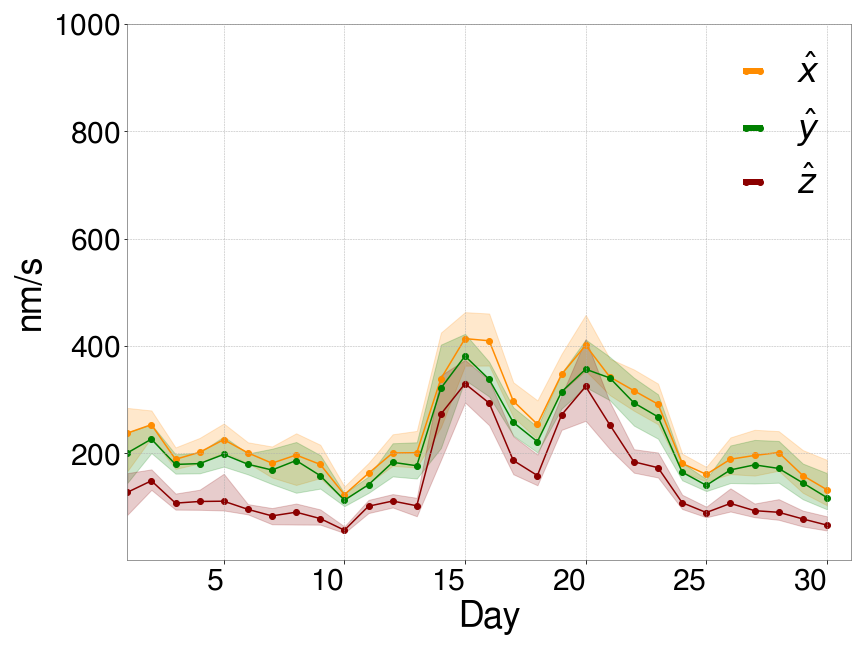}
    \end{subfigure}
    \hfill
    \begin{subfigure}[b]{0.32\textwidth}
        \caption{June: 1.0–3.0 Hz}
        \centering
        \includegraphics[width=\textwidth]{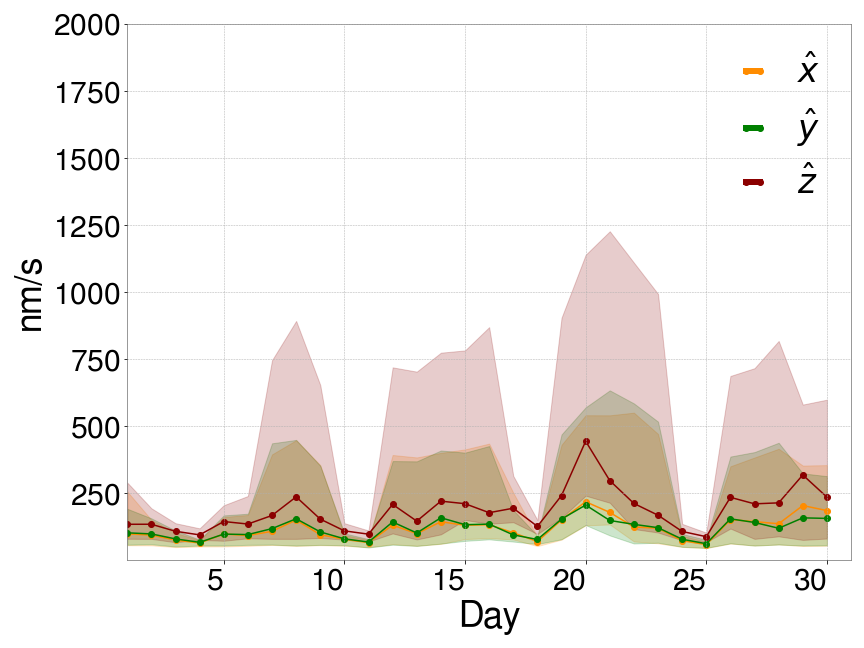}
    \end{subfigure}

    %\vskip\baselineskip

    \begin{subfigure}[b]{0.32\textwidth}
        \caption{December: 0.1–0.3 Hz}
        \centering
        \includegraphics[width=\textwidth]{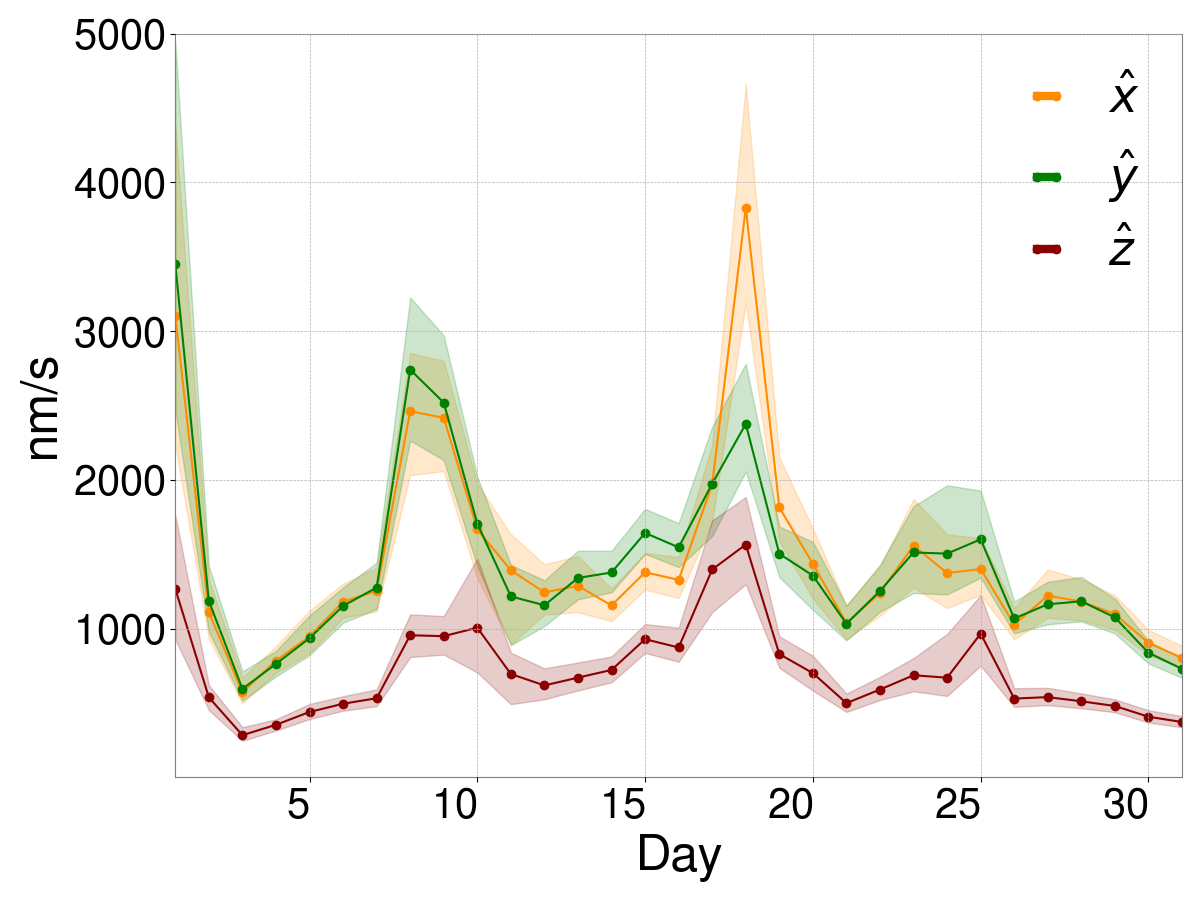}
    \end{subfigure}
    \hfill
    \begin{subfigure}[b]{0.32\textwidth}
        \caption{December: 0.3–1.0 Hz}
        \centering
        \includegraphics[width=\textwidth]{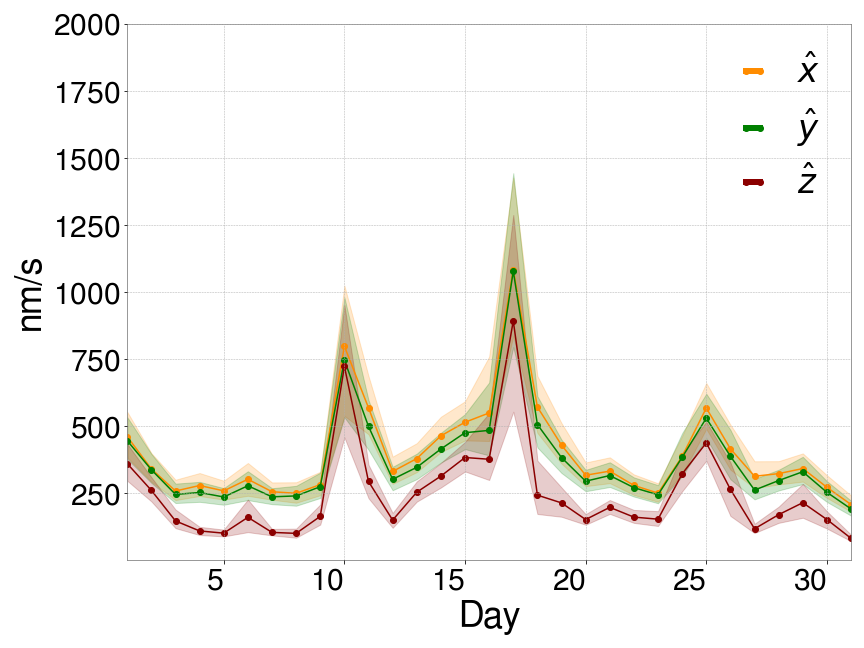}
    \end{subfigure}
    \hfill
    \begin{subfigure}[b]{0.32\textwidth}
        \caption{December: 1.0–3.0 Hz}
        \centering
        \includegraphics[width=\textwidth]{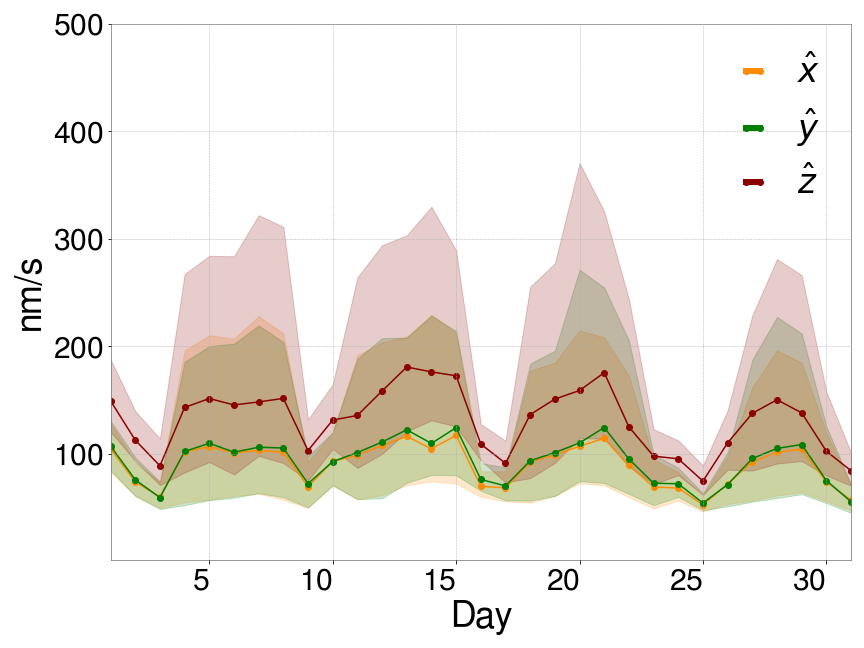}
    \end{subfigure}

    \caption{Comparison of motion directions at ITMX during June and December across three frequency bands.}
    \label{fig:comparison_itmx}
\end{figure*}

\begin{figure*}[htbp]
    \centering
    {\large\textbf{ETMX: Comparison of motion directions}\par\vspace{0.7em}}
    
    \begin{subfigure}[b]{0.32\textwidth}
        \caption{June: 0.1–0.3 Hz}
        \centering
        \includegraphics[width=\textwidth]{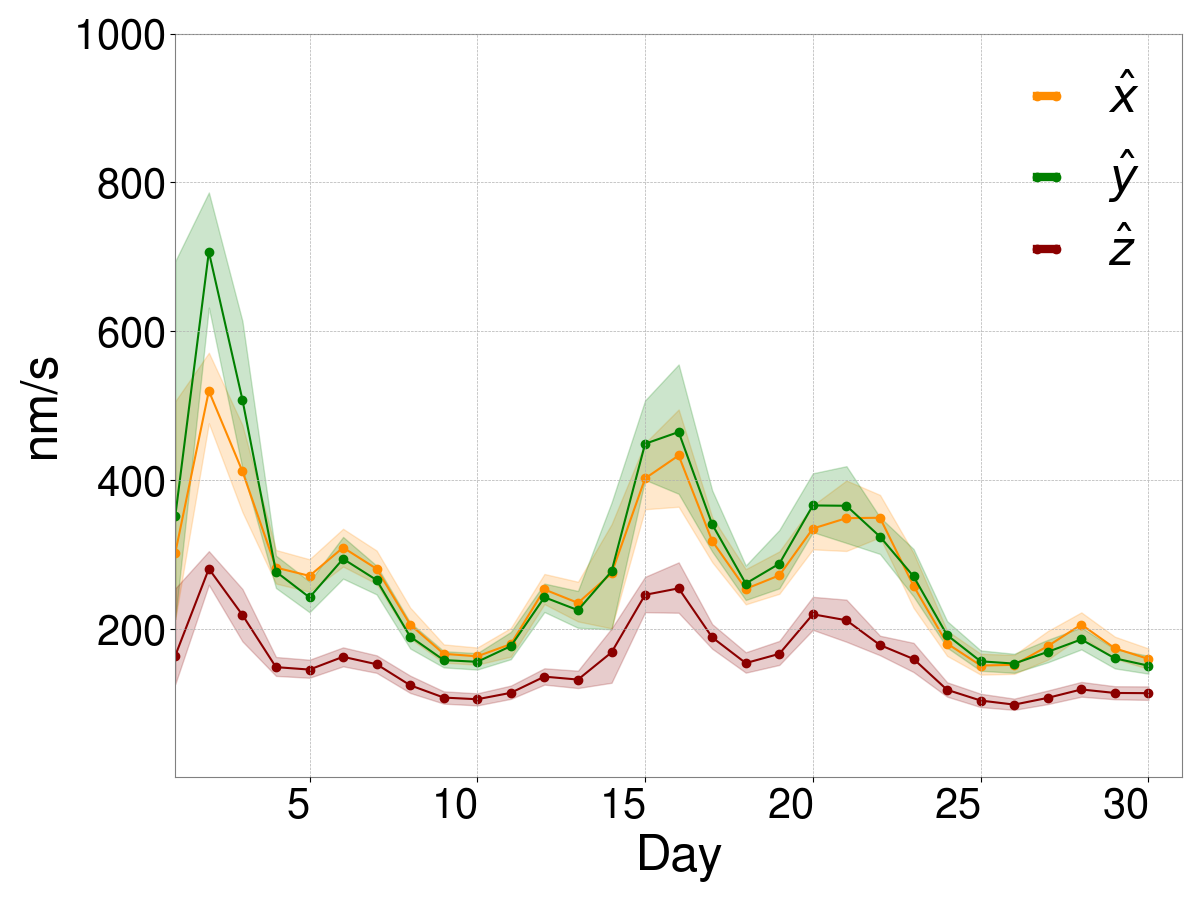}
    \end{subfigure}
    \hfill
    \begin{subfigure}[b]{0.32\textwidth}
        \caption{June: 0.3–1.0 Hz}
        \centering
        \includegraphics[width=\textwidth]{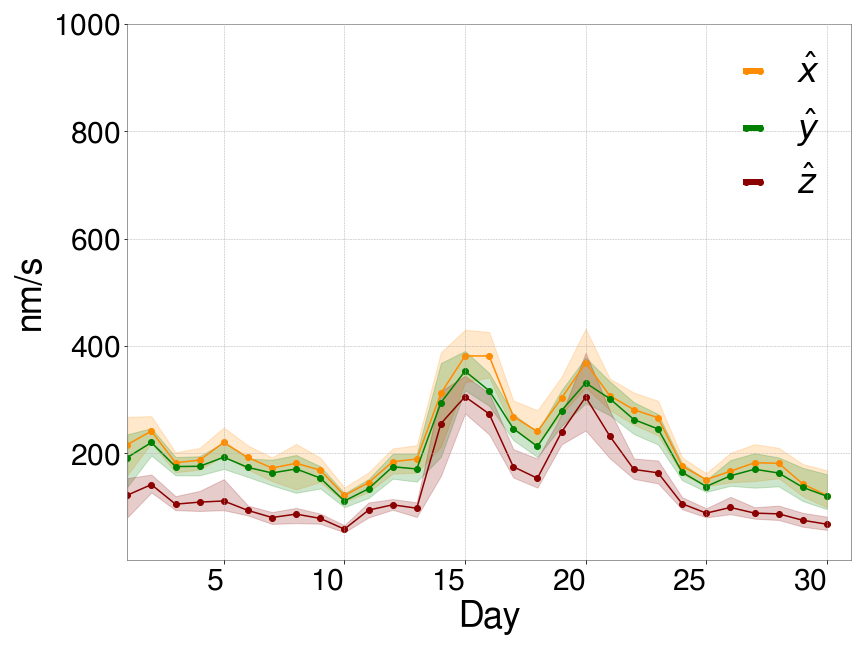}
    \end{subfigure}
    \hfill
    \begin{subfigure}[b]{0.32\textwidth}
        \caption{June: 1.0–3.0 Hz}
        \centering
        \includegraphics[width=\textwidth]{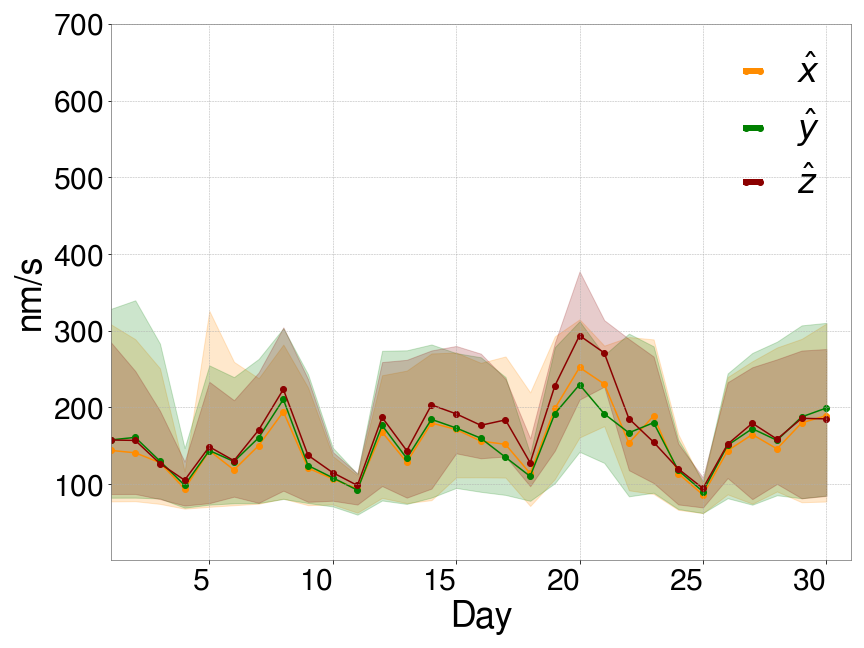}
    \end{subfigure}

    \begin{subfigure}[b]{0.32\textwidth}
        \caption{December: 0.1–0.3 Hz}
        \centering
        \includegraphics[width=\textwidth]{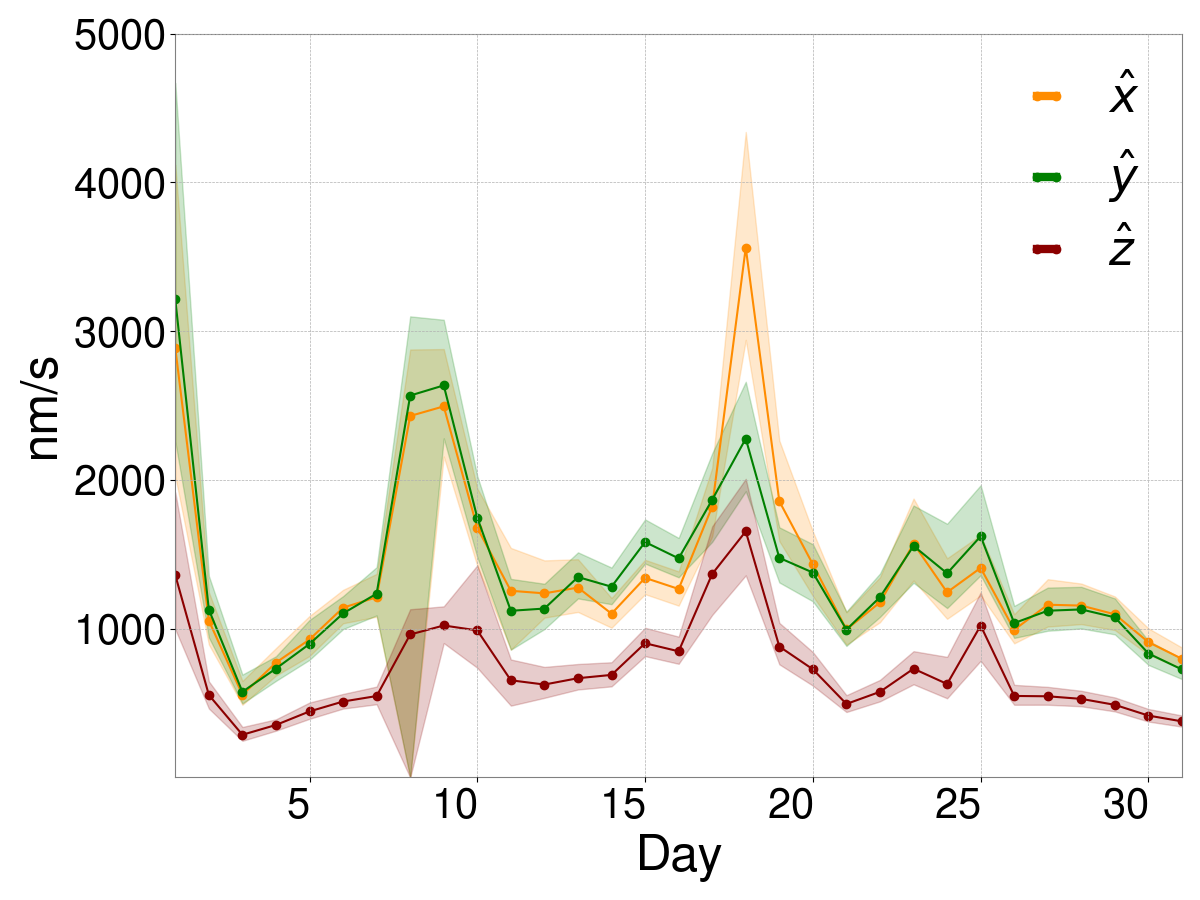}
    \end{subfigure}
    \hfill
    \begin{subfigure}[b]{0.32\textwidth}
        \caption{December: 0.3–1.0 Hz}
        \centering
        \includegraphics[width=\textwidth]{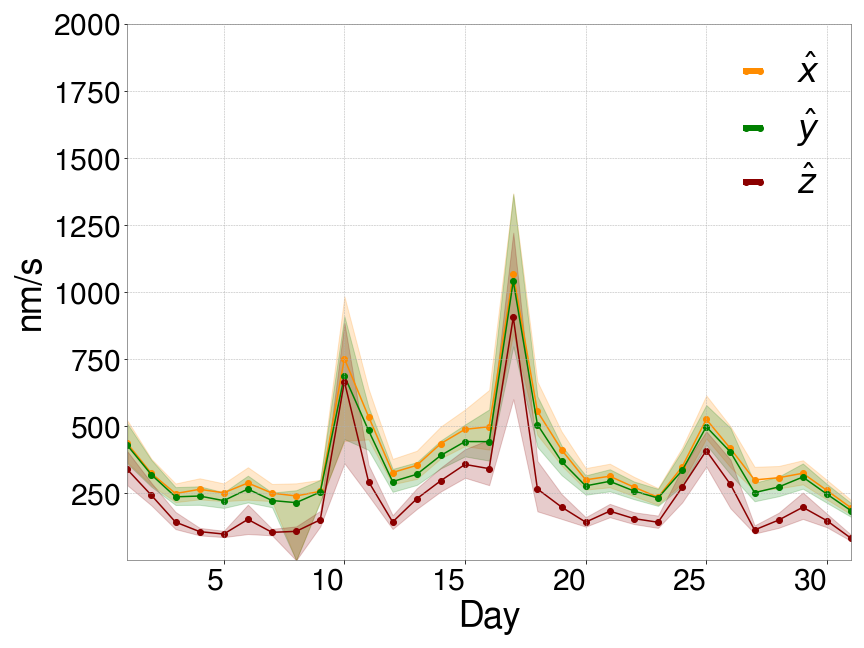}
    \end{subfigure}
    \hfill
    \begin{subfigure}[b]{0.32\textwidth}
        \caption{December: 1.0–3.0 Hz}
        \centering
        \includegraphics[width=\textwidth]{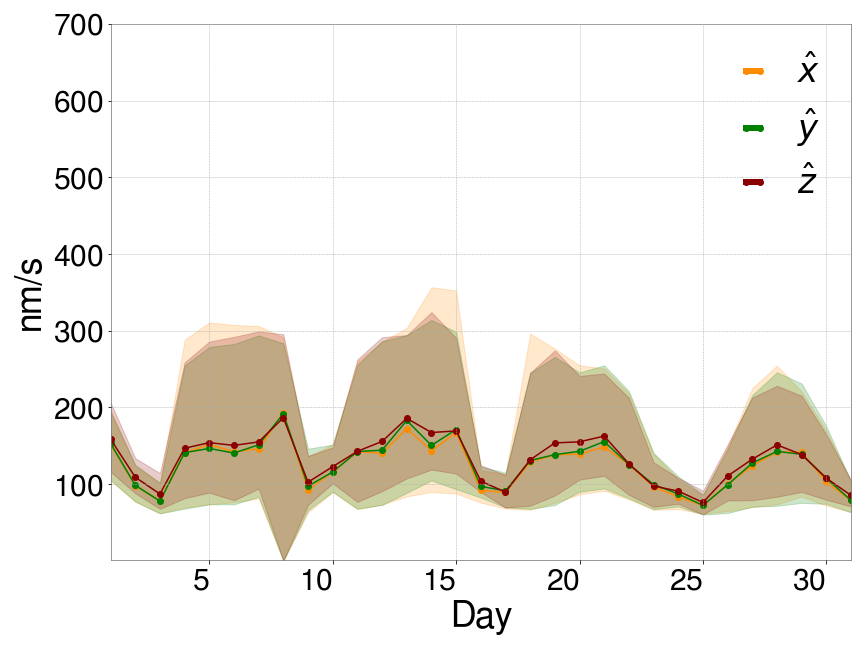}
    \end{subfigure}

    \caption{Comparison of motion directions at ETMX during June and December across three frequency bands}
    \label{fig:comparison_etmx}
\end{figure*}

\begin{figure*}[htbp]
    \centering
    {\large\textbf{ETMY: Comparison of motion directions}\par\vspace{0.7em}}

    \begin{subfigure}[b]{0.32\textwidth}
        \caption{June: 0.1–0.3 Hz}
        \centering
        \includegraphics[width=\textwidth]{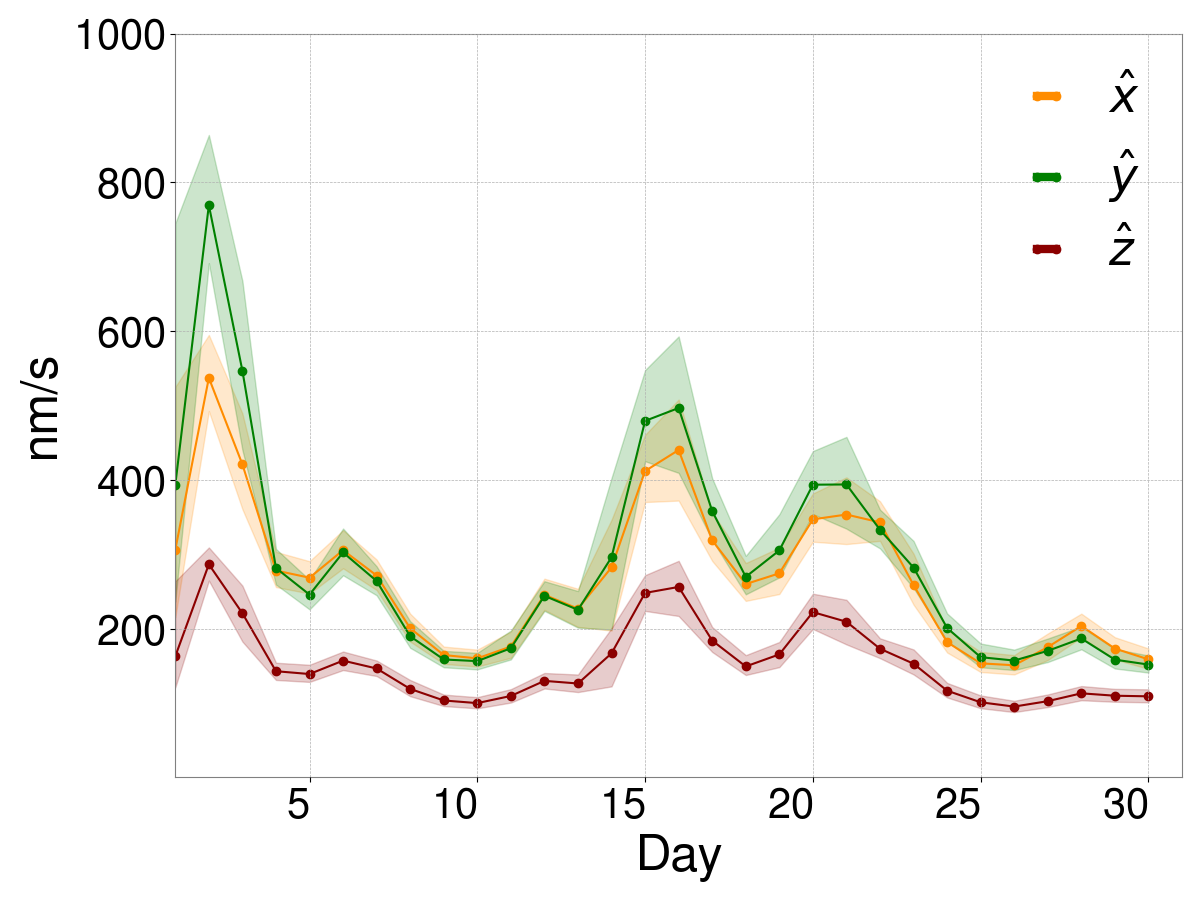}
    \end{subfigure}
    \hfill
    \begin{subfigure}[b]{0.32\textwidth}
        \caption{June: 0.3–1.0 Hz}
        \centering
        \includegraphics[width=\textwidth]{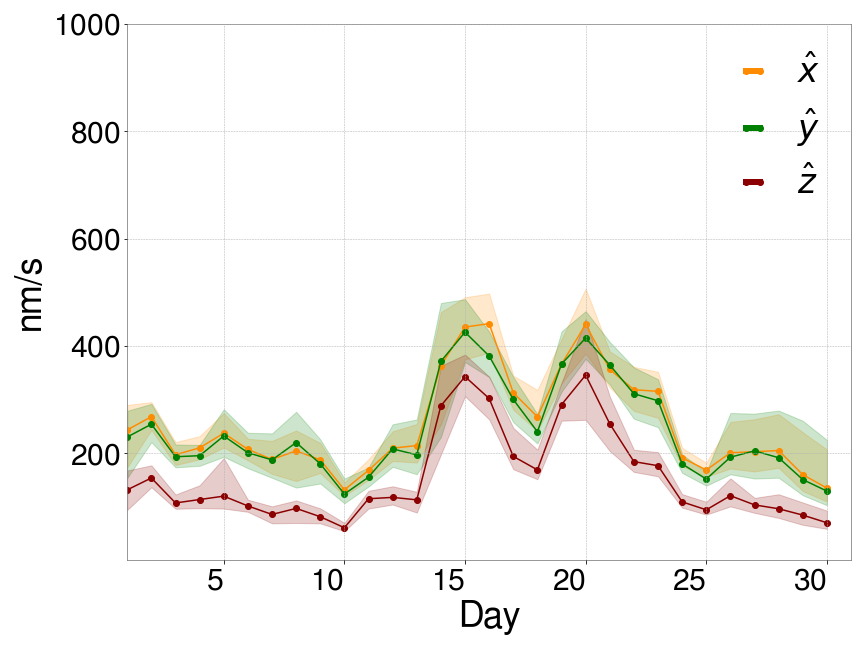}
    \end{subfigure}
    \hfill
    \begin{subfigure}[b]{0.32\textwidth}
        \caption{June: 1.0–3.0 Hz}
        \centering
        \includegraphics[width=\textwidth]{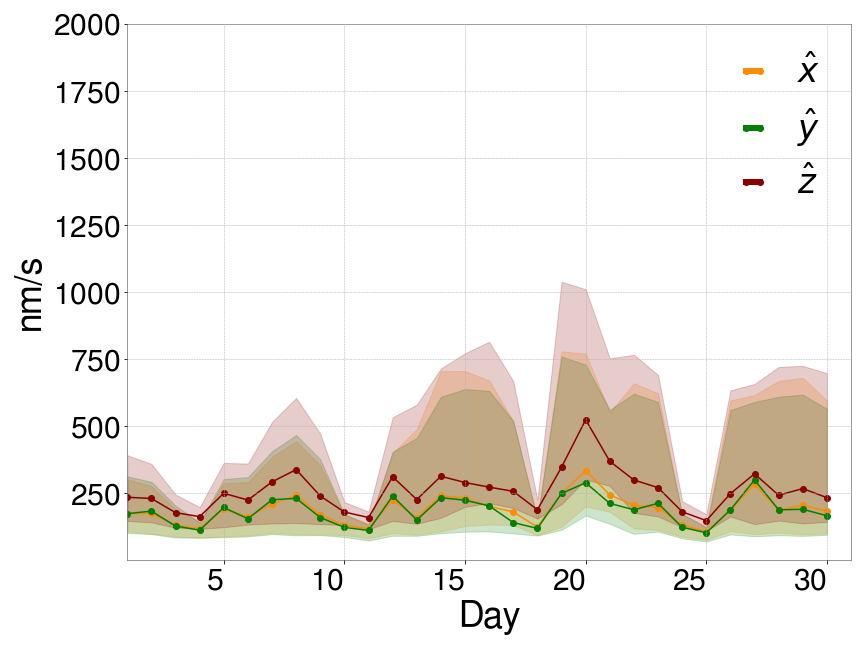}
    \end{subfigure}

    \begin{subfigure}[b]{0.32\textwidth}
        \caption{December: 0.1–0.3 Hz}
        \centering
        \includegraphics[width=\textwidth]{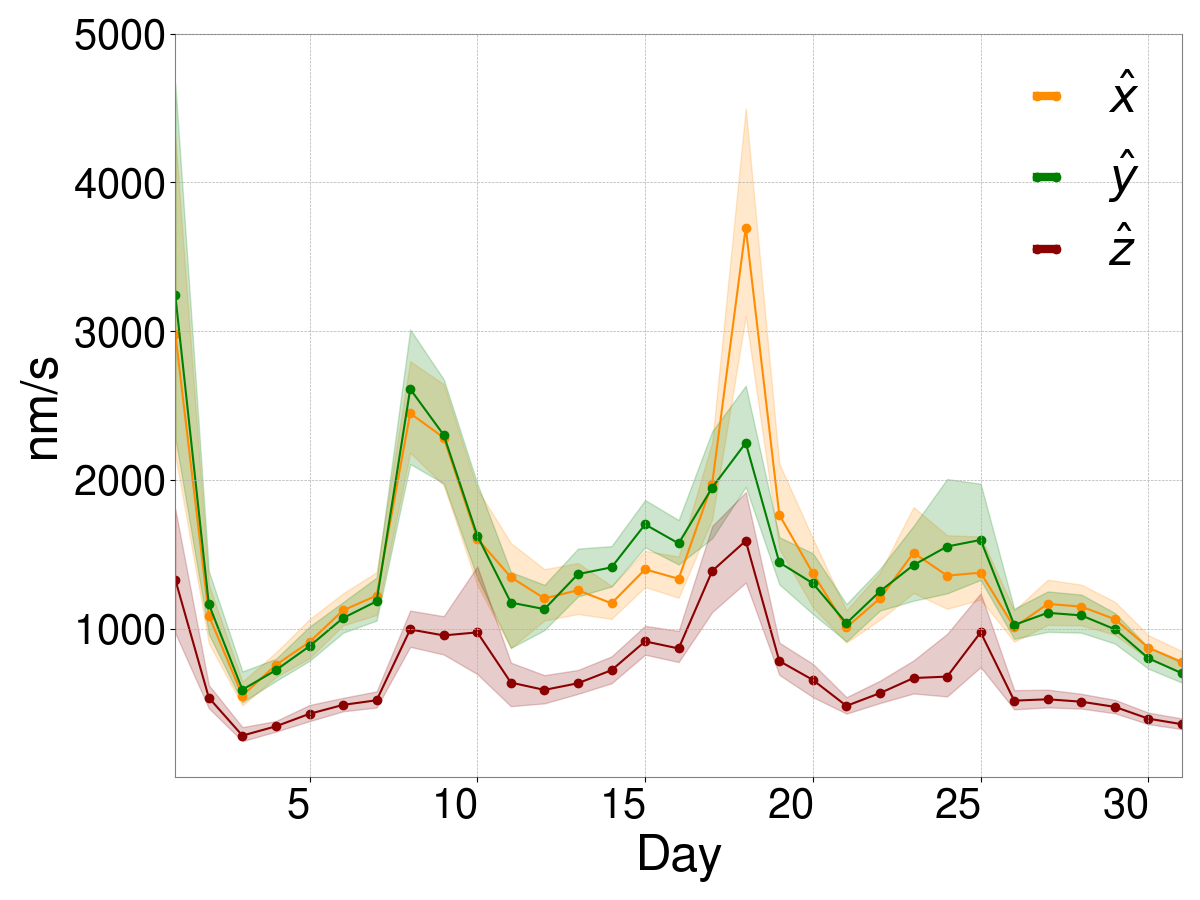}
    \end{subfigure}
    \hfill
    \begin{subfigure}[b]{0.32\textwidth}
        \caption{December: 0.3–1.0 Hz}
        \centering
        \includegraphics[width=\textwidth]{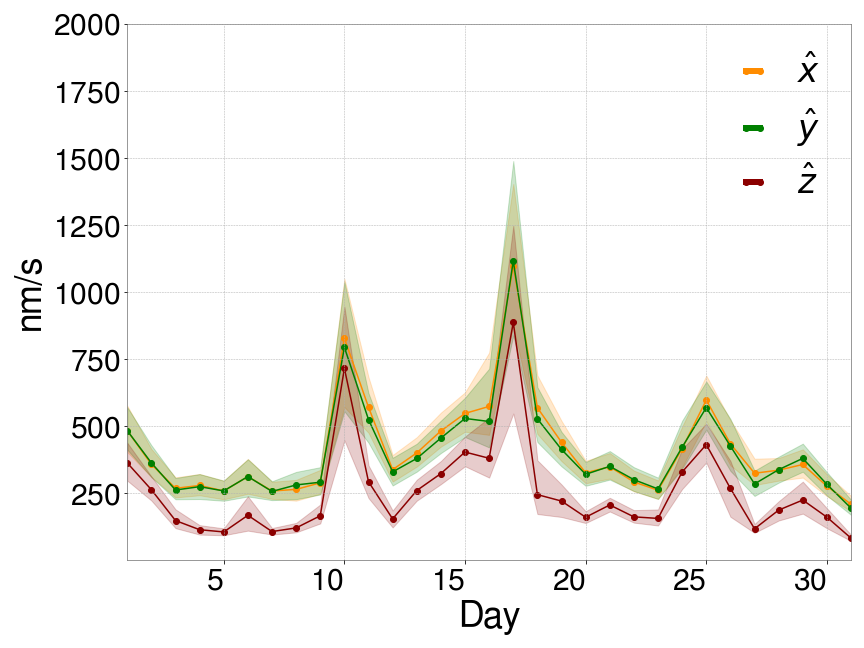}
    \end{subfigure}
    \hfill
    \begin{subfigure}[b]{0.32\textwidth}
        \caption{December: 1.0–3.0 Hz}
        \centering
        \includegraphics[width=\textwidth]{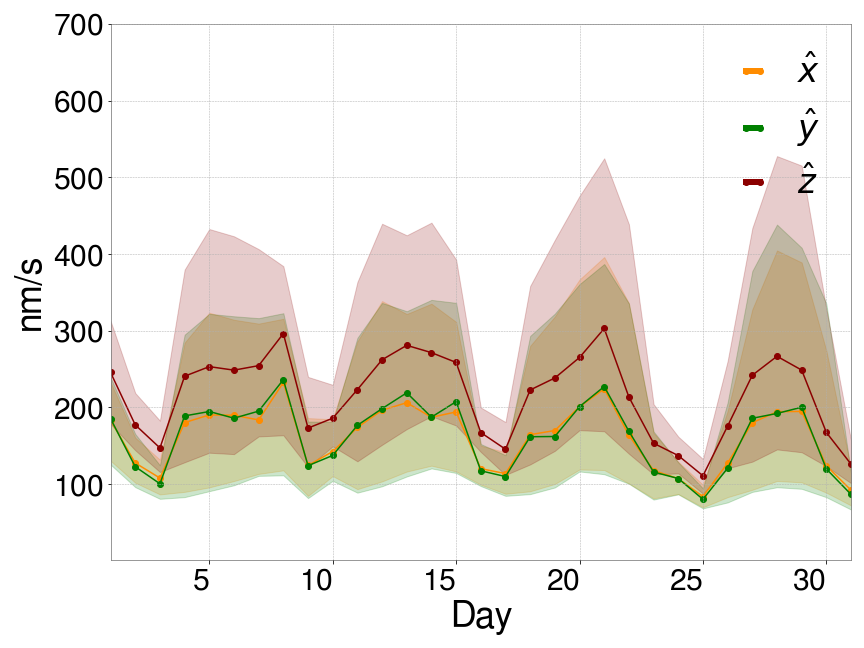}
    \end{subfigure}

    \caption{Comparison of motion directions at ETMY during June and December across three frequency bands}
    \label{fig:comparison_etmy}
\end{figure*}

Figures~\ref{fig:comparison_itmx}, \ref{fig:comparison_etmx}, and \ref{fig:comparison_etmy} show a comparison of ground motion in the three spatial directions for ITMX, ETMX, and ETMY, respectively. For each seismometer, the root mean square (RMS) velocity data are presented across three frequency bands: LMS (lower microseismic band, 0.1–0.3~Hz), HMS (higher microseismic band, 0.3–1.0Hz), and anthropogenic band (1.0–3.0~Hz). Each data point represents the daily median velocity measured by the seismometer, and the shaded regions indicate the first and third quartiles. 

By evaluating all days in both months, we observe that, in general, the motion in the three directions follows similar trends, suggesting that any directional component can be informative when analyzing behavioral variations. In terms of amplitude, for the LMS and HMS bands, motion in the $\hat{x}$ and $\hat{y}$ directions is similar and generally larger than in the $\hat{z}$ direction. Conversely, in the anthropogenic band, the $\hat{z}$ direction exhibits higher amplitudes and larger variations.

\begin{figure*}[htbp]
    \centering
    {\large\textbf{ITMX vs. ITMY}\par\vspace{0.7em}}
    \begin{subfigure}[b]{0.32\textwidth}
        \caption{June: 0.1–0.3 Hz}
        \centering
        \includegraphics[width=\textwidth]{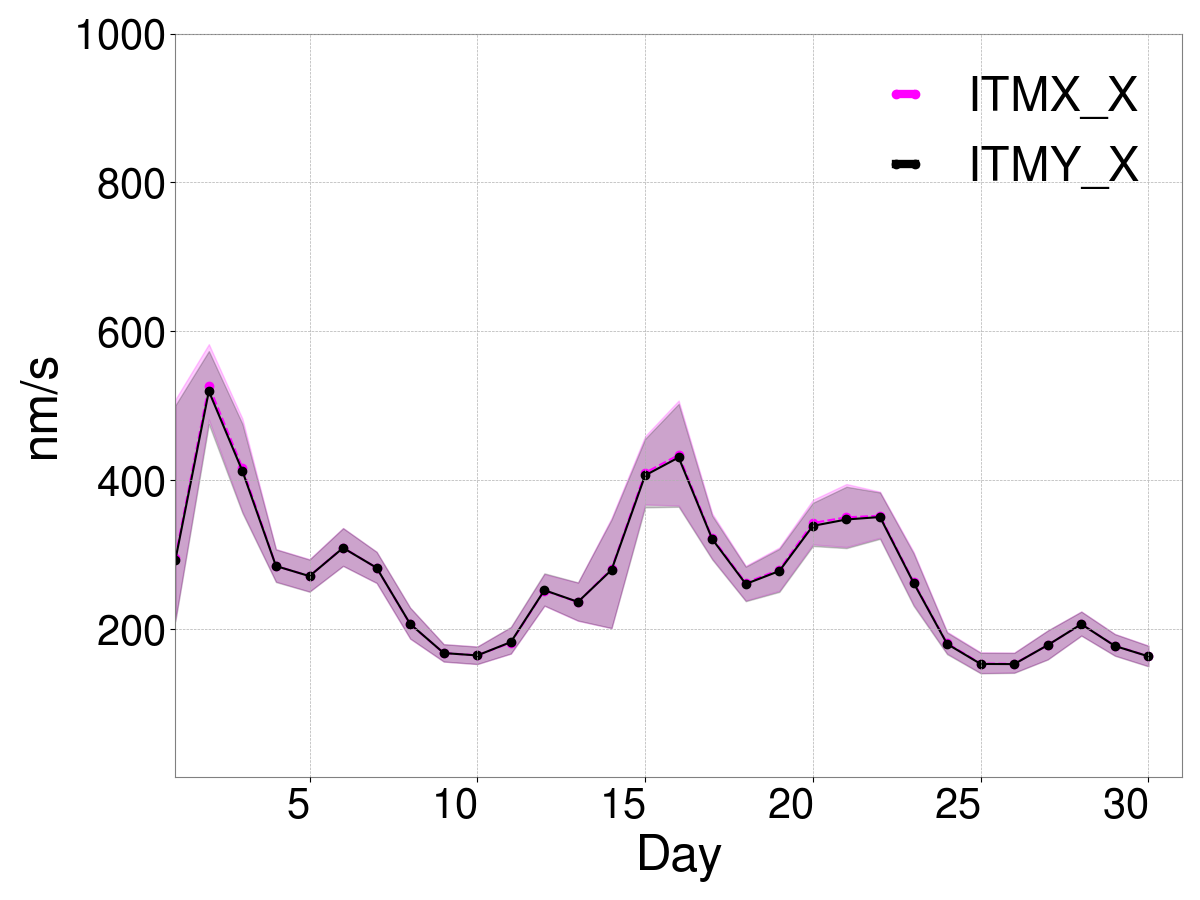}
    \end{subfigure}
    \hfill
    \begin{subfigure}[b]{0.32\textwidth}
        \caption{June: 0.3–1.0 Hz}
        \centering
        \includegraphics[width=\textwidth]{figures/comparison_TMsjune_L1_ISI-GND_STS_ITMX_X_BLRMS_100M_300M.png}
    \end{subfigure}
    \hfill
    \begin{subfigure}[b]{0.32\textwidth}
        \caption{June: 1.0–3.0 Hz}
        \centering
        \includegraphics[width=\textwidth]{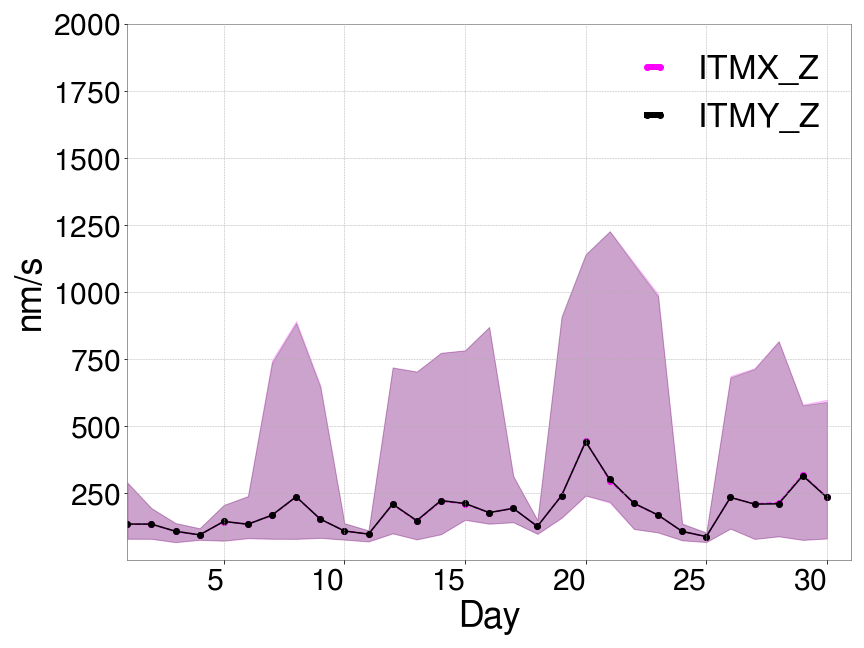}
    \end{subfigure}

    \caption{Comparison between ITMX and ITMY during June.}
    \label{fig:comparison_itmx_itmy}
\end{figure*}

\begin{figure*}[htbp]
    \centering
    {\large\textbf{ETMX vs. ETMY}\par\vspace{0.7em}}
    \begin{subfigure}[b]{0.32\textwidth}
        \caption{June: 0.1–0.3 Hz}
        \centering
        \includegraphics[width=\textwidth]{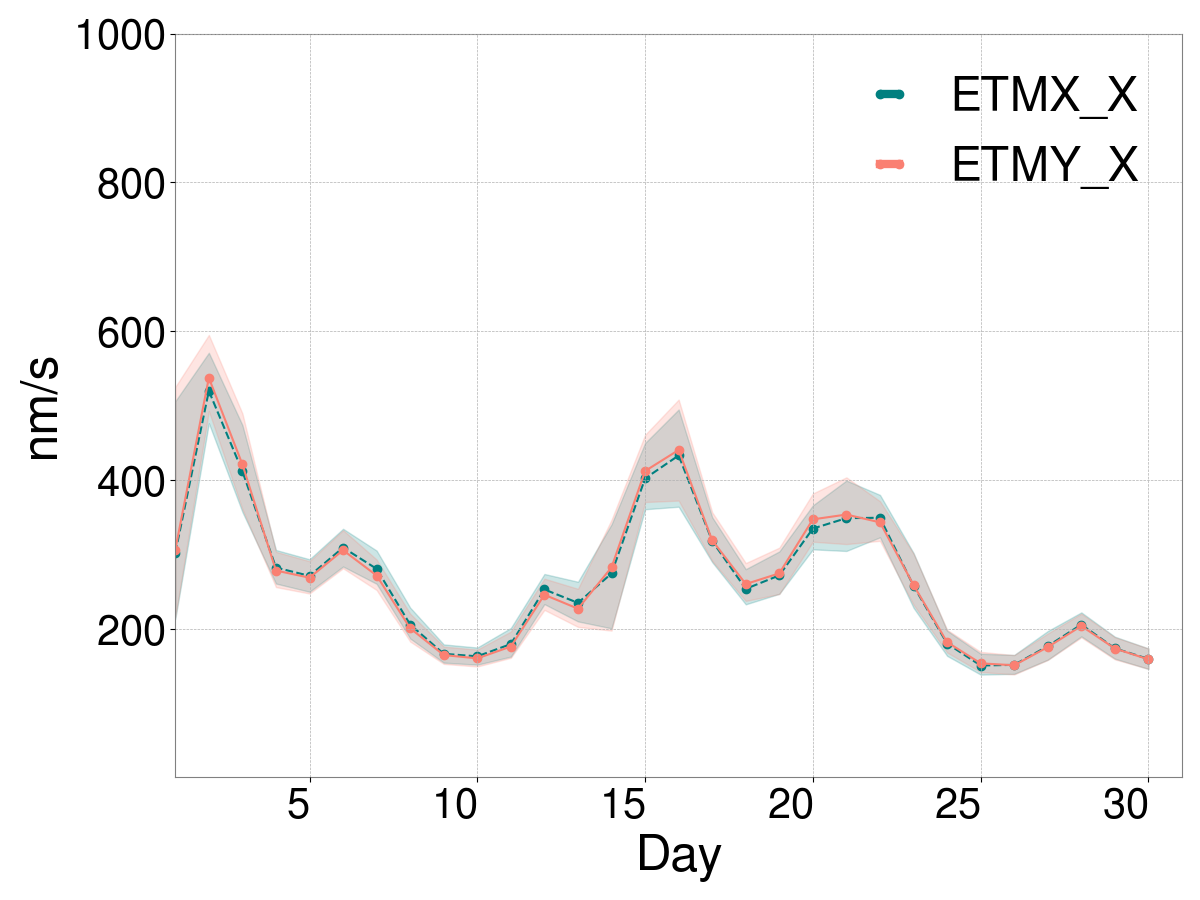}
    \end{subfigure}
    \hfill
    \begin{subfigure}[b]{0.32\textwidth}
        \caption{June: 0.3–1.0 Hz}
        \centering
        \includegraphics[width=\textwidth]{figures/comparison_TMsjune_L1_ISI-GND_STS_ETMX_X_BLRMS_100M_300M.png}
    \end{subfigure}
    \hfill
    \begin{subfigure}[b]{0.32\textwidth}
        \caption{June: 1.0–3.0 Hz}
        \centering
        \includegraphics[width=\textwidth]{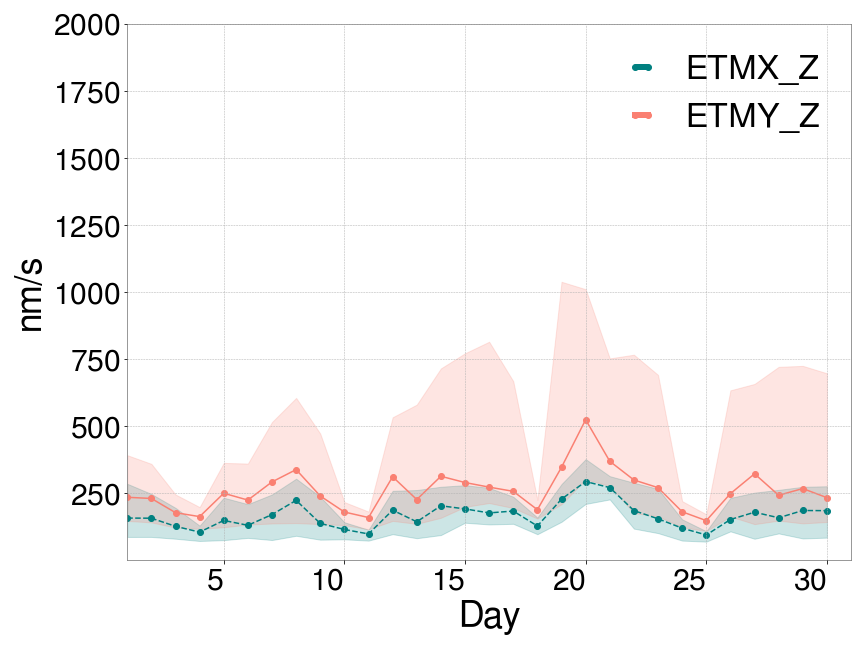}
    \end{subfigure}
    \caption{Comparison between ETMX and ETMY during June.}
    \label{fig:comparison_etmx_etmy}
\end{figure*}

We also present comparisons between seismometers located at different positions. After observing that the $\hat{x}$ and $\hat{y}$ directions show very similar behavior in both the LMS and HMS bands, we selected $\hat{x}$ as representative for these two, and $\hat{z}$ for the anthropogenic band. Only one month (June) is presented, as both months lead to the same conclusions. Figure~\ref{fig:comparison_itmx_itmy} shows a comparison between ITMX and ITMY using the same daily median data. As expected, the signals are nearly identical due to the proximity of the sensors. Figure~\ref{fig:comparison_etmx_etmy} compares ETMX and ETMY, which exhibit very similar median amplitudes and variations for the LMS and HMS bands. In the anthropogenic band, however, ETMY shows higher median values and larger variations than ETMX, likely due to its closer proximity to nearby roads.

\nocite{*}

\FloatBarrier
\bibliography{references}% Produces the bibliography via BibTeX.

\end{document}